\documentclass[12pt,letterpaper]{article}

\usepackage{verbatim}
\usepackage{amsmath}
\usepackage{amssymb}
\usepackage{graphicx}
\usepackage{cite}
\usepackage{subfig}
\usepackage{setspace}
\usepackage[percent]{overpic}

\def\Re{{\cal R \mskip-4mu \lower.1ex \hbox{\it e}\,}}
\def\Im{{\cal I \mskip-5mu \lower.1ex \hbox{\it m}\,}}
\def\ie{{\it i.e.}}
\def\eg{{\it e.g.}}

\def\sub#1{_{\lower.25ex\hbox{$\scriptstyle#1$}}}

\def\to{\rightarrow}

\def\subw{_{\rm w}}
\def\mh{\ifmmode m\sbl H \else $m\sbl H$\fi}
\def\mch{\ifmmode m_{H^\pm} \else $m_{H^\pm}$\fi}
\def\mt{\ifmmode m_t\else $m_t$\fi}
\def\mc{\ifmmode m_c\else $m_c$\fi}
\def\mz{\ifmmode M_Z\else $M_Z$\fi}
\def\mw{\ifmmode M_W\else $M_W$\fi}
\def\mws{\ifmmode M_W^2 \else $M_W^2$\fi}
\def\mhs{\ifmmode m_H^2 \else $m_H^2$\fi}   
\def\mzs{\ifmmode M_Z^2 \else $M_Z^2$\fi}
\def\mts{\ifmmode m_t^2 \else $m_t^2$\fi}
\def\mcs{\ifmmode m_c^2 \else $m_c^2$\fi}
\def\mchs{\ifmmode m_{H^\pm}^2 \else $m_{H^\pm}^2$\fi}
\def\ztwo{\ifmmode Z_2\else $Z_2$\fi}
\def\zone{\ifmmode Z_1\else $Z_1$\fi}
\def\mtwo{\ifmmode M_2\else $M_2$\fi}
\def\mone{\ifmmode M_1\else $M_1$\fi}
\def\tb{\ifmmode \tan\beta \else $\tan\beta$\fi}
\def\xw{\ifmmode x\subw\else $x\subw$\fi}
\def\ch{\ifmmode H^\pm \else $H^\pm$\fi}
\def\lum{\ifmmode {\cal L}\else ${\cal L}$\fi}
\def\inpb{\ifmmode {\rm pb}^{-1}\else ${\rm pb}^{-1}$\fi}
\def\infb{\ifmmode {\rm fb}^{-1}\else ${\rm fb}^{-1}$\fi}
\def\epem{\ifmmode e^+e^-\else $e^+e^-$\fi}
\def\ppb{\ifmmode \bar pp\else $\bar pp$\fi}
\def\pbp{\ifmmode ~^(\bar p^)p\else $~^(\bar p^)p$\fi}
\def\bsg{\ifmmode B\to X_s\gamma\else $B\to latexilaX_s\gamma$\fi}
\def\bsll{\ifmmode B\to X_s\ell^+\ell^-\else $B\to X_s\ell^+\ell^-$\fi}
\def\bstt{\ifmmode B\to X_s\tau^+\tau^-\else $B\to X_s\tau^+\tau^-$\fi}

\newskip\zatskip \zatskip=0pt plus0pt minus0pt
\def\matth{\mathsurround=0pt}
\def\lsim{\mathrel{\mathpalette\atversim<}}

\def\atversim#1#2{\lower0.7ex\vbox{\baselineskip\zatskip\lineskip\zatskip
  \lineskiplimit 0pt\ialign{$\matth#1\hfil##\hfil$\crcr#2\crcr\sim\crcr}}}

\renewcommand{\thefootnote}{\fnsymbol{footnote}}

\begin{document} \begin{titlepage} 
\rightline{\vbox{\halign{&#\hfil\cr
&SLAC-PUB-15076\cr
}}}
\vspace{1in} 
\begin{center}

{{\Large\bf The Higgs Sector and Fine-Tuning in the pMSSM}
\footnote{Work supported by the Department of 
Energy, Contract DE-AC02-76SF00515}\\}
\medskip
\medskip
\normalsize 
{\large Matthew W.~Cahill-Rowley, JoAnne L.~Hewett, Ahmed Ismail, and

Thomas G.~Rizzo\footnote{email: mrowley, hewett, aismail, rizzo@slac.stanford.edu} \\
\vskip .6cm
SLAC National Accelerator Laboratory,  \\
2575 Sand Hill Rd, Menlo Park, CA 94025, USA\\}
\vskip .5cm

\end{center} 
\vskip 0.8cm

\begin{abstract} 

The recent discovery of a 125 GeV Higgs, as well as the lack of any positive findings in searches for supersymmetry, has renewed interest in both the supersymmetric Higgs sector and fine-tuning. Here, we continue our study of the phenomenological MSSM (pMSSM), discussing the light Higgs and fine-tuning within the context of two sets of previously generated pMSSM models. We find an abundance of models with experimentally-favored Higgs masses and couplings. We investigate the decay modes of the light Higgs in these models, finding strong correlations between many final states. We then examine the degree of fine-tuning, considering contributions from each of the pMSSM parameters at up to next-to-leading-log order. In particular, we examine the fine-tuning implications for our model sets that arise from the discovery of a 125 GeV Higgs. Finally, we investigate a small subset of models with low fine-tuning and a light Higgs near 125 GeV, describing the common features of such models. We generically find a light stop and bottom with complex decay patterns into a set of light electroweak gauginos, which will make their discovery more challenging and may require novel search techniques.

\end{abstract}

\renewcommand{\thefootnote}{\arabic{footnote}} \end{titlepage}


\section{Introduction and Background}
\label{sec:intro}

With 5 fb$^{-1}$ of 7 and 8 TeV data analyzed, the LHC has begun a serious exploration of the electroweak scale. Although the parameter space for supersymmetry is being probed aggressively, direct evidence for sparticles remains elusive. As a result, our understanding of supersymmetry continues to be shaped by exclusion contours and by indirect data from a variety of observations. However, the LHC is currently opening a new window on supersymmetry by achieving sensitivity to the supersymmetric Higgs sector. Specifically, the recent discovery of a Standard Model (SM)-like Higgs boson near 125 GeV~\cite{discovery} could provide valuable information about the MSSM or suggest a new direction for theoretical investigation. As a result, several studies of the phenomenology of the MSSM Higgs sector, using both analytic results and parameter scans, have recently been performed~\cite{Higgs125}.  In this work, we consider the light Higgs and associated fine-tuning (FT) within the framework of the phenomenological MSSM (pMSSM)~\cite{pmssm}, and discuss the origin of correlations between the various observables related to the light Higgs. Our analysis makes use of the two model sets recently generated in~\cite{CahillRowley:2012cb}, one requiring a neutralino Lightest Supersymmetric particle (LSP) and the other requiring a gravitino LSP.

At tree level, the Higgs sector of the MSSM can be completely described by only two parameters. 
However, the tree-level prediction for the light CP-even Higgs mass, $m_h < m_Z$, is a clear indication that radiative corrections play a crucial role in determining the properties of the Higgs sector~\cite{hmass}. Since radiative corrections couple the Higgs sector to the rest of the SUSY spectrum, a larger number of parameters may have important effects on the properties of the lightest Higgs boson. As we will see below, the pMSSM, with its large parameter freedom, allows for such large radiative corrections.  Many pMSSM models predict significant deviations from the SM Higgs, and could be excluded by the experimental verification of a SM Higgs boson. On the other hand, the large parameter space of the pMSSM can also
easily accommodate a relatively heavy $\sim 125$ GeV SM-like Higgs boson, and we will see below that thousands of models from our previously generated sample {\cite{CahillRowley:2012cb}
are consistent with the current indications from the LHC and predict a spectrum of properties for such a Higgs boson.

In addition to providing knowledge about the viability of particular SUSY models, discovery of a SM-like Higgs boson would also yield valuable information about the fine-tuning of the MSSM. As has been pointed out by many authors~\cite{FT+Higgs}, the SUSY parameter values necessary to generate a $\sim 125$ GeV Higgs boson, particularly the requirement of heavy stops and/or large stop mixing, can lead to large fine-tuning. Discovery of a Higgs in this mass range therefore poses a challenge for natural electroweak-scale supersymmetry, the severity of which is strongly model-dependent. In addition to the Higgs sector, model-independent LHC searches for the sparticles whose mass parameters contribute strongly to fine-tuning, particularly stops and sbottoms, are beginning to set important limits, and the combination of these searches with data on the Higgs sector is expected to either lead to the discovery of SUSY or to disfavor the natural MSSM within the next year. 
We will see that the pMSSM contains a corner of parameter space that has an acceptable amount of fine-tuning and that such models contain light stops with complex decay patterns thus evading
the LHC direct searches so far.

In particular, we analyze the consequences of a Higgs with a mass near 125 GeV for the pMSSM using the two model sets previously generated in~\cite{CahillRowley:2012cb}. We also consider the consequences of requiring the $ g g \to h \to \gamma \gamma$ rate to be larger than its SM value, as may be slightly favored by current data. Working within the pMSSM allows us to consider a tremendous variety of viable models, with spectra that can differ significantly from those predicted in more constrained scenarios such as mSUGRA. The contents of this paper are as follows: A brief  summary of the model generation procedure, particularly as it relates to the Higgs sector, is given in Section~\ref{sec:modset}.  In Section~\ref{sec:higgs}, we consider the frequency and characteristics of models with Higgs properties in the experimentally favored region, paying particular attention to the diphoton channel. We also describe correlations between various Higgs decay  channels, and examine the origin of deviations from SM predictions. In Section~\ref{sec:FT}, we describe fine-tuning in the pMSSM and show how it is affected by the discovery of a 125 GeV  Higgs. Although the pMSSM allows greater flexibility than constrained SUSY models, we will see that models with a Higgs mass in the favored range still suffer from significant fine tuning. We also examine the origin of the fine-tuning and discuss the parameters which give the largest contributions. Finally, in Section~\ref{sec:imp}, we identify and describe the characteristics of a small set of models with relatively low fine-tuning, and discuss how they are likely to be affected by future results from the LHC and dark matter direct detection. In all cases, we present results for both the neutralino and gravitino LSP model sets and discuss the origin of any differences in their predictions.

\section{Model Set Generation and Higgs Sector Calculations}
\label{sec:modset}

Our study uses the pMSSM model sets generated in~\cite{CahillRowley:2012cb} by scanning over a 19 (20) dimensional space in the case of the neutralino (gravitino) LSP model set. The $> 100$ free parameters of the MSSM are reduced to the 19 (20) parameters of the pMSSM by applying phenomenologically-motivated assumptions including:

-No new sources of CP violation

-Minimal Flavor Violation

-Degenerate 1st and 2nd generation scalar masses

-A-terms (and Yukawa couplings) for the 1st and 2nd generations set to zero.

The parameters are chosen randomly with flat priors within a set range, with an absolute value less than 4 TeV (see~\cite{CahillRowley:2012cb} for exact scan ranges). The exception is the gravitino mass, which is scanned logarithmically between 1 eV and 1 TeV, and is indirectly relevant for the Higgs sector through effects on sparticle mass distributions described in ~\cite{CahillRowley:2012cb}. The resulting randomly-generated models are tested against constraints from flavor physics, rare decays, cosmology, SUSY and Higgs searches at LEP and the Tevatron, and the LHC. The neutralino LSP model set is also subjected to the 7 TeV 1 fb$^{-1}$ and 5 fb$^{-1}$ ATLAS SUSY searches, and both model sets are confronted with the latest results from non-MET searches for heavy stable charged particles, the pseudoscalar Higgs, and the rare decay $B_s \to \mu^+ \mu^-$. The resulting models are in agreement with all current experimental data, although the data set is rapidly evolving. We note that our scan does not provide comprehensive coverage of the pMSSM, so that the absence of a model with certain properties does not mean that such a model is not viable.

Of particular relevance to the current study are the masses and couplings of the Higgs sector. All masses were calculated using SOFTSUSY 3.1.7~\cite{Allanach:2001kg}, and compared to SuSpect 2.41~\cite{Djouadi:2002ze}; models in which the calculated mass of any sparticle differed by more than 25\% between the two generators were discarded. We note that the theoretical uncertainty in the mass of the lightest Higgs boson is expected to be $\sim 3$ GeV, arising mainly from both higher order corrections and the small uncertainty in $m_t$~\cite{Heinemeyer:2004gx}. 

The decay widths and branching fractions of all Higgs sector particles were calculated using SUSY-HIT 1.3~\cite{Djouadi:2006bz}, which includes the program HDECAY 3.4~\cite{Djouadi:1997yw}. We note that the partial widths given by HDECAY for a Standard Model Higgs boson differ slightly from the values published by the Higgs Working Group~\cite{Dittmaier:2011ti}. In order to separate the effects of SUSY from uncertainties in the SM Higgs properties resulting from {\it e.g.}, the value of $\alpha_s$, we use the SM Higgs properties calculated in HDECAY when making comparisons between a SUSY Higgs and a SM Higgs with the same mass. 

\section{Results}
\label{sec:res}

In this section we describe the properties of the CP-even light Higgs boson in our two pMSSM model sets.  We will examine the range of Higgs mass values and then focus on the characteristics of the models that contain a Higgs in the mass range $125\pm 2$ GeV.  We then examine the amount of fine-tuning present in these model samples.

\subsection{Light Higgs Properties}
\label{sec:higgs}

The searches for the Higgs boson at the LHC~\cite{ATLAS:2012ae,Chatrchyan:2012tx} and at the Tevatron~\cite{TEVNPH:2012ab} indicate the possible existence of such a particle in the mass range $m_h \sim 125\pm 2$ 
GeV with roughly the properties anticipated in the SM. Both the neutralino and gravitino LSP pMSSM model sets are strongly affected by the discovery of a Higgs-like object. Figure~\ref{fig:higgspec} shows the spectrum of the predicted Higgs masses in these two model samples. We observe that $19.4~(9.0)\%$ of models in the neutralino (gravitino) LSP model set have $m_h=125 \pm 2$ GeV. As discussed in~\cite{CahillRowley:2012cb}, the preference for somewhat lighter $h$ masses in the gravitino LSP model set results from a statistical preference for lighter stops, which in turn tend to give smaller radiative corrections to $m_h$. Our goal in this section is to carefully examine the properties of the light Higgs boson in the pMSSM, and then to compare and contrast the specific results for the neutralino and gravitino LSP model sets.  

\begin{figure}
\centering
\includegraphics[width=7.5in]{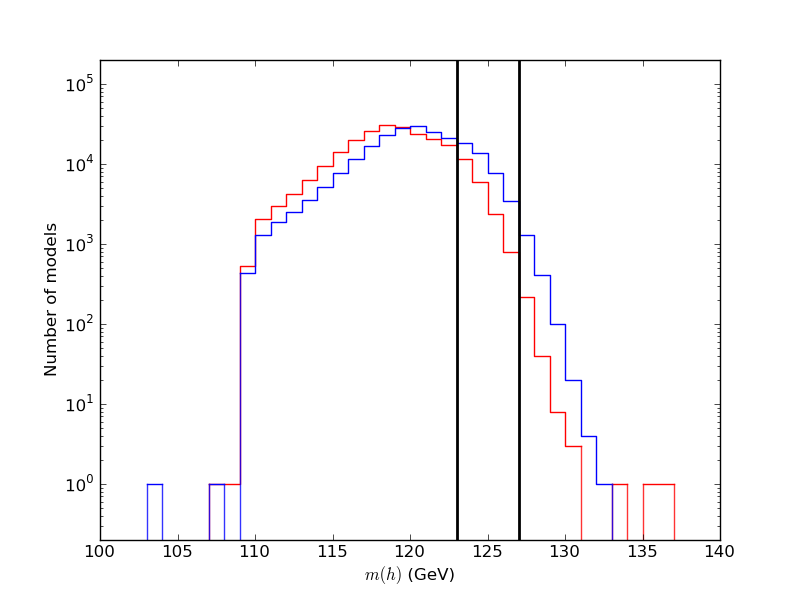}
\vspace*{0.5cm}
\caption{Distribution of the lightest CP-even Higgs mass for the neutralino (blue) and gravitino (red) LSP pMSSM model sets, highlighting the $m_h=125\pm 2$ GeV region.}
\label{fig:higgspec}
\end{figure}

Since ATLAS and CMS both report the highest statistical significance in the $gg\to h\to \gamma\gamma$ mode, we begin our discussion with this diphoton final state. 
For our study, we define the general set of ratios $R_{XX}$, which describe the relative signal strength for producing the final state $XX$ through the process $gg\to h\to XX$:
\begin{equation}
R_{XX}={{\Gamma(h\to gg) B(h\to XX)}\over {\Gamma(h_{SM}\to gg) B(h_{SM}\to XX)}}\,, 
\end{equation}
where $h_{SM}$ corresponds to a SM Higgs boson with the same mass as the Supersymmetric $h$ and $XX$ labels a specific final state. Figure~\ref{fig:higgs1} shows 
the values of $R_{\gamma\gamma}$ as a function of $m_h$ obtained in our neutralino and gravitino LSP model sets. We have highlighted two special regions of interest with $m_h=125 \pm 2$ 
GeV: $0.5 \leq R_{\gamma\gamma} < 1$ and  $1 \leq R_{\gamma\gamma} \leq 1.5$, which define subsets of models within the favored mass range. We make this special distinction since the current data on $R_{\gamma\gamma}$ may slightly favor values larger than unity, although 
this is by no means definitive given the present statistical uncertainty. In the analysis that follows we 
will make the requirement that the more `interesting' models satisfy $R_{\gamma\gamma}>0.5$. Considering the pMSSM models with $m_h$ in the favored mass range, only $23.1~(5.3)\%$ of those in the neutralino (gravitino) LSP model set also have $R_{\gamma\gamma}>1$. Thus, only $\simeq 4.5~(0.5)\%$ of the entire neutralino (gravitino) model set lies in the desired mass range and {\it also} predicts $R_{\gamma\gamma}>1$. We note that models with these properties are substantially less common in the gravitino model set.

\begin{figure}
\centering
\subfloat{
     \begin{overpic}[height=3.5in]{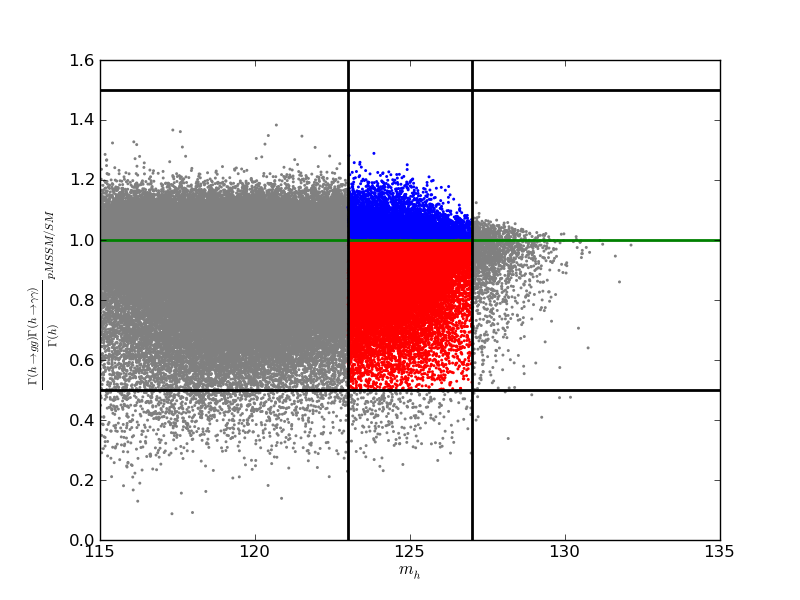}
     \put(40,70){Neutralino LSP}
     \end{overpic}
     } \\
\subfloat{
     \begin{overpic}[height=3.5in]{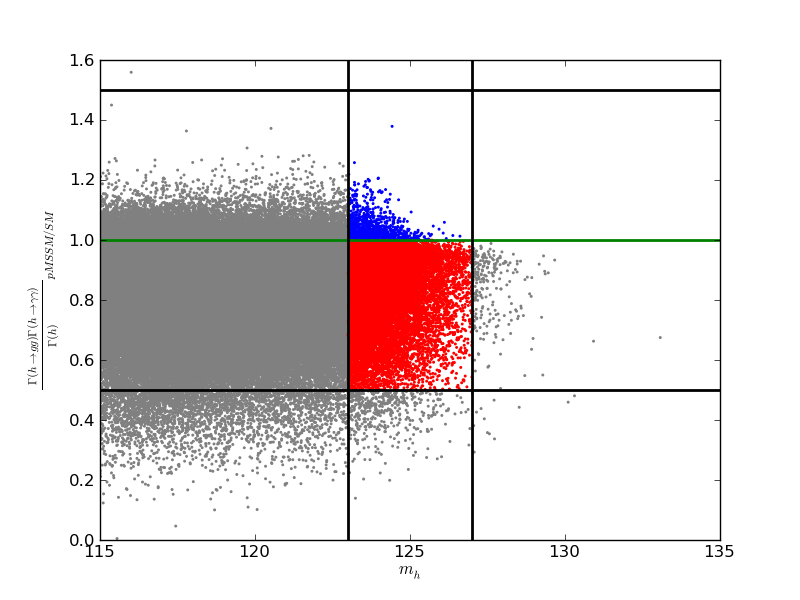}
     \put(40,70){Gravitino LSP}
     \end{overpic}
     }
\vspace*{0.5cm}
\caption{The ratio $R_{\gamma\gamma}$, defined in the text, is shown as a function of the $h$ mass in the neutralino (top) and gravitino (bottom) LSP model sets. The subset of models in the mass range $m_h=125\pm 2$ GeV and with $1 \leq R_{\gamma\gamma} \leq 1.5$ are highlighted in blue and with $0.5 \leq R_{\gamma\gamma} < 1$ in red.}
\label{fig:higgs1}
\end{figure}

\begin{figure}
\centering
\subfloat{
     \begin{overpic}[height=3.5in]{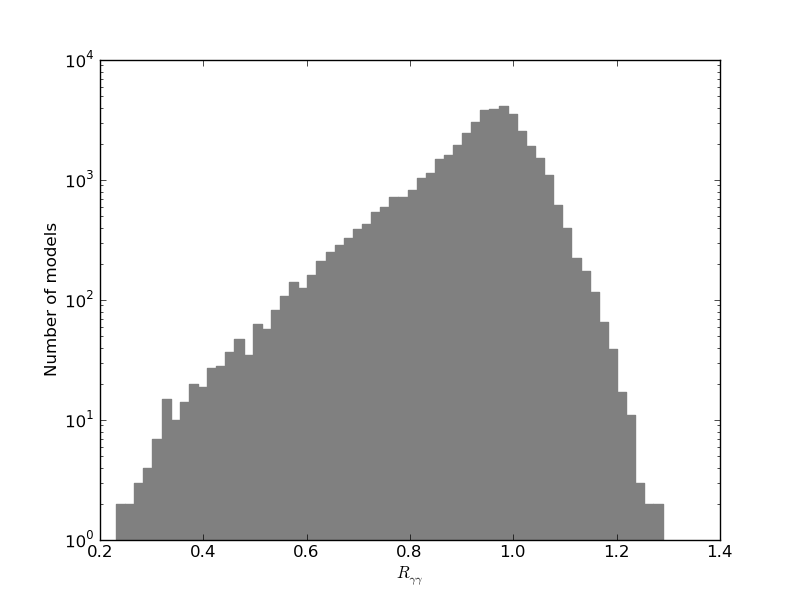}
     \put(40,70){Neutralino LSP}
     \end{overpic}
     } \\
\subfloat{
     \begin{overpic}[height=3.5in]{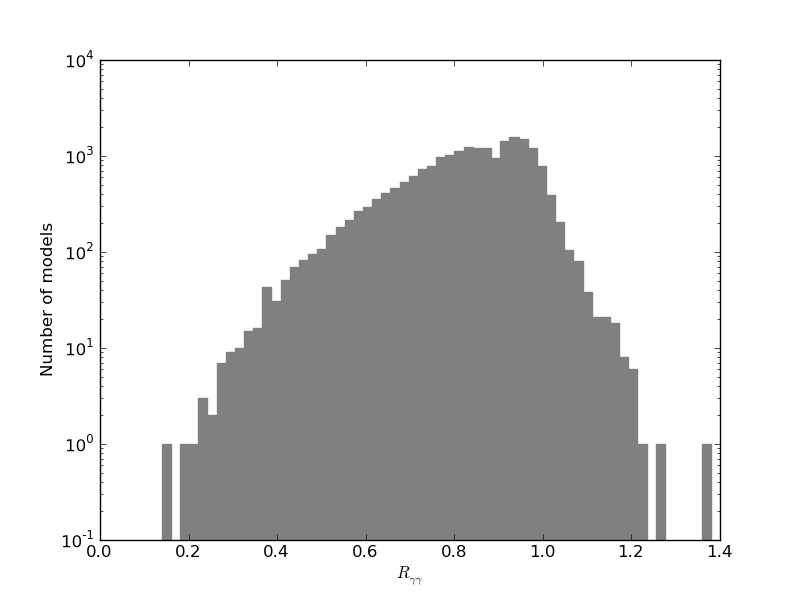}
     \put(40,70){Gravitino LSP}
     \end{overpic}
     }
\vspace*{0.5cm}
\caption{Histograms of the values of $R_{\gamma\gamma}$ for models with $m_h = 125 \pm 2$ GeV in the neutralino (top) and gravitino (bottom) LSP model sets.}
\label{fig:higgs2}
\end{figure}

In Fig.~\ref{fig:higgs2} we see that the ratio $R_{\gamma\gamma}$ is strongly peaked just below unity for both model sets, especially for the neutralino LSP case, and falls off quite rapidly for larger or smaller values. However, the detailed shape of the $R_{\gamma\gamma}$ distribution differs significantly between the two model sets, with the gravitino distribution appearing somewhat more broad. This difference originates from the somewhat lighter sparticle spectra in the gravitino LSP model set, leading to different predictions for observables such as $R_{\gamma \gamma}$ which are sensitive to these masses through radiative corrections. 
We again observe that fewer models lie near or above $R_{\gamma\gamma}=1$ in the gravitino LSP model set than in the neutralino LSP model set.

\begin{figure}
\centering
\subfloat{
     \begin{overpic}[width=3.5in]{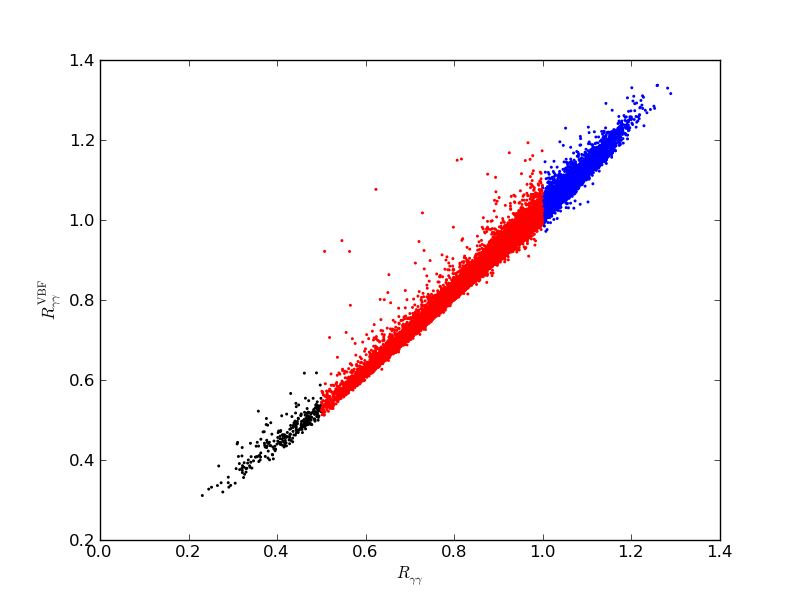}
     \put(37,70){Neutralino LSP}
     \end{overpic}
     } ~
\subfloat{ 
     \begin{overpic}[width=3.5in]{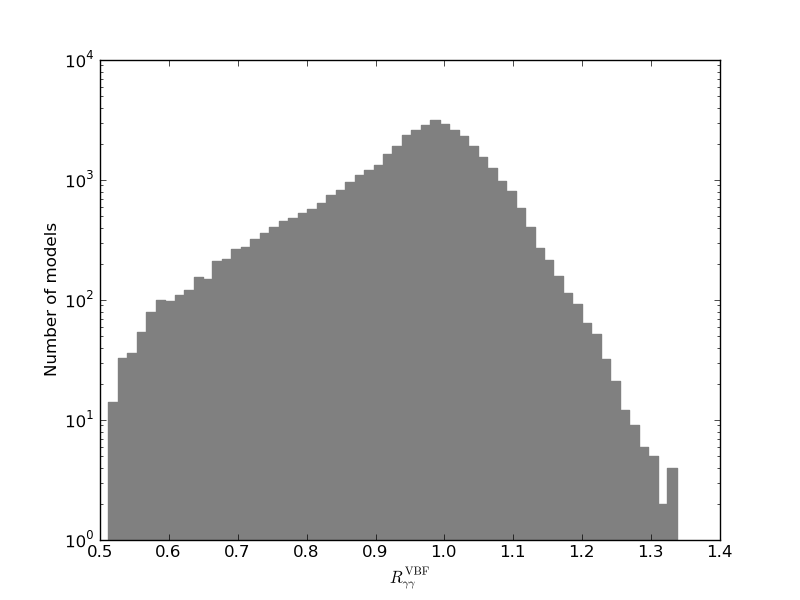}
     \put(37,70){Neutralino LSP}
     \end{overpic}
     } \\
\subfloat{
     \begin{overpic}[width=3.5in]{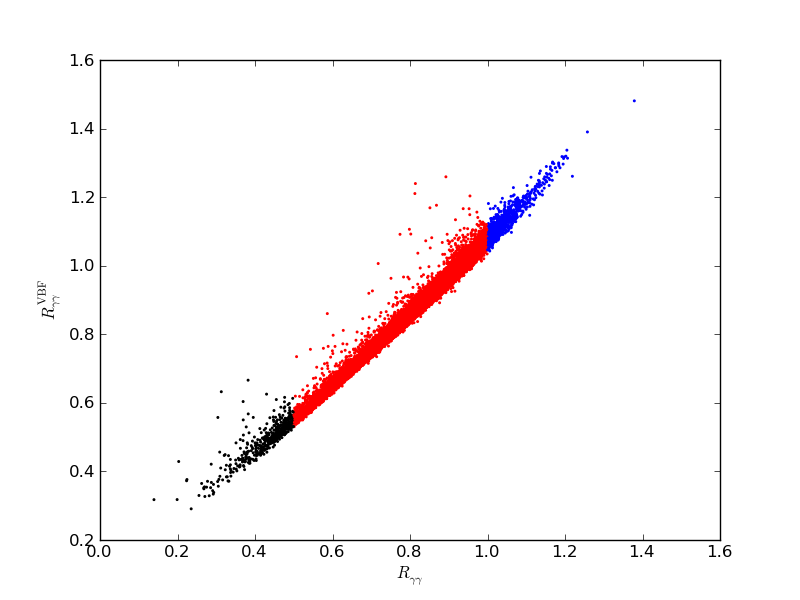}
     \put(37,70){Gravitino LSP}
     \end{overpic}
     } ~
\subfloat{
     \begin{overpic}[width=3.5in]{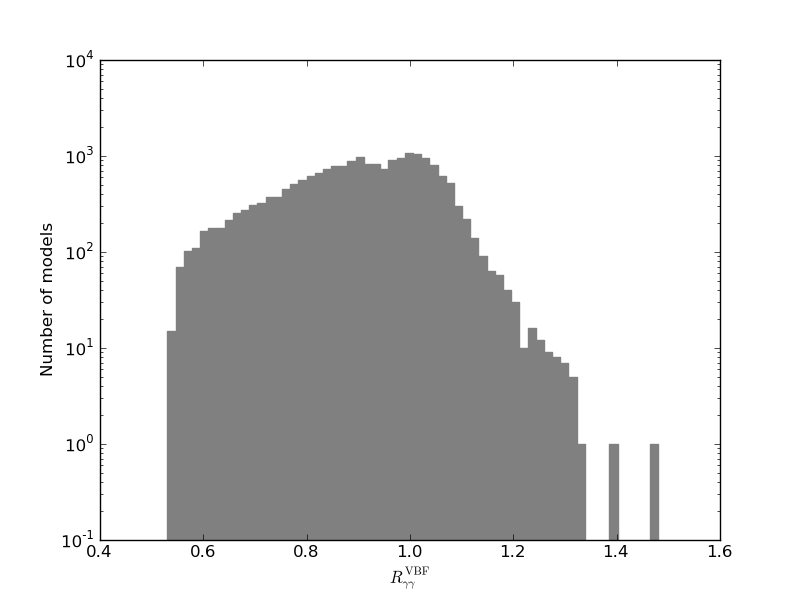}
     \put(37,70){Gravitino LSP}
     \end{overpic}
     }
\vspace*{0.5cm}
\caption{Correlation of $R_{\gamma\gamma}^{VBF}$ with $R_{\gamma\gamma}$ and histograms of the values of $R_{\gamma\gamma}^{VBF}$ in the neutralino (top) and gravitino (bottom) LSP model sets. The models shown satisfy $m_h = 125 \pm 2$ GeV and the histograms are made with the additional requirement $R_{\gamma\gamma}>0.5$.  The color coding is as described in Fig.~\ref{fig:higgs1}.}
\label{fig:higgs3}
\end{figure}

The CMS Higgs search in the $\gamma\gamma$ channel~\cite{Chatrchyan:2012tw} seems to indicate that at least a fraction of the putative Higgs signal arises from vector boson fusion (VBF), $W^+W^-/ZZ 
\to h \to \gamma\gamma$, in addition to the dominant gluon fusion production channel. It is therefore interesting to investigate the analog of $R_{\gamma\gamma}$ in this channel,
\begin{equation}
R_{\gamma\gamma}^{VBF}={{\Gamma(h\to WW) B(h\to \gamma\gamma)}\over {\Gamma(h_{SM}\to WW) B(h_{SM}\to \gamma\gamma)}}\,, 
\end{equation}
as well as its correlation with $R_{\gamma\gamma}$. These results are shown for both model sets in Fig.~\ref{fig:higgs3}, where we see that these two quantities are quite 
highly (and almost linearly) correlated, with few outliers. We see that the distributions of values for $R_{\gamma\gamma}^{VBF}$ are also qualitatively similar in shape to those 
obtained for $R_{\gamma\gamma}$ itself. As can be seen in Fig.~\ref{fig:higgs3}, only a handful of models have $R_{\gamma\gamma}^{VBF}/R_{\gamma\gamma}$ appreciably larger than unity, corresponding to models where gluon fusion production of the Higgs is suppressed.

\begin{figure}
\centering
\subfloat{
     \begin{overpic}[width=3.5in]{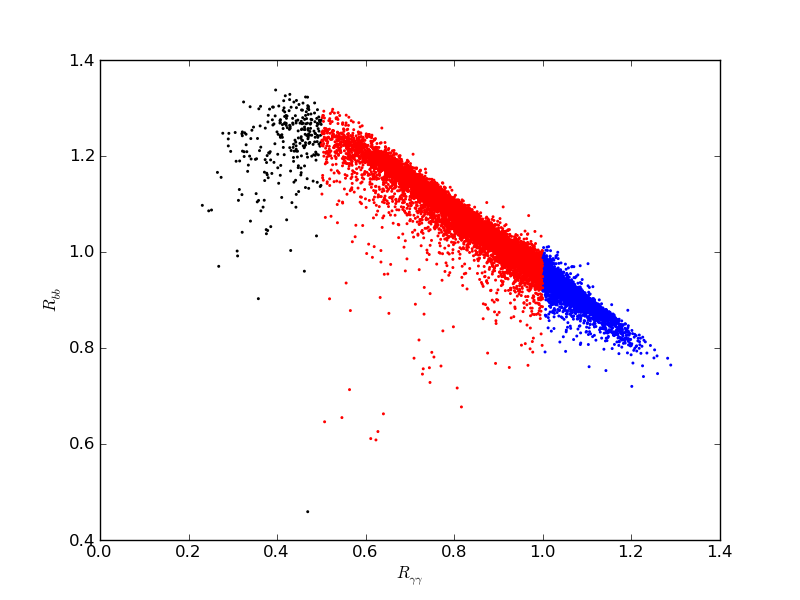}
     \put(37,70){Neutralino LSP}
     \end{overpic}
     } ~
\subfloat{ 
     \begin{overpic}[width=3.5in]{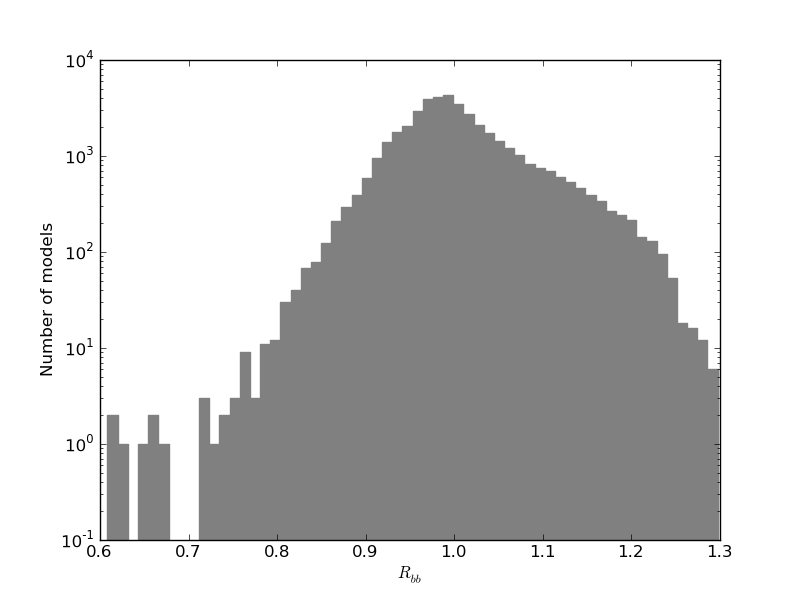}
     \put(37,70){Neutralino LSP}
     \end{overpic}
     } \\
\subfloat{
     \begin{overpic}[width=3.5in]{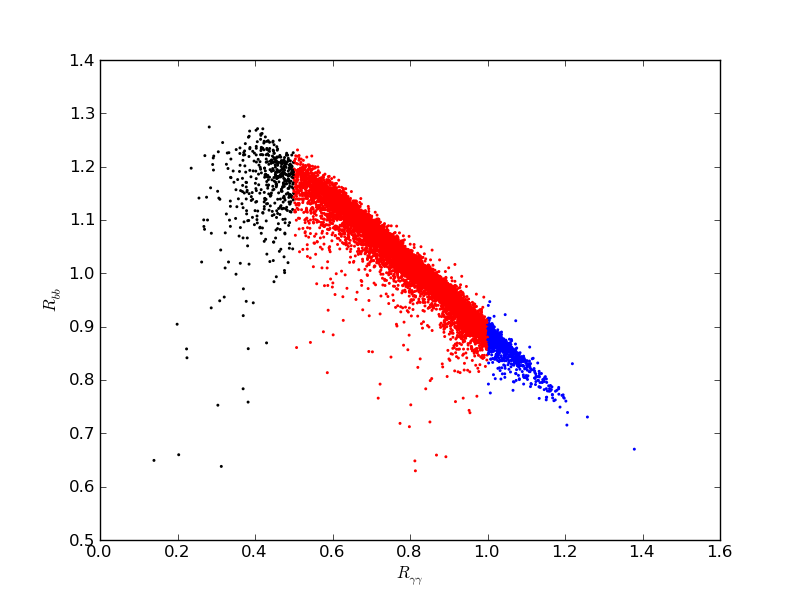}
     \put(37,70){Gravitino LSP}
     \end{overpic}
     } ~
\subfloat{
     \begin{overpic}[width=3.5in]{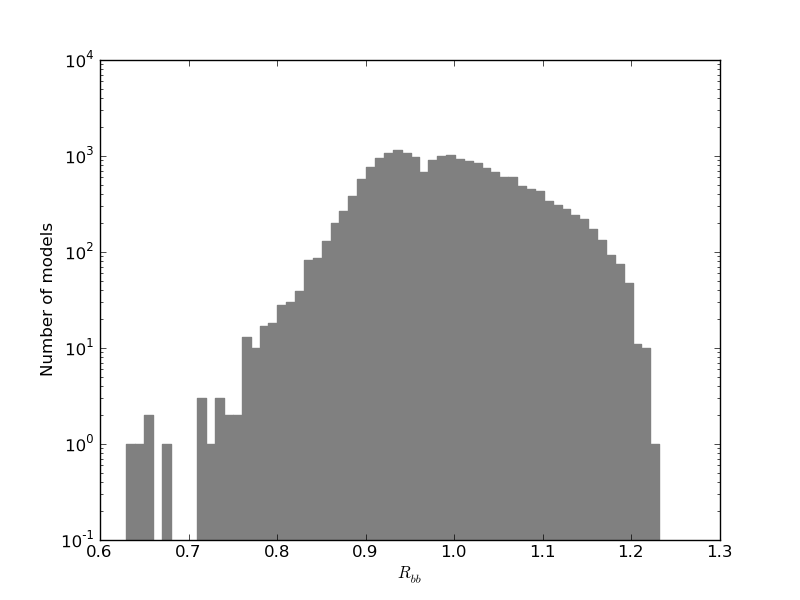}
     \put(37,70){Gravitino LSP}
     \end{overpic}
     }
\vspace*{0.5cm}
\caption{Same as the previous Figure but now for $R_{bb}$.}
\label{fig:higgs4}
\end{figure}

Although observation of the decay $h\to b\bar b$, and therefore determination of $R_{bb}$, is unlikely at the LHC in the near future, the recent Tevatron measurements~\cite{tevatronjuly} do provide some information on 
the $h b\bar b$ coupling itself. Within the MSSM in the decoupling limit, with $m_h=125 \pm 2$ GeV, $h\to b\bar b$ is the dominant decay mode for the light Higgs. As a result, changes to the $h\to b\bar b$ partial width alter the branching fractions (and therefore the signal strengths) for the other Higgs decay modes~\cite{hbbimp}. Interestingly, the $hb\bar b$ coupling can get very large radiative corrections due to sbottom and gluino loops~\cite{RADCOR}, particularly when various pMSSM mass parameters obtain large values. Since our parameters are scanned up to 4 TeV, large corrections are quite common in both model sets.
Examination of the ratio $R_{bb}$ is therefore interesting, and may allow us to probe these corrections. Given that the Tevatron Higgs searches relying on the $b \bar{b}$ decay mode of the Higgs observe an excess in the same region as the LHC discovery, $R_{bb}$ is unlikely to be very small. Fig.~\ref{fig:higgs4} displays both the distribution of values for $R_{bb}$ (assuming that 
$R_{\gamma\gamma}>0.5)$ and its correlation with $R_{\gamma\gamma}$ for both pMSSM model sets. As we can see from these figures, $R_{bb}$ is {\it anti-correlated} with $R_{\gamma\gamma}$, in sharp contrast with the positive correlation seen for $R_{\gamma\gamma}^{VBF}$ (as well as the other $R_{XX}$ observables, as we will see below).
We also note that a fairly significant set of models are found to lie away from the dominant, almost linear, anti-correlation region in either model set. Here we see the rather general result that large values of $R_{\gamma\gamma}$ are obtained by reducing $R_{bb}$ through large corrections to the $hb\bar b$ coupling discussed above. We will continue our discussion of the $h\to b\bar b$ partial width and its impact on the distribution of the non-$b$ $R_{XX}$ ratios below.

The $h\to \tau^+\tau^-$ decay mode probes the $h\tau^+\tau^-$ coupling which is indirectly tested by the ratio $R_{\tau\tau}$. Like the $hb\bar b$ coupling, the $h\tau^+\tau^-$ coupling can also receive loop corrections enhanced by pMSSM mass parameters, although they are not as large as for the former due to the smaller gauge couplings that appear in these loops (see, e.g.~\cite{Carena:2002es}). Fig.~\ref{fig:higgs5} displays both 
the distribution of values of $R_{\tau\tau}$ (assuming that $R_{\gamma\gamma}>0.5)$ and its correlation with $R_{\gamma\gamma}$ for both pMSSM model sets. 
Although the correlation is again observed to be almost linear (as it is in all cases except for $R_{bb}$), significant deviations from $R_{\tau\tau} \approx R_{\gamma\gamma}$ occur for a small but non-negligible set of models.

\begin{figure}
\centering
\subfloat{
     \begin{overpic}[width=3.5in]{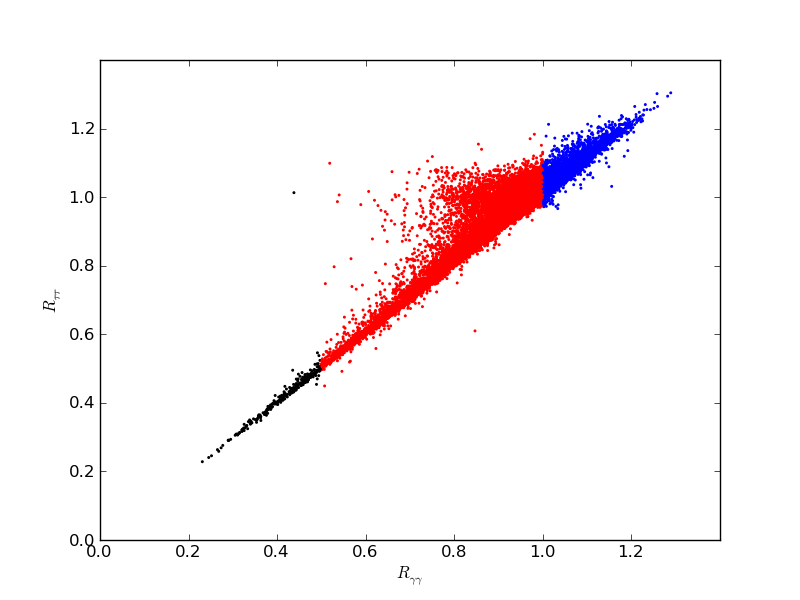}
     \put(37,70){Neutralino LSP}
     \end{overpic}
     } ~
\subfloat{ 
     \begin{overpic}[width=3.5in]{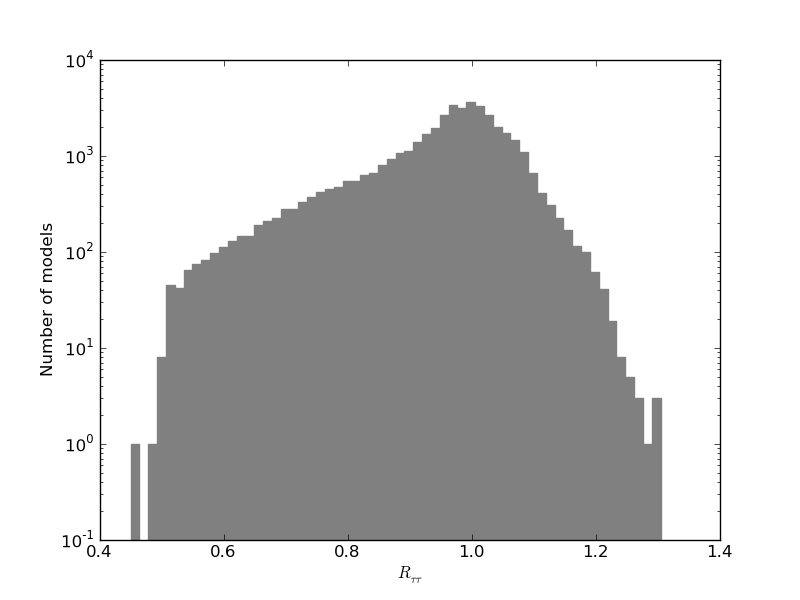}
     \put(37,70){Neutralino LSP}
     \end{overpic}
     } \\
\subfloat{
     \begin{overpic}[width=3.5in]{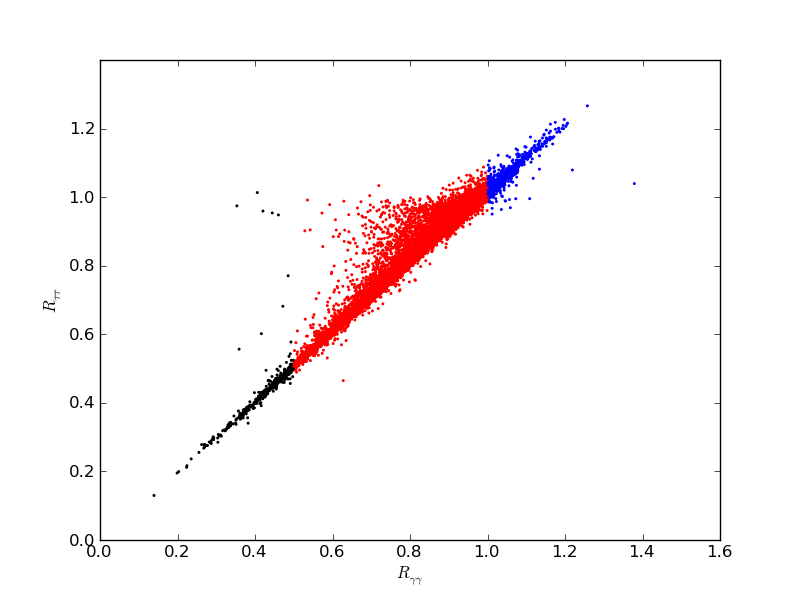}
     \put(37,70){Gravitino LSP}
     \end{overpic}
     } ~
\subfloat{
     \begin{overpic}[width=3.5in]{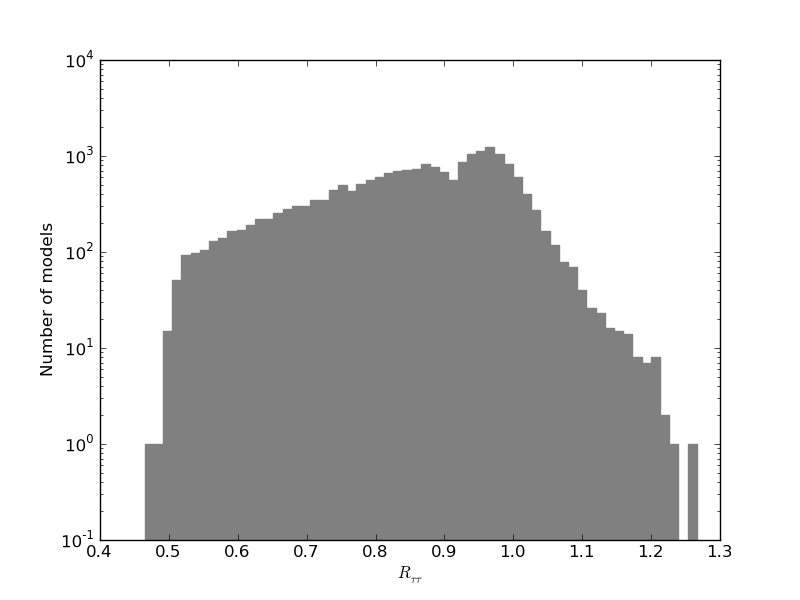}
     \put(37,70){Gravitino LSP}
     \end{overpic}
     }
\vspace*{0.5cm}
\caption{Same as the previous Figure but now for $R_{\tau\tau}$.}
\label{fig:higgs5}
\end{figure}

In the decoupling limit, the decay rates for both $h\to W^+W^-,ZZ$ at tree-level are essentially fixed once the Higgs mass is known. The ratio of these two partial widths is also 
completely determined up to (small) radiative corrections by the Higgs mass due to the custodial symmetric two-doublet nature of the MSSM, so anything we say about the ratio $R_{ZZ}$ will also be applicable to the 
ratio $R_{W^+W^-}$ at the percent level.{\footnote {We have explicitly verified the validity of this statement within our two pMSSM model sets.}}  Since $h\to ZZ$ is the cleaner mode and is 
important for determining the light Higgs mass, for brevity we will only discuss the quantity $R_{ZZ}$.  Fig.~\ref{fig:higgs6} displays both the distribution of values 
for $R_{ZZ}$ (again assuming that $R_{\gamma\gamma}>0.5)$ and its correlation with $R_{\gamma\gamma}$ for both pMSSM model sets. We again see that 
these two ratios are highly correlated and that the histograms of the values of $R_{ZZ}$ are quite similar to those obtained for the other non-b $R_{XX}$ in both LSP model sets.

\begin{figure}
\centering
\subfloat{
     \begin{overpic}[width=3.5in]{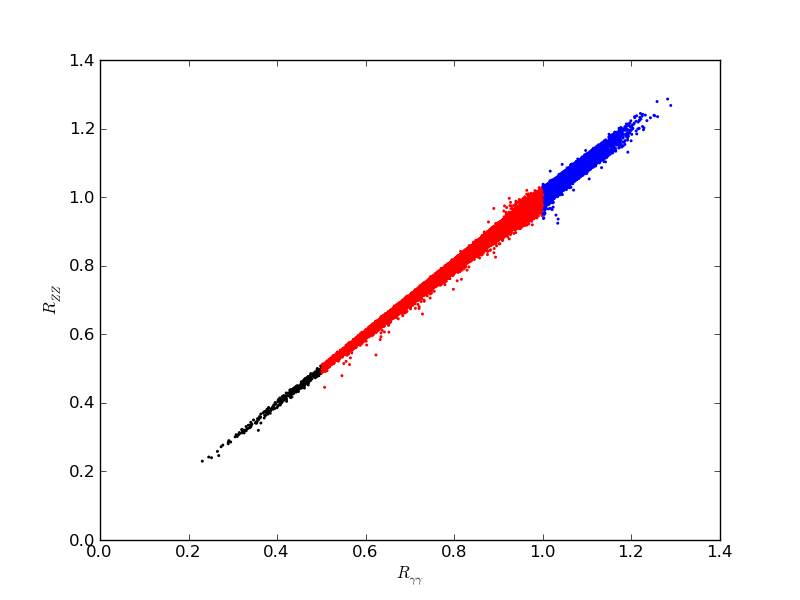}
     \put(37,70){Neutralino LSP}
     \end{overpic}
     } ~
\subfloat{ 
     \begin{overpic}[width=3.5in]{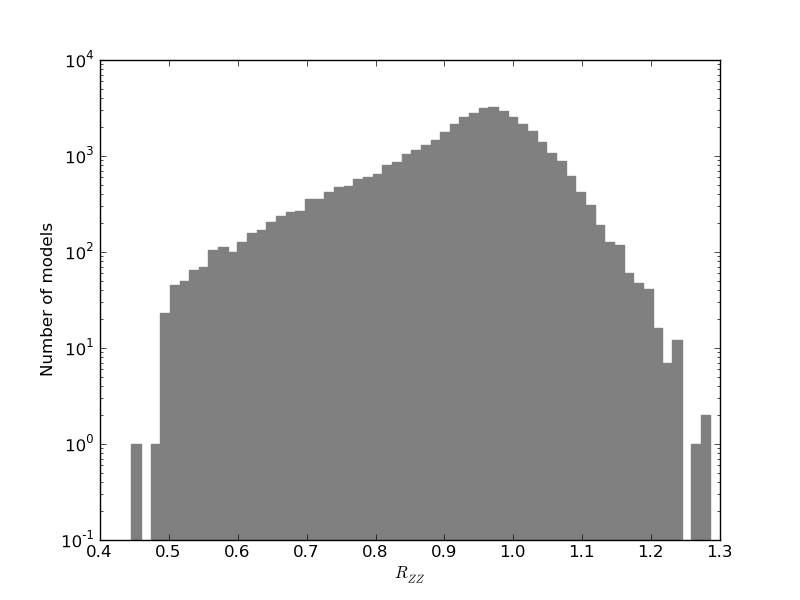}
     \put(37,70){Neutralino LSP}
     \end{overpic}
     } \\
\subfloat{
     \begin{overpic}[width=3.5in]{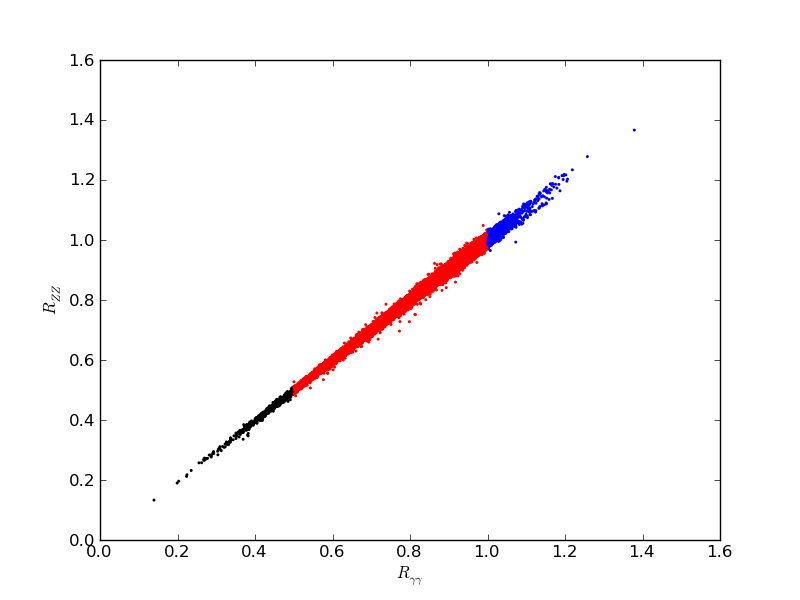}
     \put(37,70){Gravitino LSP}
     \end{overpic}
     } ~
\subfloat{
     \begin{overpic}[width=3.5in]{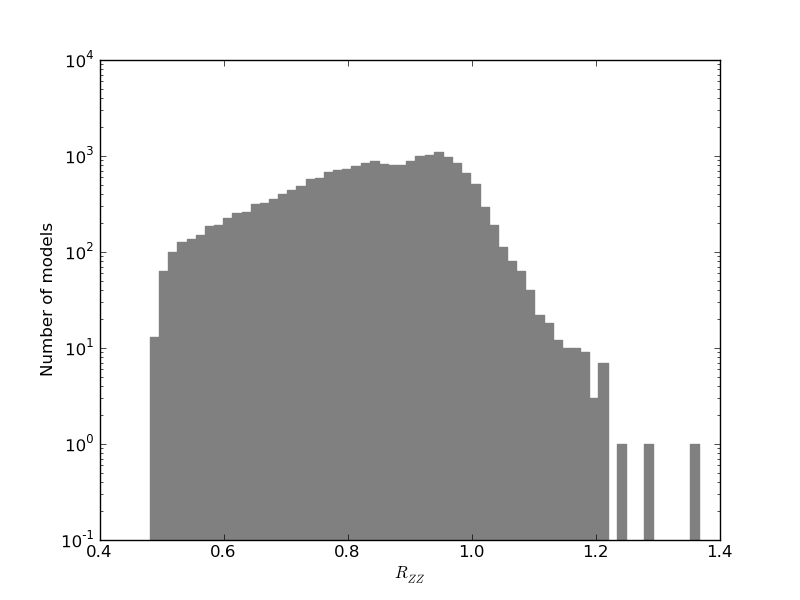}
     \put(37,70){Gravitino LSP}
     \end{overpic}
     }
\vspace*{0.5cm}
\caption{Same as the previous Figure but now for $R_{ZZ}$.}
\label{fig:higgs6}
\end{figure}

In principle, the loop-induced decay $h\to Z\gamma$ probes a different combination of pMSSM model parameters than does the more familiar $h\to \gamma\gamma$ mode. It is therefore interesting to consider whether the ratio $R_{Z\gamma}$ can differ significantly from $R_{\gamma\gamma}$ and thereby provide orthogonal constraints on pMSSM models. Unfortunately, as can be seen 
in Fig.~\ref{fig:higgs7}, these 
two observables are as highly correlated as the other observables described above (with the exception of $R_{bb}$), and the distribution of $R_{Z\gamma}$ values is therefore seen to be quite similar to the distributions of the other $R_{XX}$.

\begin{figure}
\centering
\subfloat{
     \begin{overpic}[width=3.5in]{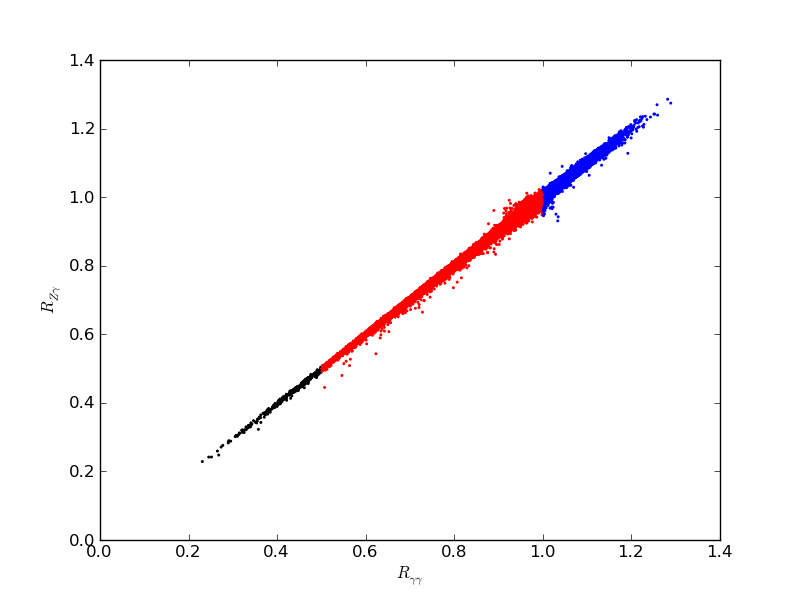}
     \put(37,70){Neutralino LSP}
     \end{overpic}
     } ~
\subfloat{ 
     \begin{overpic}[width=3.5in]{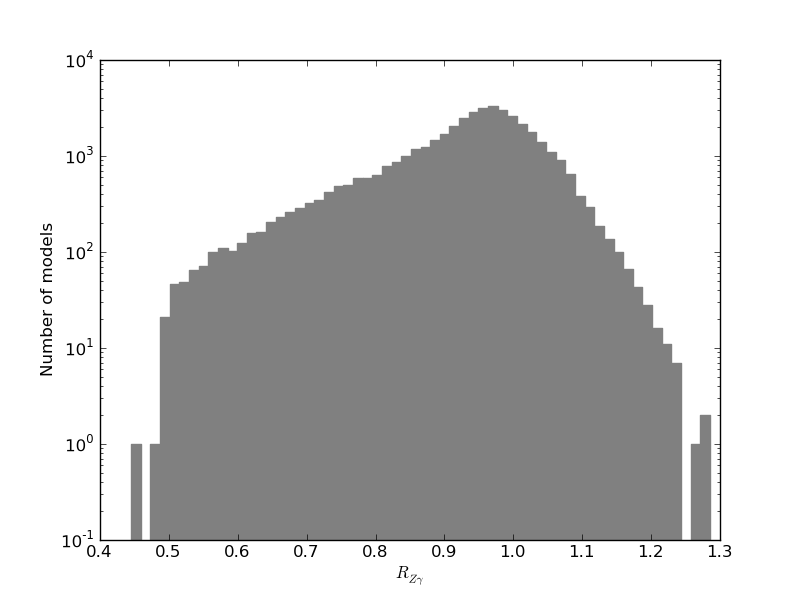}
     \put(37,70){Neutralino LSP}
     \end{overpic}
     } \\
\subfloat{
     \begin{overpic}[width=3.5in]{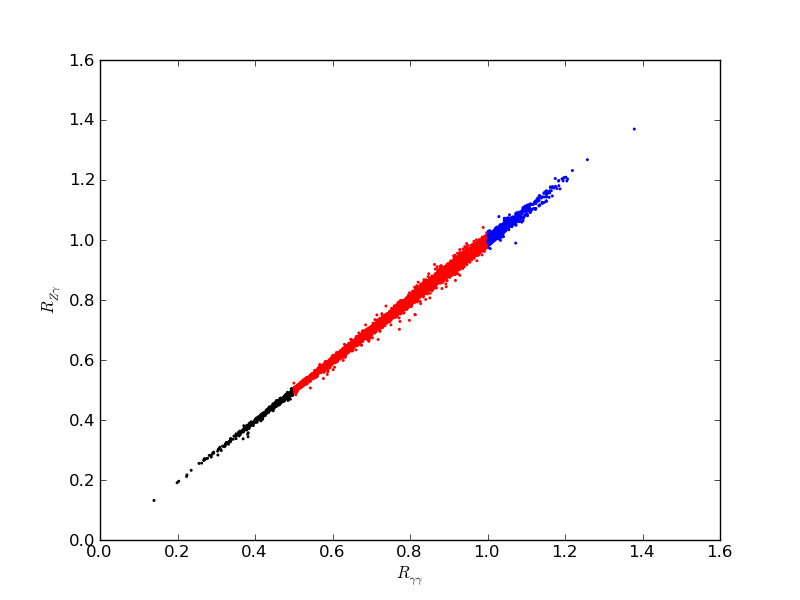}
     \put(37,70){Gravitino LSP}
     \end{overpic}
     } ~
\subfloat{
     \begin{overpic}[width=3.5in]{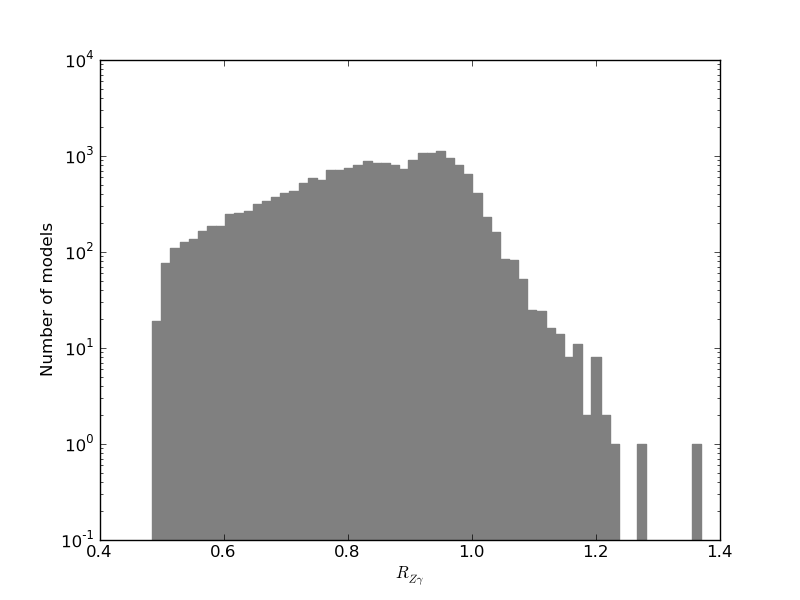}
     \put(37,70){Gravitino LSP}
     \end{overpic}
     }
\vspace*{0.5cm}
\caption{Same as the previous Figure but now for $R_{Z\gamma}$.}
\label{fig:higgs7}
\end{figure}

In all cases, except for $R_{bb}$, the $R_{XX}$ are highly correlated with $R_{\gamma \gamma}$ and therefore have nearly identical distributions in our model sets. To understand this trend, we decompose the $R_{XX}$ into their component parts, looking particularly at the partial widths for the various decay channels. Again, we analyze the ratios to the corresponding SM values, defining the variable
\begin{equation}
r_X=\Gamma(h\to XX)/\Gamma(h_{SM} \to XX).
\end{equation}
We note that in the decoupling limit at tree level, all the $r_X$'s are identically 1. Any deviations from $r_X = 1$ are therefore the result of radiative corrections. Unsurprisingly, the distributions of $r_X$ values are peaked near unity in both model sets. In most cases, the tails of these distributions are at the few percent level, only marginally larger than the uncertainty in their calculation (see Figs.~\ref{fig:higgs8a} and~\ref{fig:higgs8b} for two examples of this, $r_{\gamma}$ and $r_g$). The distributions for $r_b$ (and to a lesser extent $r_{\tau}$), however, have significant tails; after requiring $m_h=125\pm 2$ GeV, 25\% (46\%) of neutralino (gravitino) LSP models deviate from $r_b=1$ by more than 20\% and 0.4\% (2\%) of neutralino (gravitino) LSP models deviate from $r_{\tau}=1$ by more than 20\% (Fig.~\ref{fig:higgs9} displays histograms of $r_b$ for each model set). In particular, we observe that the histogram of $r_b$ values extends beyond $\sim 3$ for both model sets, although it is slightly more sharply peaked near unity in the neutralino case. The large variability in the $h\to b \bar{b}$ width, and therefore the total width, relative to the SM value (with the other $r_X \simeq 1$) means that the $h\to b \bar{b}$ width determines to a good approximation the shape of the various $R_{XX}$ distributions we examined above. Varying the $h\to b \bar{b}$ partial width while approximately holding the other partial widths fixed at their SM value explains to a large extent the strong correlations between the non-$b$ $R_{XX}$ described above, as well as the shapes of their distributions. A small portion of the correlated variation of the $R_{XX}$ is also explained by changing the $h\to g g$ partial width, altering the production cross-section for the light Higgs. However, this variation is highly sub-dominant due to the small spread of values for $r_g$ within our model sets. The extent to which $r_b$ controls the signal strength in the diphoton channel is seen in Fig~\ref{fig:rbRyy}. In particular, we see that a few models have a low value of $R_{\gamma \gamma}$ despite having $r_b \approx 1$, resulting from suppression of the gluon fusion production channel. However, {\it all} of our models with a significant enhancement of $R_{\gamma \gamma}$ have a suppressed $hb\bar b$ coupling.

\begin{figure}
\centering
\subfloat{
     \begin{overpic}[height=3.5in]{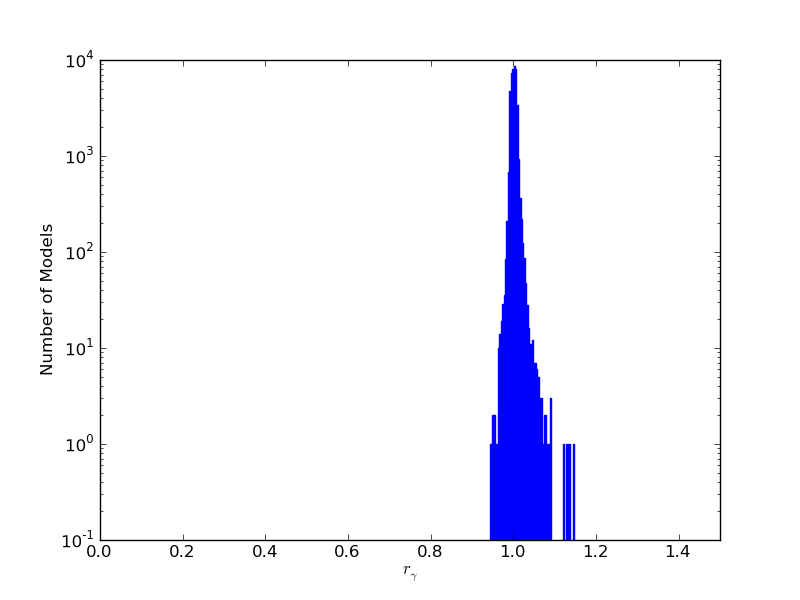}
     \put(40,70){Neutralino LSP}
     \end{overpic}
     } \\
\subfloat{
     \begin{overpic}[height=3.5in]{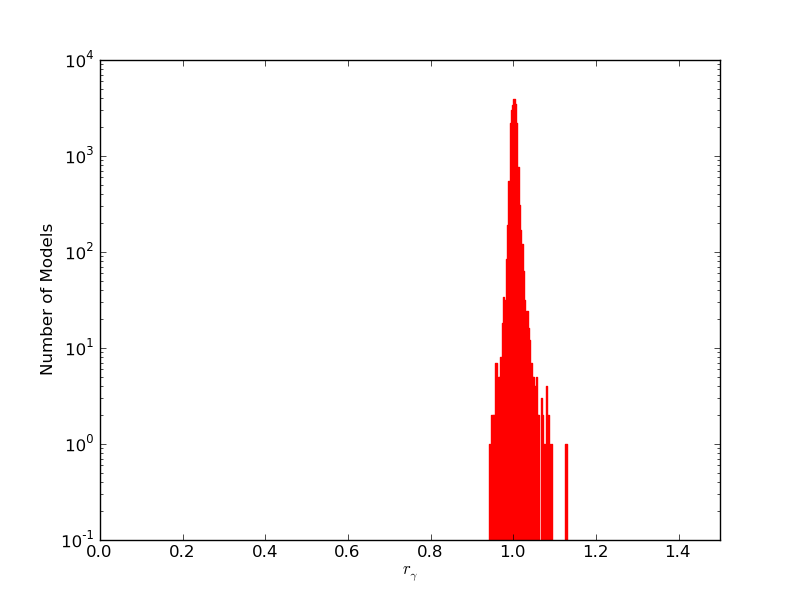}
     \put(40,70){Gravitino LSP}
     \end{overpic}
     }
\vspace*{0.5cm}
\caption{Histograms of the values of $r_\gamma$ for models with $m_h = 125 \pm 2$ GeV in the neutralino (top) and gravitino (bottom) LSP model sets.}
\label{fig:higgs8a}
\end{figure}

\begin{figure}
\centering
\subfloat{
     \begin{overpic}[height=3.5in]{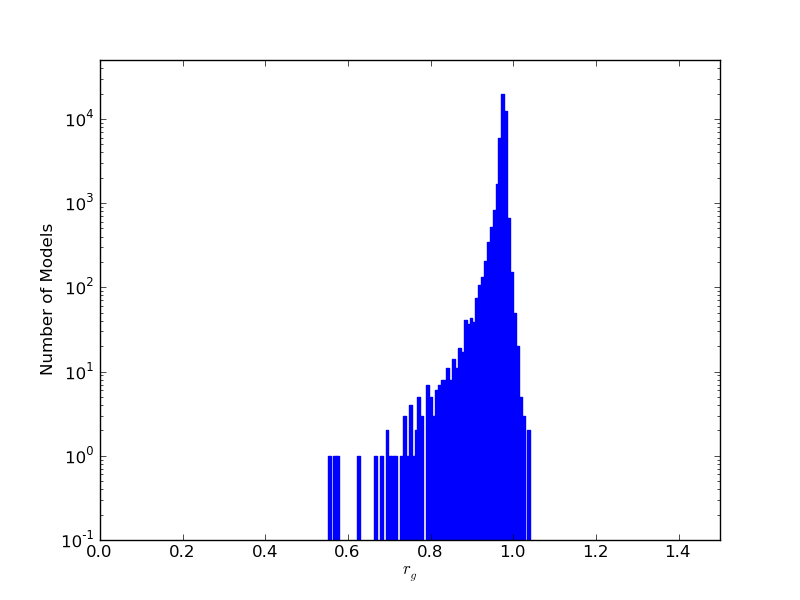}
     \put(40,70){Neutralino LSP}
     \end{overpic}
     } \\
\subfloat{
     \begin{overpic}[height=3.5in]{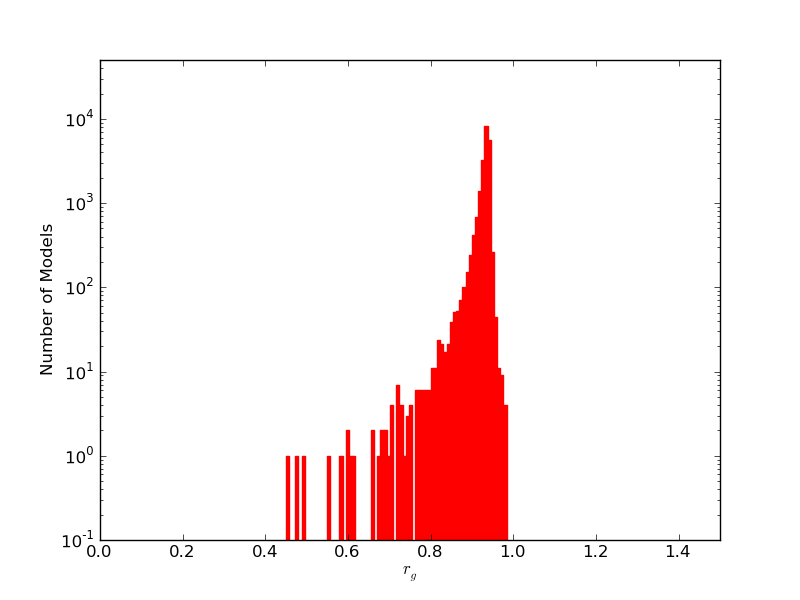}
     \put(40,70){Gravitino LSP}
     \end{overpic}
     }
\vspace*{0.5cm}
\caption{Histograms of the values of $r_g$ for models with $m_h = 125 \pm 2$ GeV in the neutralino (top) and gravitino (bottom) LSP model sets.}
\label{fig:higgs8b}
\end{figure}

\begin{figure}
\centering
\subfloat{
     \begin{overpic}[height=3.5in]{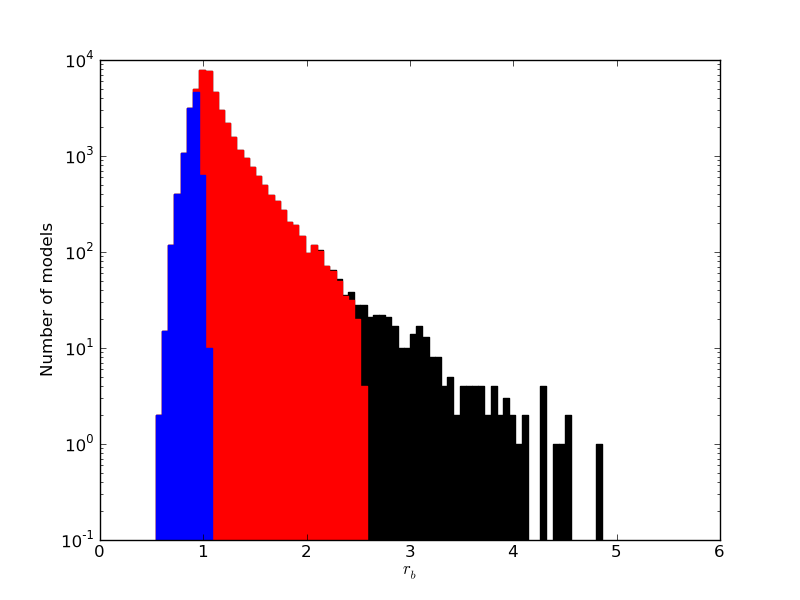}
     \put(40,70){Neutralino LSP}
     \end{overpic}
     } \\
\subfloat{
     \begin{overpic}[height=3.5in]{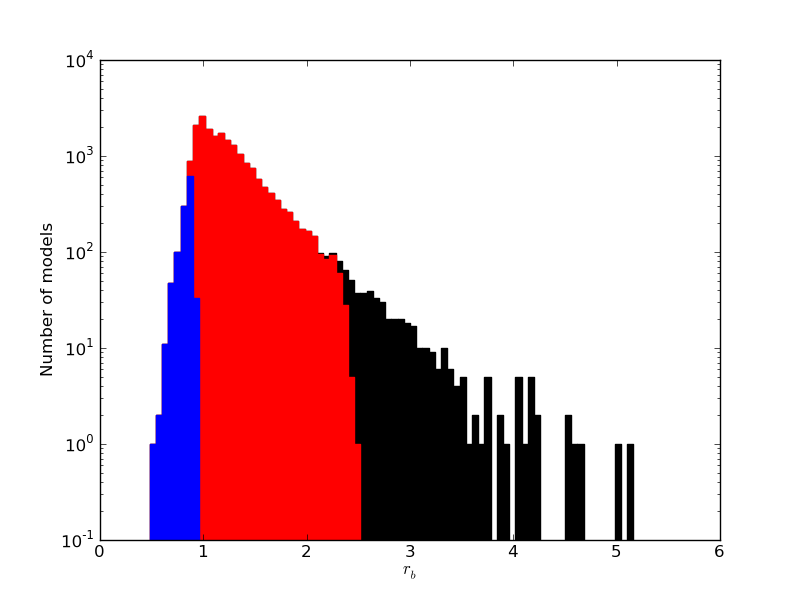}
     \put(40,70){Gravitino LSP}
     \end{overpic}
     }
\vspace*{0.5cm}
\caption{Histograms of the values of $r_b$ for models with $m_h = 125 \pm 2$ GeV in the neutralino (top) and gravitino (bottom) LSP model sets. Colors represent the ranges of $R_{\gamma \gamma}$ defined in Fig.~\ref{fig:higgs1}.}
\label{fig:higgs9}
\end{figure}

\begin{figure}
\centering
\subfloat{
     \begin{overpic}[height=3.5in]{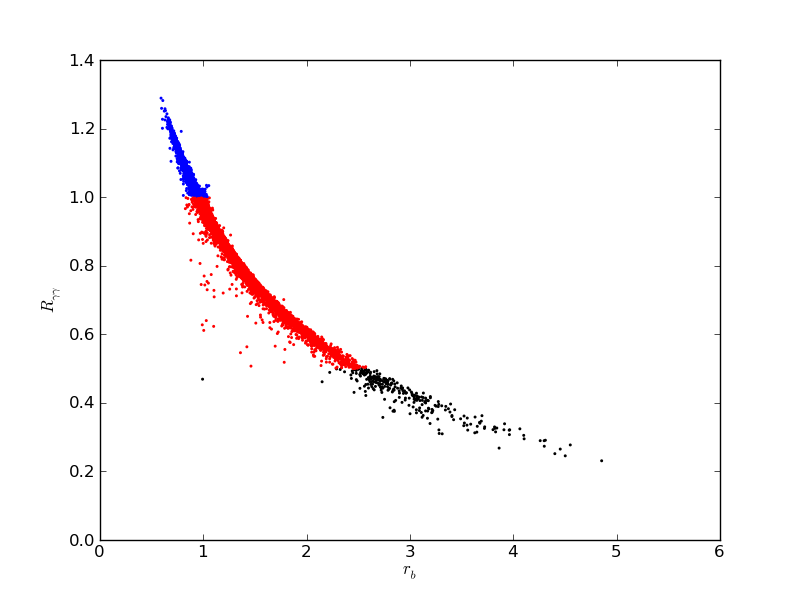}
     \put(40,70){Neutralino LSP}
     \end{overpic}
     } \\
\subfloat{
     \begin{overpic}[height=3.5in]{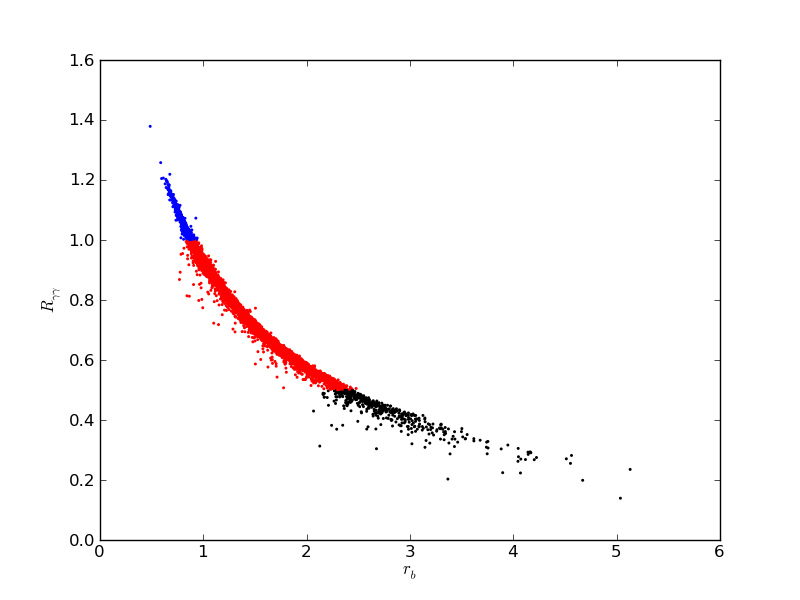}
     \put(40,70){Gravitino LSP}
     \end{overpic}
     }
\vspace*{0.5cm}
\caption{Correlation of $R_{\gamma\gamma}$ with the normalized partial width for $h\to b\bar b$, $r_b$, for models with $m_h=125\pm 2$ GeV. The upper (lower) panel is for the neutralino (gravitino) LSP model set. The color code is the same as in previous figures.}
\label{fig:rbRyy}
\end{figure}

Since slowly decoupling~\cite{Haber:2000kq} radiative corrections to the $hb\bar b$ coupling are the dominant factor determining the diphoton signal strength, it is interesting to ask which parameters result in large corrections. In our model sets, the large radiative corrections to the $hb\bar b$ coupling are strongly correlated with the bottom squark mixing, specifically the off-diagonal 
term in the sbottom mass matrix, $X_b=A_b-\mu \tan \beta$. To appreciate the strength of this correlation, we show in Fig.~\ref{fig:higgsabc} the relationship between $r_b$, the relative partial width for $h\to b\bar b$, as a function of the ratio $X_b/m(\tilde b_2)$ for $m_h=125\pm 2$ GeV in both model sets. Once again, the colors of the points correspond to the value of $R_{\gamma\gamma}$ for a given model. We observe that large values of the partial width for $h\to b\bar b$ (small values of $R_{\gamma \gamma}$) occur when the sbottom mixing is large and negative. Similarly, we see that values of $R_{\gamma\gamma} > 1$ occur when a large positive mixing suppresses the $h\to b\bar b$ partial width in comparison with the SM.

\begin{figure}
\centering
\subfloat{
     \begin{overpic}[height=3.5in]{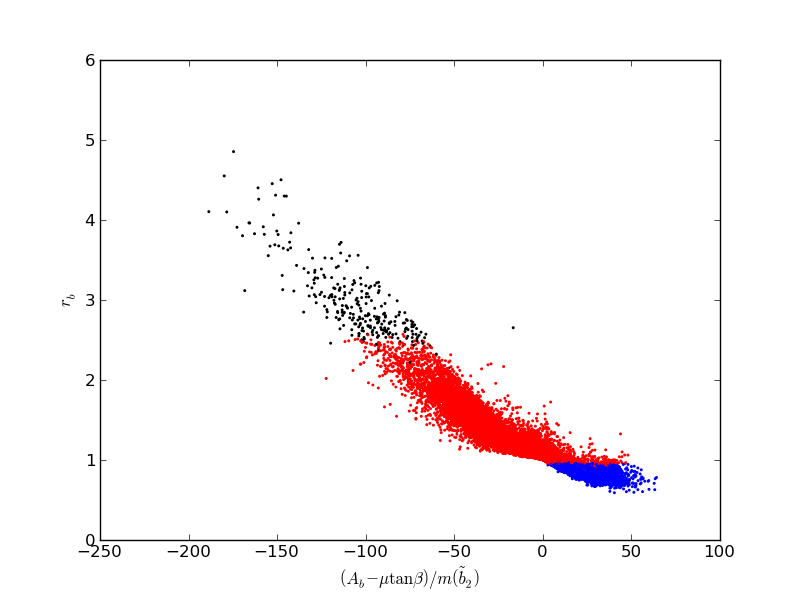}
     \put(40,70){Neutralino LSP}
     \end{overpic}
     } \\
\subfloat{
     \begin{overpic}[height=3.5in]{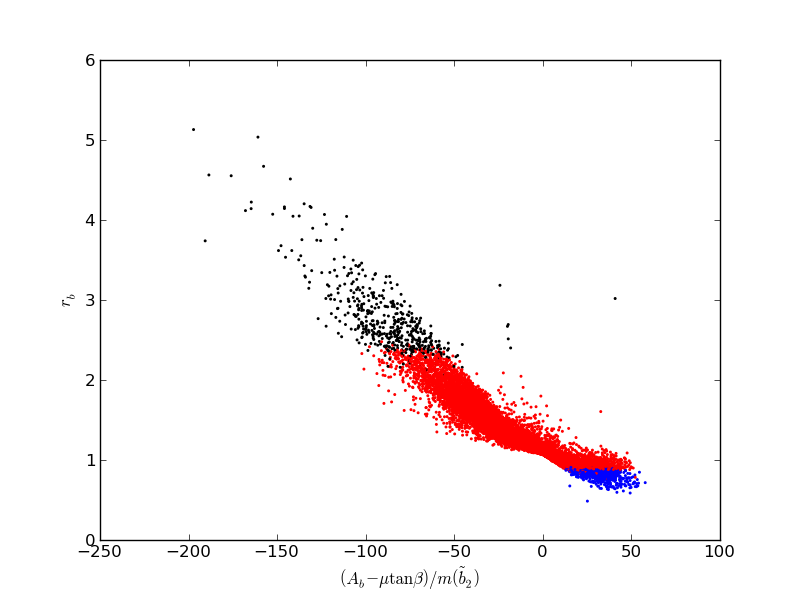}
     \put(40,70){Gravitino LSP}
     \end{overpic}
     }
\vspace*{0.5cm}
\caption{The relative $h\to b\bar b$ partial width, as a function of the ratio $X_b/m(\tilde b_2)$ , with $X_b$ as defined in the text. Plotted models have $m_h=125\pm 2$ GeV. The 
upper (lower) panel is for the neutralino (gravitino) model set. The color code is the same as in previous figures.}
\label{fig:higgsabc}
\end{figure}

How are the {\it other} properties of the pMSSM models influenced by requiring a light Higgs boson in the 123-127 GeV mass range? To address this problem we show in Fig.~\ref{fig:higgs10} the values of $X_t=(A_t-\mu \cot \beta)/M_S$ (with $M_S^2=m_{t_1}m_{t_2}$) as a 
function of either $m_{t_1}$ or $m_h$ for both model sets with the ranges of $R_{\gamma\gamma}$ color-coded as described above. Here we observe the well-known result that larger $X_t$ values are generally selected by requiring a Higgs mass in this range~\cite{Martin:1997ns}, with the maximal contribution to the Higgs mass from stop mixing occurring at $X_t = \sqrt{6} M_S$. Note that the blue points, corresponding to values of the ratio $1 < R_{\gamma\gamma} \leq 1.5$, are roughly evenly distributed within the preferred 
$X_t$ zones. Note also that as $m_{t_1}$ gets smaller the required value of $\frac{X_t}{M_S}$ grows. We observe that models with $m_{t_1}$ as low as $\sim 250$ GeV can still produce a Higgs mass in the desired range. Fig.~\ref{fig:higgs11} shows the effect of applying the Higgs mass cut on the distributions of $X_t$, $m_{t_1}$, $m_{t_2}$, and $\tan \beta$. We see that large (but not dangerously large in the sense that the stops are non-tachyonic) values of $X_t$ are preferred. We also notice that models with a very large $m_{t_1}$ are depleted whereas the opposite is true for $m_{t_2}$. Additionally, we see that models with low values of $\tan \beta$ 
are also relatively depleted by imposing the $m_h=125 \pm 2$ GeV requirement.

\begin{figure}
\centering
\subfloat{
     \begin{overpic}[width=3.5in]{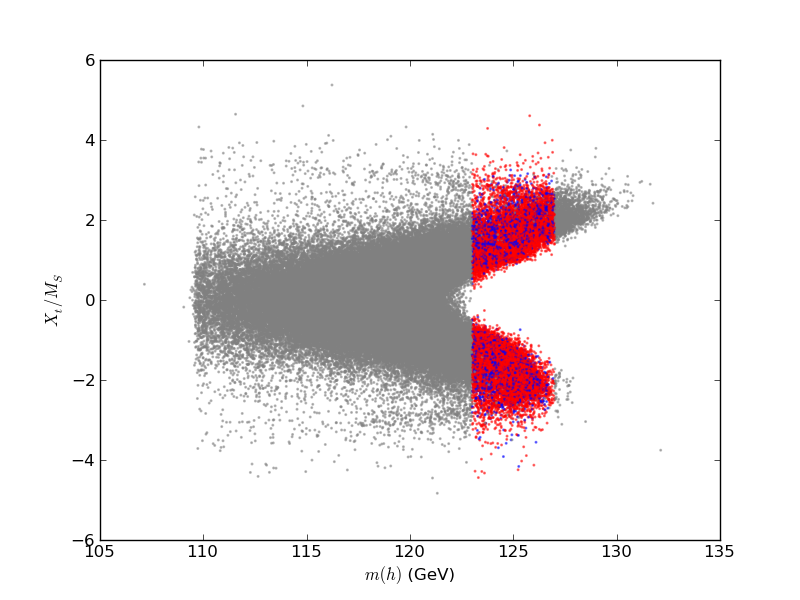}
     \put(37,70){Neutralino LSP}
     \end{overpic}
     } ~
\subfloat{ 
     \begin{overpic}[width=3.5in]{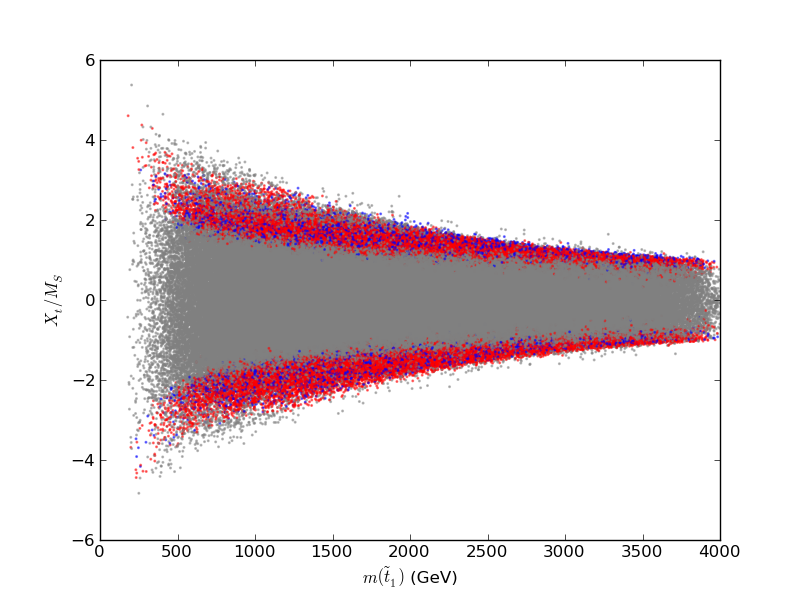}
     \put(37,70){Neutralino LSP}
     \end{overpic}
     } \\
\subfloat{
     \begin{overpic}[width=3.5in]{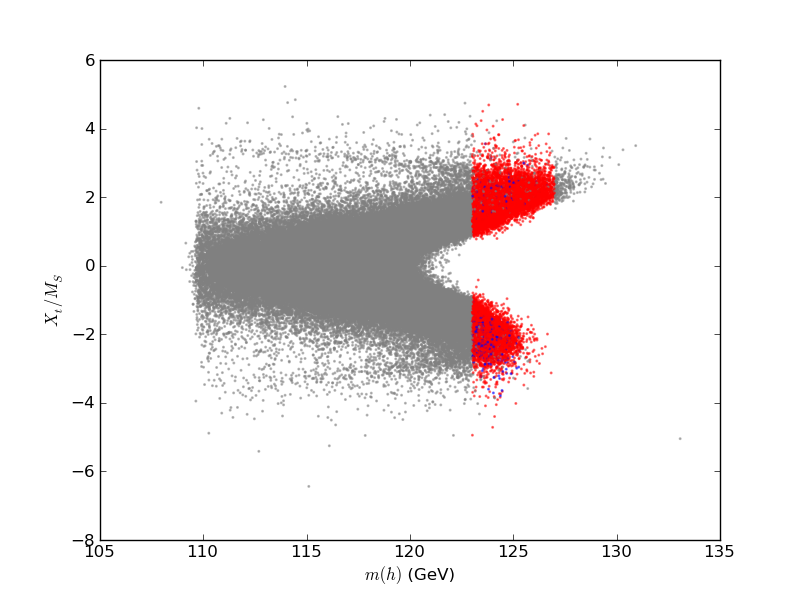}
     \put(37,70){Gravitino LSP}
     \end{overpic}
     } ~
\subfloat{
     \begin{overpic}[width=3.5in]{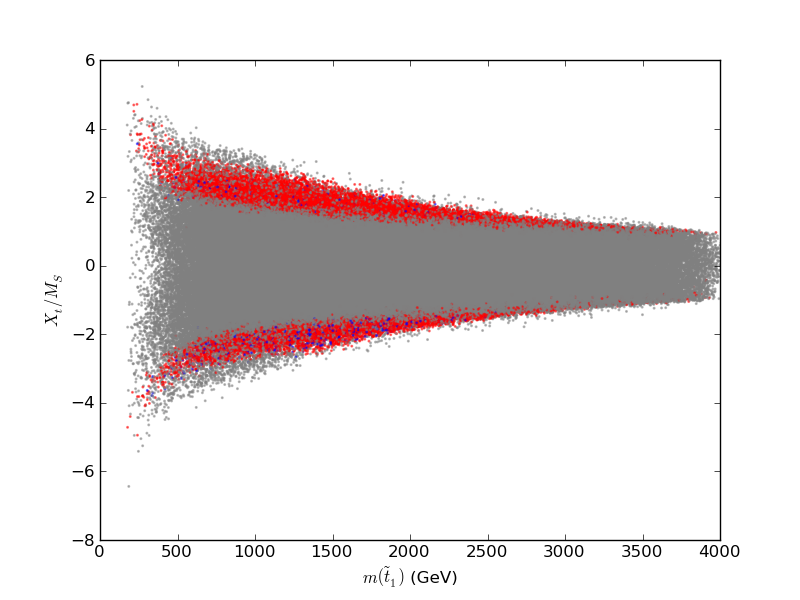}
     \put(37,70){Gravitino LSP}
     \end{overpic}
     }
\vspace*{0.5cm}
\caption{$X_t=(A_t-\mu/\tan \beta)/M_S$ (with $M_S^2=m_{t_1}m_{t_2}$) as a function of either $m_{t_1}$ or $m_h$ for the neutralino (top) and gravitino (bottom) model sets. The color code is the same as in previous figures.}
\label{fig:higgs10}
\end{figure}

\begin{figure}
\centering
\subfloat{
     \begin{overpic}[width=3.5in]{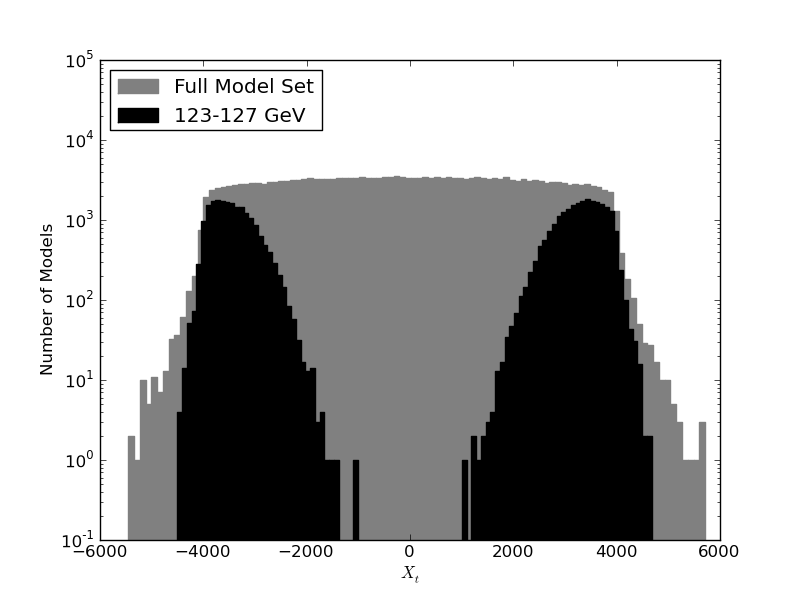}
     \put(37,70){Neutralino LSP}
     \end{overpic}
     } ~
\subfloat{ 
     \begin{overpic}[width=3.5in]{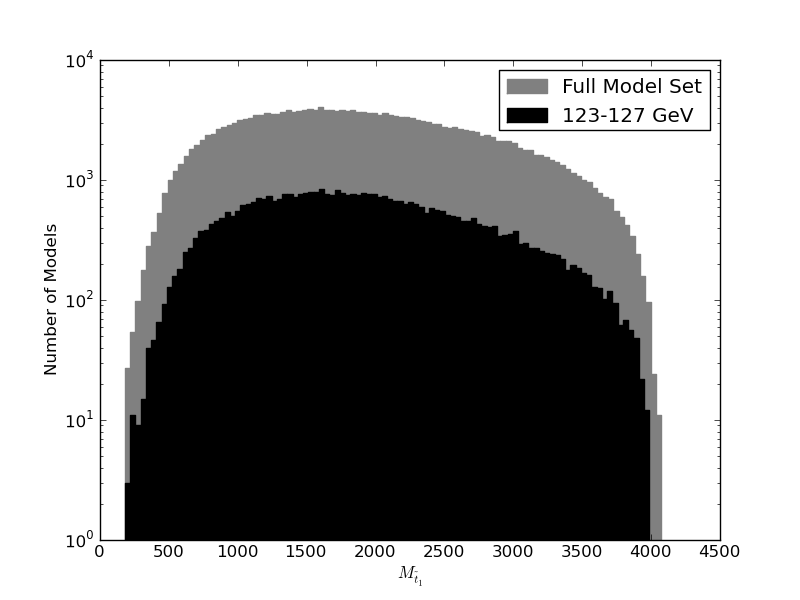}
     \put(37,70){Neutralino LSP}
     \end{overpic}
     } \\
\subfloat{
     \begin{overpic}[width=3.5in]{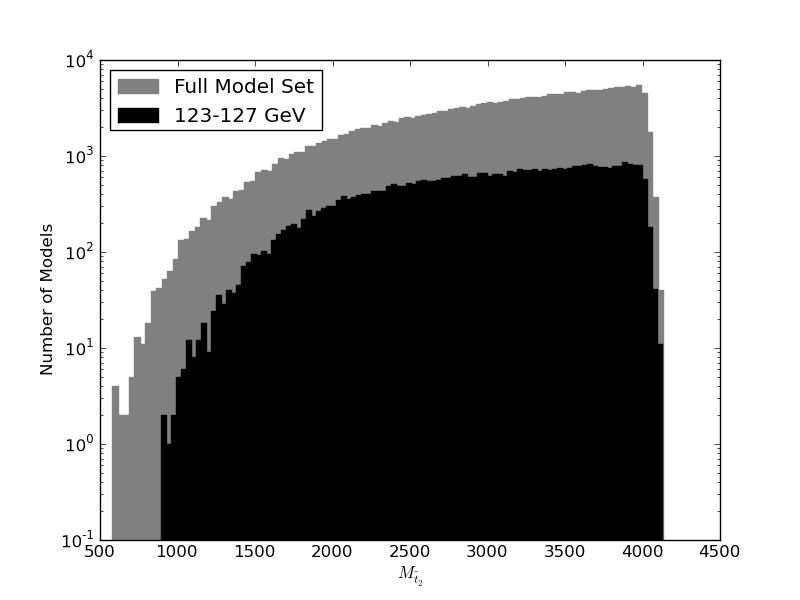}
     \put(37,70){Neutralino LSP}
     \end{overpic}
     } ~
\subfloat{
     \begin{overpic}[width=3.5in]{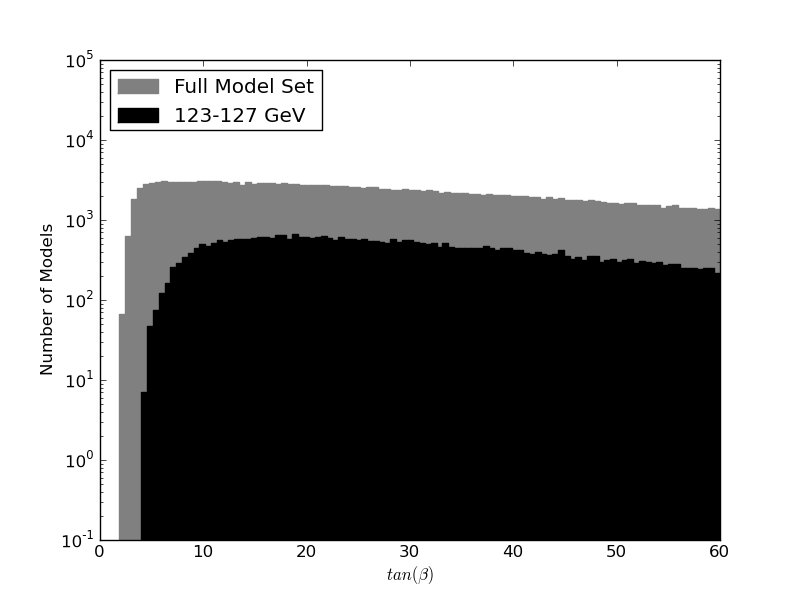}
     \put(37,70){Neutralino LSP}
     \end{overpic}
     }
\vspace*{0.5cm}
\caption{Comparison of the allowed values of $X_t, ~m_{t_1}$ and $m_{t_2}$ (all in GeV) as well as $\tan \beta$ both before and after applying the $m_h=125\pm 2$ GeV requirement in the 
neutralino pMSSM model set.}
\label{fig:higgs11}
\end{figure}

\subsection{Fine-Tuning in the pMSSM}
\label{sec:FT}

As has recently been discussed by several groups of authors~\cite{FT+Higgs}, the possible observation of a SM-like Higgs in the $\simeq 123-127 $ GeV mass range, combined 
with the lack of any sparticle signatures at the 7 TeV LHC, appears to indicate that the general MSSM is reasonably (or even substantially) fine-tuned (FT). Furthermore, the amount of 
FT suffered by specific models of SUSY breaking can be even more substantial (due to the correlations between weak-scale parameters resulting from the small number of input parameters at a high scale). It is therefore interesting to consider the range of fine tuning values found in both our neutralino and gravitino LSP model sets, and see how it is influenced by requiring 
a light Higgs mass of $125 \pm 2$ GeV. In the analysis that follows, we will employ the standard fine-tuning approach as discussed in Refs.~\cite{Ellis:1986yg, Barbieri:1987fn}, 
but now extended to the pMSSM.  Our analysis is based on the fundamental relationships between the 19\footnote{For the gravitino LSP model set, the effect of $m_{3/2}$ on the fine-tuning is completely negligible.} weak scale soft SUSY-breaking parameters of the pMSSM, denoted here as $p_i~(1 \leq i \leq 19)$, the mass 
of the $Z$ boson and the effective scalar mass parameters in the Higgs potential. Specifically, we consider the relation  
\begin{equation}
M_Z^2 = -2\mu^2 +2~ {{m_{H_d}^2-t_\beta^2 ~m_{H_u}^2}\over {t_\beta^2-1}}\,,
\end{equation}
where $t_\beta=\tan \beta$ and $m_{H_{d,u}}^2$ are the usual doublet mass terms in the Higgs potential. This relationship is assumed to hold beyond tree-level and  
include well-known radiative corrections. Since the masses $m_{H_{d,u}}^2$ themselves depend upon the various $p_i$ via these loop corrections, the usual quantities    
\begin{equation}
Z_i = {{\partial (\log M_Z^2)}\over {\partial (\log p_i)}} = ~{{p_i}\over {M_Z^2}} ~{{\partial M_Z^2}\over {\partial p_i}}
\end{equation}
can then be directly calculated. We then define the overall amount of FT in a given pMSSM model via the single parameter~\cite{Ellis:1986yg, Barbieri:1987fn} 
\begin{equation}
\Delta = max (|Z_i|)\,, 
\end{equation}
although an alternative definition of fine-tuning,
\begin{equation}
\delta =  \big (\sum_i Z_i^2\big ) ^{1/2}\,, 
\end{equation}
will also be considered briefly in the discussion below. Clearly in the limit that only one of the $Z_i$ dominates in this sum these two definitions will yield essentially identical 
results. In practice, this need not be the case, although the contributions to both fine-tuning measures are indeed dominated by only a few of the $Z_i$. Generally we expect that in a given model, 
$\delta$ will be somewhat larger (by factors of a few) than $\Delta$. Thus requiring $\delta$ to lie below a specific value will place a stronger fine-tuning constraint than requiring $\Delta$ to be below that same value.

In performing our calculations of fine-tuning we employ the same assumptions used during the generation of our two model sets (in particular, that the masses and Yukawa couplings and, for consistency, 
the associated $A$-terms of the SM fermions of the first two generations are zero). In this case, the 1-loop, leading-log (LL) contributions to the $Z_i$ arising from the  
five pMSSM Lagrangian parameters $M_{Q1,2}, ~M_{L1,2}, ~M_{u1,2}, ~M_{d1,2}$ and  $M_{e1,2}$ are all identically zero and, in addition, the corresponding 2-loop, next-to-leading-log (NLL) contributions  
from these same parameters are very highly suppressed and can be safely ignored. 

For a generic $p_i$, contributions to the corresponding $Z_i$ may {\it first} appear at tree-level, LL or NLL order. Although in most cases we will keep only the leading term, in some cases the numerics warrant including the higher order contribution as well. All of the various LL and NLL contributions can be directly obtained using the expressions for the 1- and 2-loop $\beta$ functions for the full set of MSSM parameters as given in detail in Ref.~\cite{Martin:1993zk}, and by introducing a cutoff scale, $X= log (\Lambda/M_S)$, where $M_S^2=m_{\tilde t_1}m_{\tilde t_2}$ defines the SUSY scale as usual. All LL(NLL) contributions are then proportional to this parameter $X(X^2)$. Conventionally in numerical calculations~\cite{FT+Higgs}, it is assumed that 
$X=3$ and we will follow this convention in our numerical analysis below.

Some of the $p_i$ lead to relatively small values for the corresponding $Z_i$ due to, \eg, a large value of $\tan^2 \beta$ or the presence of small gauge couplings. Alternatively, they may have typical LL contributions to $Z_i$ but have suppressed NLL
contributions. In such cases, it is sufficient to consider only their LL order fine-tuning contributions (the tree-level term being absent). The six parameters for which this occurs are: $A_{b,\tau},~M_{1,2}$ and $M_{L3,e3}$. As examples, at LL order 
we obtain for the two $A-$terms 
\begin{equation}
Z_{A_{b(\tau)}}^{LL} = {{3(1)X}\over {2\pi^2}} ~y_{b(\tau)}^2 ~{{A_{b(\tau)}^2}\over {M_Z^2}}~{{1}\over {t_\beta^2-1}}\,, 
\end{equation}
whereas for the electroweak gaugino mass parameters one has 
\begin{equation}
Z_{M_{2(1)}}^{LL} = {{X}\over {2\pi^2}} ~3g^2(g_Y^2) ~{{M_{2(1)}^2}\over {M_Z^2}}~\,. 
\end{equation}
Note that $Z_{M_2}$ can be significant when $M_2$ approaches the 1 TeV mass scale; for $\Delta < 100~(10)$ this would require $M_2 < 2070~(654)$ GeV in LL order. 
Further note that in the special case of $M_2$, a few of the models have NLL contributions that are as large as $\sim 10\%$ of the LL contribution. In general, however, we can safely ignore these NLL contributions.

For some parameters, such as $\mu$, the fine-tuning arises at tree level~\cite {Barbieri:1987fn}, with
\begin{equation}
Z_{\mu}^{TL} = {{4\mu^2}\over {M_Z^2}}~ \Big( 1 +{{M_A^2+M_Z^2}\over {M_A^2}} \tan^2 2\beta\Big)\,, 
\end{equation}
from which we can immediately read off the tree-level constraint on $\mu$ in the decoupling, large $\tan \beta$ regime, \ie, $Z_{\mu}^{TL}\simeq 4\mu^2/M_Z^2 < 100~(10)$ implies $\mu < 455~(145)$ GeV 
and, hence, the favored scenario of light Higgsinos. Given our parameter scan ranges we would expect $Z_\mu$ to play a dominant role in our fine-tuning calculations. Since the tree-level terms are so important 
in this case, the LL 
contributions are also included in our calculation of the fine-tuning arising from $\mu$, $Z_\mu^{LL}$ (see the Appendix for this expression), since they can have a significant numerical 
impact. The parameters $p_i=M_A$ and $\tan \beta$, on the other hand, require only the dominant tree-level contributions, where we use the expressions given in 
Ref.~\cite {Barbieri:1987fn}. As we will see, the contribution of $M_A$ to FT can be important for small and moderate values of $\tan \beta$.  

For the three parameters $A_t, M_{Q3}$ and $M_{u3}$, the corresponding $Z_i$ contributions obtained at LL are potentially large and the resulting constraints 
strong so that including the corresponding NLL contributions is necessary to obtain a reliable estimate of their true impact. This is particularly important as these parameters are also crucial for generating large Higgs masses in the range of current interest, $m_h=125\pm 2$ GeV. As an example, consider $Z_{A_t}$ at LL (those for $M_{Q3}$ and $M_{u3}$ are similar and are given in the Appendix), where we obtain: 
\begin{equation}
Z_{A_t}^{LL} = {{3X}\over {2\pi^2}} ~y_t^2 ~{{A_t^2}\over {M_Z^2}}~{{-t_\beta^2}\over {t_\beta^2-1}}\,, 
\end{equation}
where $y_t$ is the top quark Yukawa coupling. Here we see a good example of the somewhat general result that the LL value of each $Z_i$ directly constrains a single pMSSM parameter. 
For large $\tan \beta \gg 1$ and with $y_t^2 \sim 1/2$, demanding $|Z_{A_t}^{LL}|<100~(10)$ implies $A_t \lsim 1.91~(0.60)$ TeV. This constraint gets somewhat more complicated when the NLL 
contribution is also included, as several of the other pMSSM parameters become involved in the analytic expression. We find that in this case the NLL contribution is given by 
\begin{equation}
Z_{A_t}^{NLL} = {{24X^2}\over {(16\pi^2)^2}} ~{{A_t}\over {M_Z^2}} ~{{-y_t^2}\over {t_\beta^2-1}} ~\Big [T_1+t_\beta^2(-T_2+T_3) \Big]\,, 
\end{equation}
where $T_1=y_b^2(A_t+A_b)$, $T_2=12y_t^2A_t+y_b^2(A_t+2A_b)$ and $T_3=(4/3)[4g_s^2(A_t-M_3)+(g_Y^2/3)(A_t-M_1)]$, with $g_{s(Y)}$ being the SM strong(hypercharge) coupling constant and 
$y_b$ being the $b-$quark Yukawa coupling. Scanning the pMSSM parameter space of our two model sets, we find that this NLL contribution can generally soften the LL constraint on the value of 
$A_t$ by $10-15\%$, with values as large as $\sim 2.2~(0.7)$ TeV now being allowed if we still require that $|Z_{A_t}^{LL+NLL}|<100~(10)$. This result is quite general: We find that including 
these important NLL contributions to the $Z_i$ tends to slightly soften the FT constraints obtained at LL. Thus, considering such terms somewhat decreases the overall amount of 
FT in our model sets. Similar NLL results can be obtained (and are particularly important) for both $M_{Q3}$ and $M_{u3}$, and can be found in the Appendix.  

In the case of the gluino mass parameter $M_3$, the contribution to the corresponding $Z_{M_3}$ first appears at NLL. However, it can be numerically significant since it scales roughly 
proportional to $(\frac{\alpha_s}{\pi})(\frac{M_3^2}{M_Z^2})$. More explicitly, at NLL we find the result
\begin{equation}
Z_{M_3}^{NLL} = {{2\alpha_s X^2}\over {(3\pi^3)}(t_\beta^2-1)} ~{{M_3}\over {M_Z^2}} ~\Big [-y_b^2(2M_3-A_b)+t_\beta^2 y_t^2(2M_3-A_t) \Big]\,, 
\label{eq:M3}
\end{equation}
which shows that a significant cancellation is possible when $A_t \simeq 2M_3$. In practice, we will see that $Z_{M_3}$ will not be too important in determining the overall amount 
of FT.

The complete expressions for all non-zero $Z_i$ contributions not given above are provided in the Appendix.

\begin{figure}
\centering
\subfloat{
     \begin{overpic}[height=3.5in]{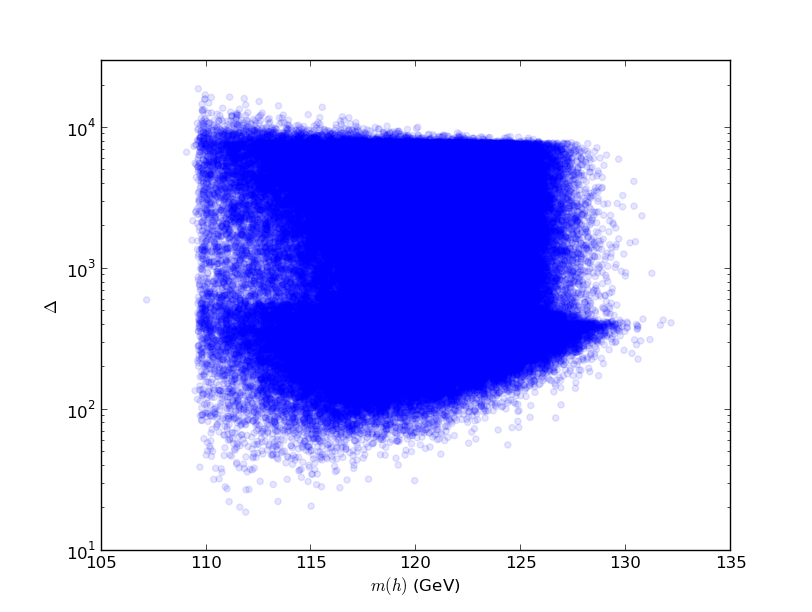}
     \put(40,70){Neutralino LSP}
     \end{overpic}
     } \\
\subfloat{
     \begin{overpic}[height=3.5in]{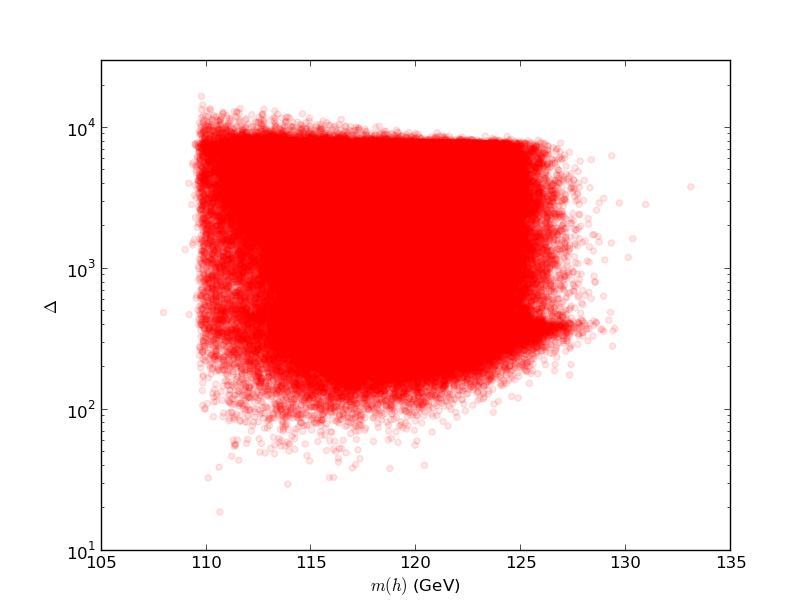}
     \put(40,70){Gravitino LSP}
     \end{overpic}
     }
\vspace*{0.5cm}
\caption{Fine-tuning as a function of the light Higgs mass in the neutralino (top blue) and gravitino (bottom red) LSP model sets.}
\label{fig:ft1}
\end{figure}

Figure~\ref{fig:ft1} shows the values of $\Delta$ obtained for our two model sets as functions of the light Higgs mass. Here, one sees that in either set we obtain the well-known result 
that pMSSM models with lower (higher) values of $m_h$ are less (more) fine-tuned. However, even without restricting the value of $m_h$ to be $125 \pm 2$ GeV, the number of 
models with low FT values is not large (although neutralino LSP models are somewhat more successful in this regard). The small number of Low-FT models ultimately results from the large scan range for $\mu$, since the large majority of $\mu$ values in the scan range will lead directly to large fine-tuning.
Interestingly, for values of $m_h$ above the LEP limit of $\sim 115$ GeV, we observe that the smallest 
obtained values of $\Delta$ grow essentially exponentially with increasing $m_h$ in both model sets. 

To get a better overall impression of the amount of fine-tuning in the two model sets and to emphasize their differences, we show in Fig.~\ref{fig:ft2} histograms of the number of models with fine-tuning 
{\it below} some fixed value of $\Delta$. Several different results are compared: the distribution of $\Delta$ for the full neutralino and gravitino LSP model sets, as well as the corresponding 
results obtained after imposing the $m_h=125 \pm 2$ GeV requirement. The kinks in the cumulative distributions occur when $\mu$ takes over as the dominant source of FT, as will be discussed below. Here we clearly see that demanding a large Higgs mass forces large values of $\Delta$. For $m_h=125 \pm 2$ GeV, very few models have $\Delta \leq 100$: We find only 15 (1) in the case of the neutralino (gravitino) LSP model set. Of the 15 neutralino LSP models, 13 pass the various LHC MET and stable sparticle searches ({\it i.e.,} are not yet excluded) \cite{CahillRowley:2012cb}. The single gravitino LSP model has a chargino NLSP that would decay in the LHC detector, so determining its viability requires a dedicated study which we leave to a future work~\cite{future}. Of course, as we can see from Fig.~\ref{fig:ft2}, the number of `satisfactory' 
models in either case grows quite rapidly as we soften our requirement on the value of $\Delta$, especially in the range $100 \lsim \Delta \lsim 500$.

\begin{figure}
\centerline{\includegraphics[height=6in]{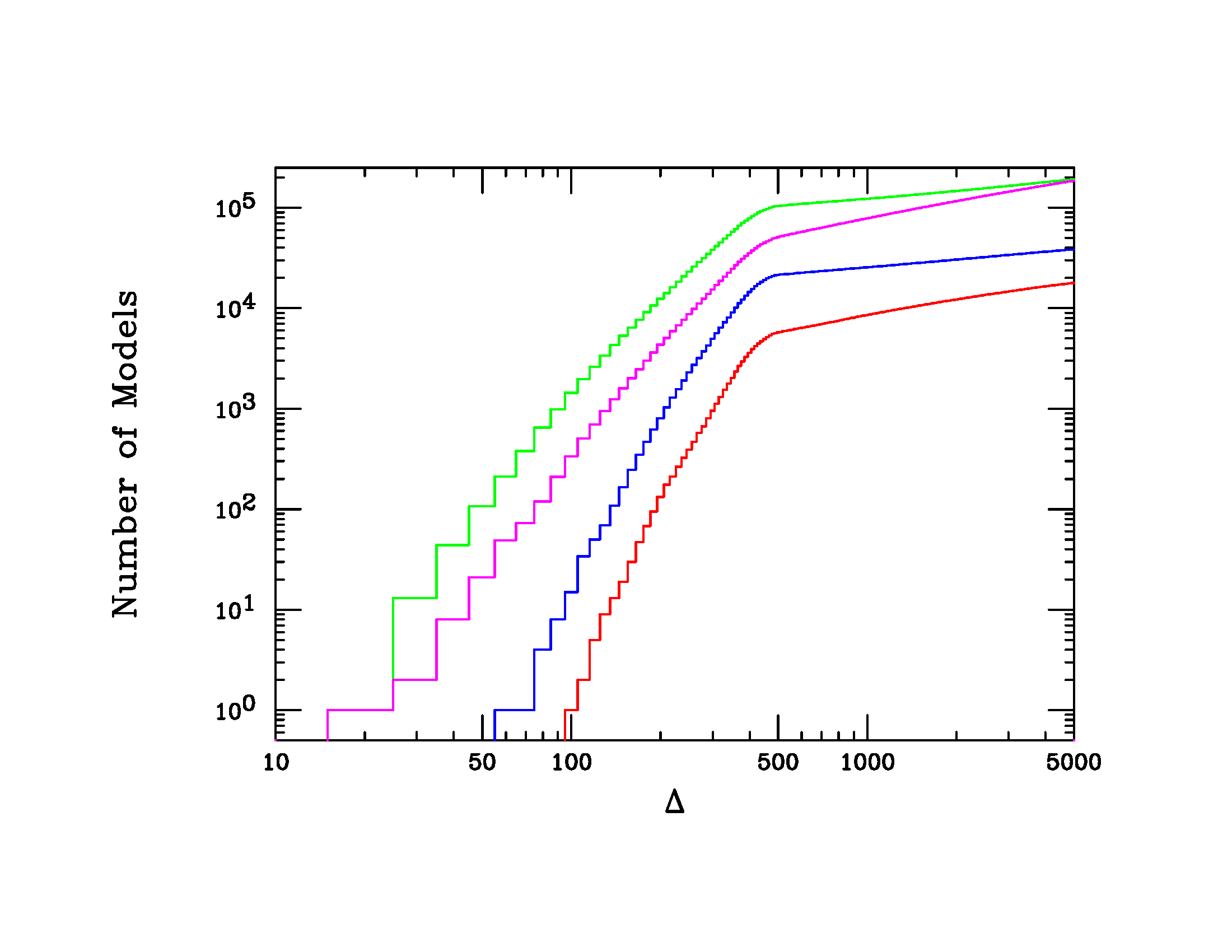}}
\vspace*{0.5cm}
\caption{Number of models with FT $\leq \Delta$. From top to bottom the histograms are for the full neutralino LSP model set (green), the full gravitino set (magenta), the neutralino set after requiring $m_h=125 \pm 2$ GeV (blue) and the gravitino set with $m_h=125 \pm 2$ GeV (red).}
\label{fig:ft2}
\end{figure}

For purposes of comparison, Fig.~\ref{fig:ft3} shows histograms of the analogous values for the FT parameter $\delta$. Since in any model we must have $\delta \geq \Delta$, we would 
expect the pMSSM models with $\delta$ below a given limit to be less frequent than those with $\Delta$ below the same value. In this figure, we see that this expectation is clearly realized for both model sets.

\begin{figure}
\centerline{\includegraphics[height=6in]{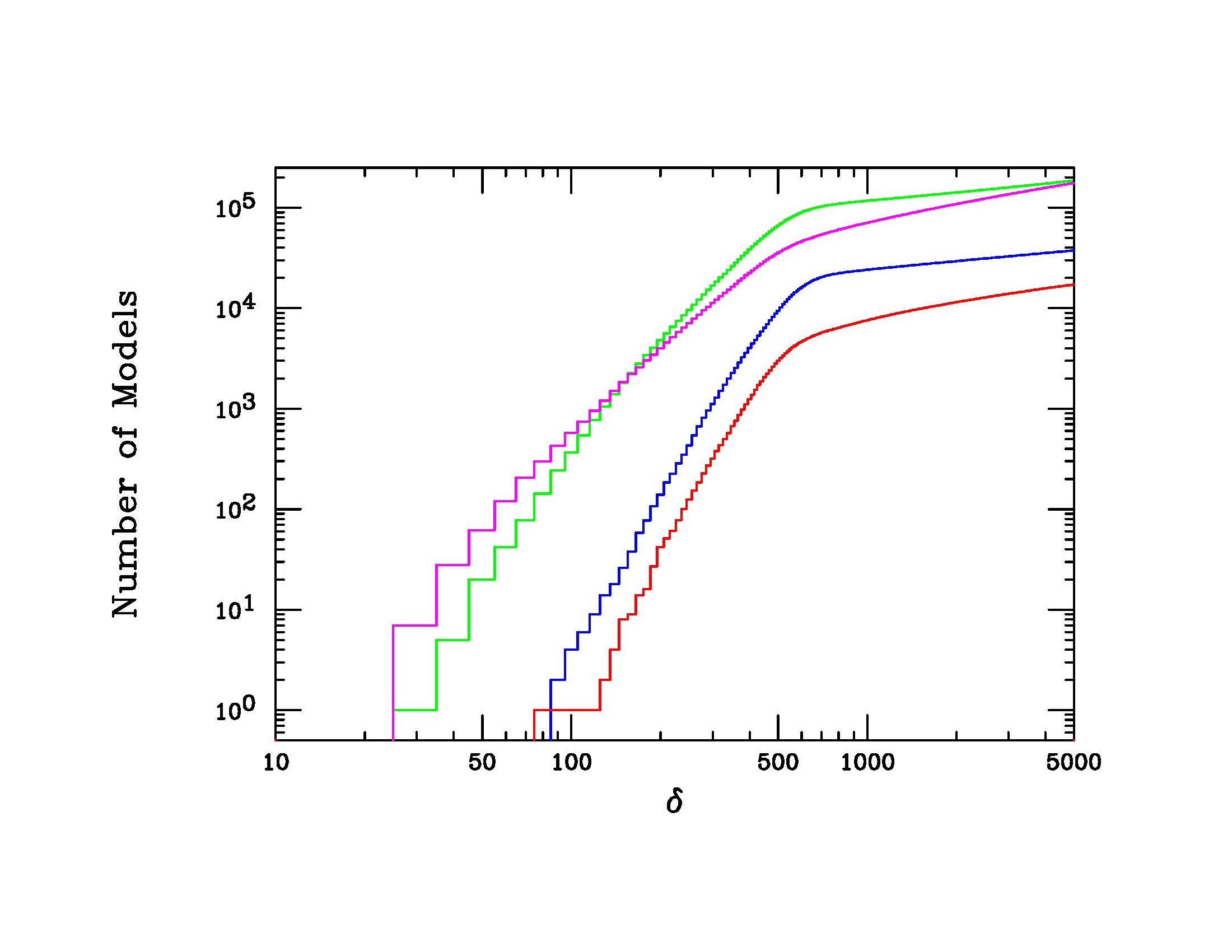}}
\vspace*{0.5cm}
\caption{Same as the previous Figure but now for the FT parameter $\delta$.}
\label{fig:ft3}
\end{figure}

It is interesting to ask which of the $Z_i$ dominates the FT; Figs.~\ref{fig:ft4} and ~\ref{fig:ft5} address this question for both LSP model sets, before and after including the Higgs boson mass 
constraint of $m_h=125\pm 2$ GeV. In all cases we see that the contribution from $\mu$ is dominant, which should not be too surprising given its tree-level nature and the discussion above. Specifically, $\mu$ is scanned over a wide range of values, most of which lead to large fine-tuning regardless of the other model parameters. After $\mu$, the parameters $M_{Q3},~M_{u3}$ and $A_t$ are next in importance in generating significant fine-tuning. Unsurprisingly, their 
importance increases substantially when large Higgs masses are required, since these parameters are involved in the necessary large radiative corrections. We see that the fine-tuning arising from these parameters is significantly larger in the neutralino LSP model set; this is 
a result of the somewhat heavier (on average) sparticle mass spectra in this model sample. Note that the gluino mass parameter, $M_3$, is never among the most dominant contributors due to the relatively small numerical coefficient in the NLL expression (see Eq.~\ref{eq:M3}). (This does {\it not}, however, mean that large values of $M_3$ can't produce large fine-tuning, only that they are never the dominant fine-tuning source).
Lastly, we note that there are a reasonable number of cases where $M_2$ provides the dominant contribution to FT (this occurs when $M_2$ is heavier than $\sim 1$ TeV as discussed above).

\begin{figure}
\centering
\subfloat{
     \begin{overpic}[height=3.5in]{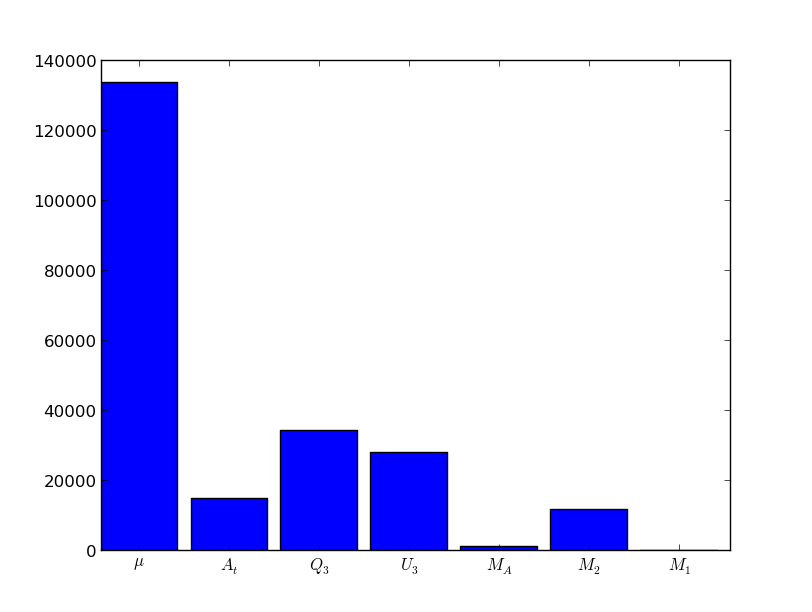}
     \put(40,70){Neutralino LSP}
     \end{overpic}
     } \\
\subfloat{
     \begin{overpic}[height=3.5in]{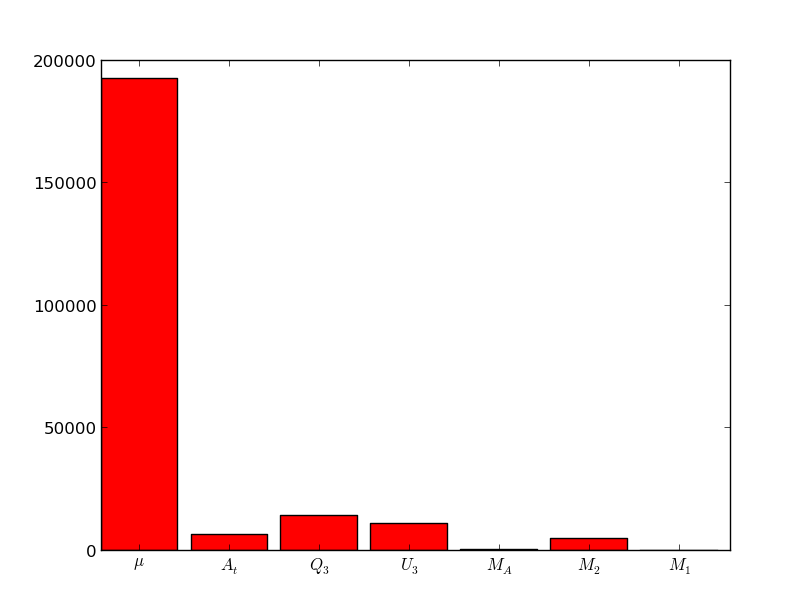}
     \put(40,70){Gravitino LSP}
     \end{overpic}
     }
\vspace*{0.5cm}
\caption{Histograms of the identities of the largest $Z_i$ for the full neutralino (top) and gravitino (bottom) LSP model sets.}
\label{fig:ft4}
\end{figure}

\begin{figure}
\centering
\subfloat{
     \begin{overpic}[height=3.5in]{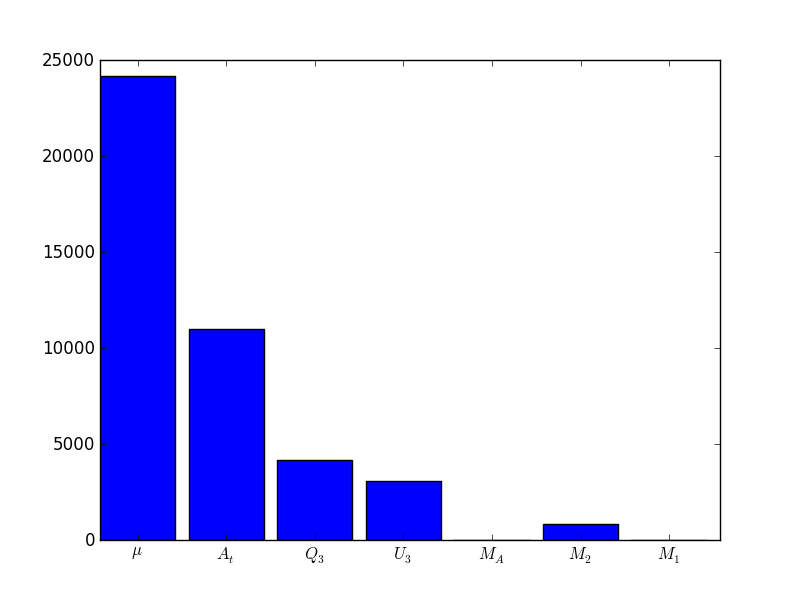}
     \put(40,70){Neutralino LSP}
     \end{overpic}
     } \\
\subfloat{
     \begin{overpic}[height=3.5in]{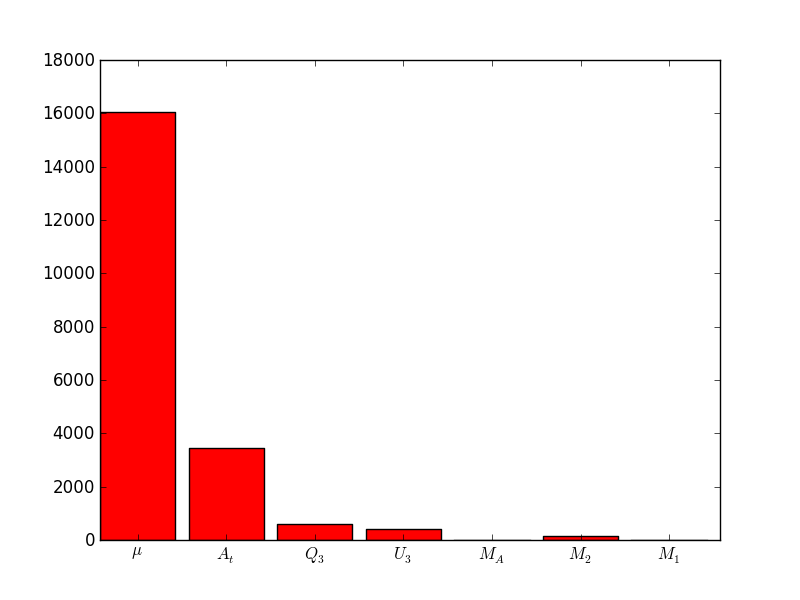}
     \put(40,70){Gravitino LSP}
     \end{overpic}
     }
\vspace*{0.5cm}
\caption{Same as the previous Figure but now requiring $m_h=125\pm 2$ GeV.}
\label{fig:ft5}
\end{figure}

Further information about the impact of specific parameters on FT can be obtained by examining the fractional contribution of the various $Z_i$ to the parameter $\delta$ and how these are influenced by the $m_h=
125 \pm 2$ GeV Higgs mass constraint. These distributions for both LSP model sets can be found in Figs.~\ref{fig:ft6} and~\ref{fig:ft7}. Note that while these results for the neutralino 
and gravitino LSP model sets are qualitatively quite similar, they differ in some details. In both cases we see that while the value of $\mu$ is most commonly the dominant fine-tuning contribution, it always fails to make up more than $90\%$ of the total. However, while  $M_{Q3},~M_{u3}$ and $A_t$ are subdominant, they can essentially saturate the $\delta$ sum in some cases. When the Higgs mass constraint is imposed, the relative importance of the $A_t$ contribution to FT is seen to significantly increase in both model sets (due to the large stop mixing requirement) but even more so in the 
neutralino LSP set. In all cases $M_3$ is seen to play a sub-dominant role.

\begin{figure}
\centering
\subfloat{
     \begin{overpic}[height=3.5in]{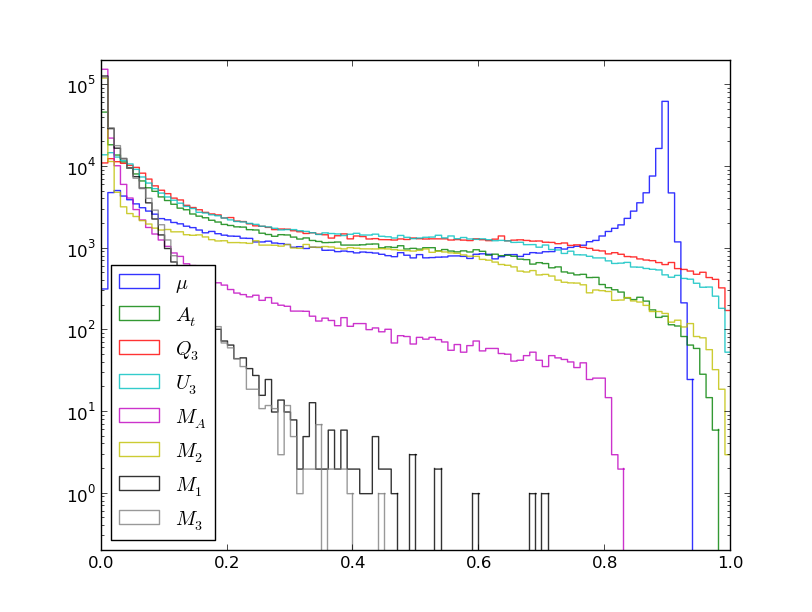}
     \put(40,70){Neutralino LSP}
     \end{overpic}
     } \\
\subfloat{
     \begin{overpic}[height=3.5in]{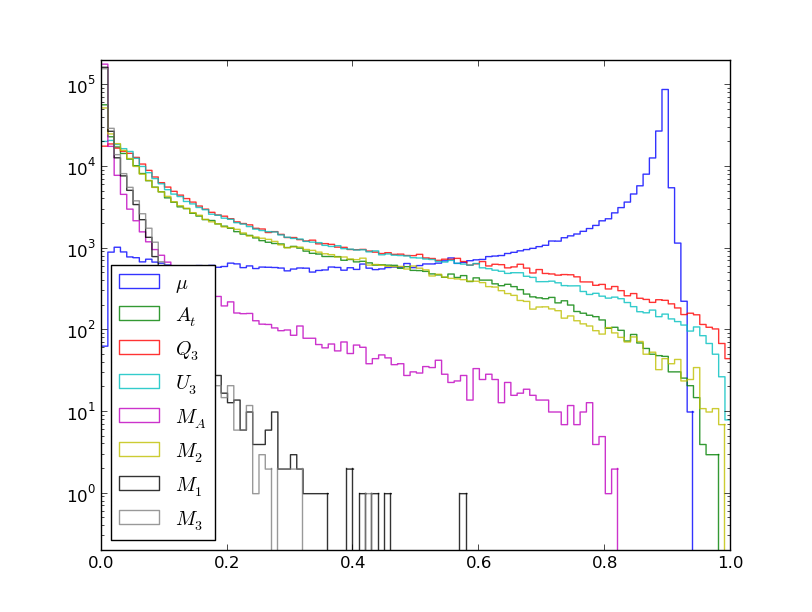}
     \put(40,70){Gravitino LSP}
     \end{overpic}
     }
\vspace*{0.5cm}
\caption{Histograms of the number of models as a function of the fractional contributions of the various $Z_i$ to the parameter $\delta$ for both the neutralino (top) and gravitino (bottom) 
LSP model sets. Larger values on the $x$-axis correspond to the greater dominance of a particular $Z_i$ in the sum $\delta$.}
\label{fig:ft6}
\end{figure}

\begin{figure}
\centering
\subfloat{
     \begin{overpic}[height=3.5in]{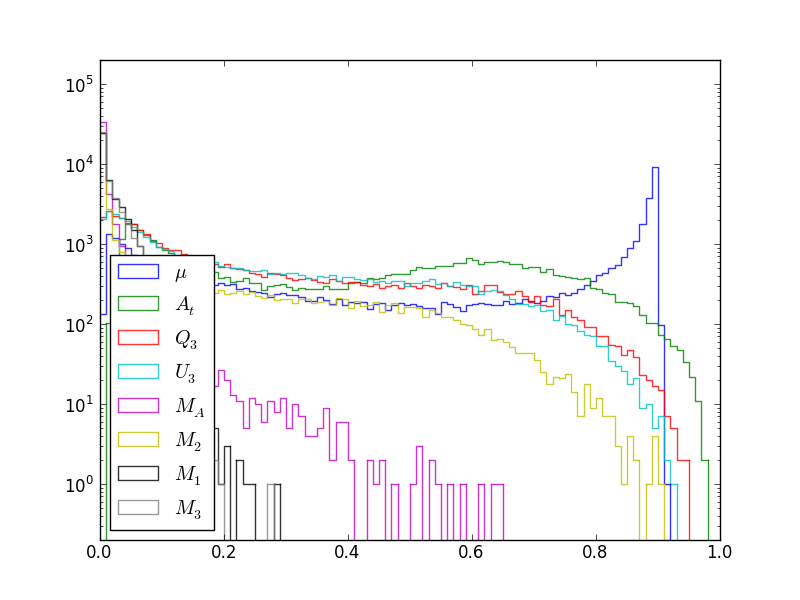}
     \put(40,70){Neutralino LSP}
     \end{overpic}
     } \\
\subfloat{
     \begin{overpic}[height=3.5in]{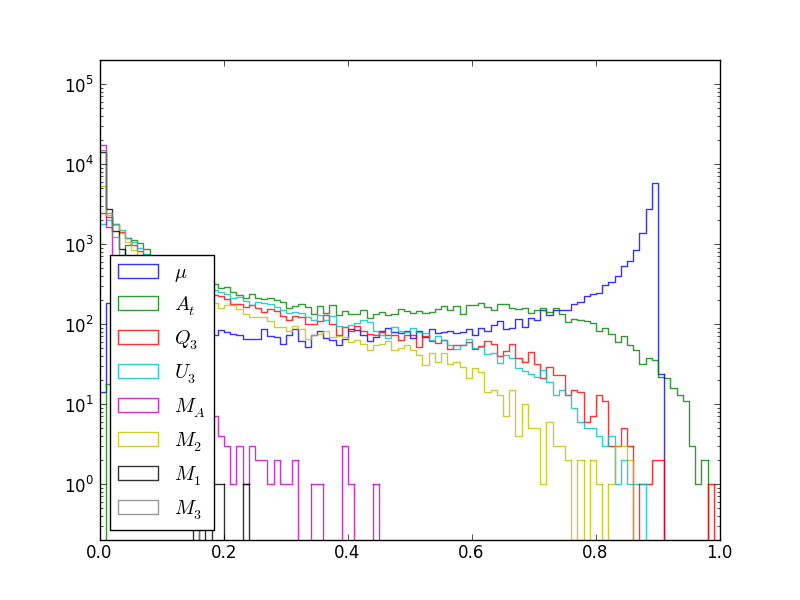}
     \put(40,70){Gravitino LSP}
     \end{overpic}
     }
\vspace*{0.5cm}
\caption{Same as the previous Figure but now requiring $m_h=125\pm 2$ GeV.}
\label{fig:ft7}
\end{figure}

\section{Implications of Low Fine-Tuning and Higgs Mass Requirements} 

\label{sec:imp}

If we simultaneously require low FT and a light Higgs mass of $m_h=125\pm 2$ GeV in the pMSSM, we find some rather drastic constraints on the structure 
of the SUSY mass spectrum~\cite{FT+Higgs}. To see some examples of what such spectra may be like in the pMSSM, we consider the 13 neutralino LSP models with $\Delta < 100$ that have $m_h=125\pm 2$ GeV and have not yet been excluded by the LHC SUSY searches as described in~\cite{CahillRowley:2012cb}. Although this is a rather small 
sample upon which to base any final conclusions, examining these models in detail can give us an idea of the characteristics of a viable natural sparticle spectrum, as well as the
potential challenges upcoming LHC 3$^{rd}$ generation SUSY searches may face if the MSSM with low fine tuning is realized in nature.

Figures~\ref{fig:spec0},~\ref{fig:spec1},~\ref{fig:spec2} and~\ref{fig:spec3} show the sparticle mass spectra for these 13 models in this low FT subset. As one might expect, these model spectra share a number of common features: ($i$) The first and second generation squarks and the gluino (mostly) lie above 1.25 TeV, except in model 1477135 which has a light $\tilde d_R$ at $\sim 560$ GeV. In this model, $\tilde d_R$ mostly decays to $\tilde \chi_2^0$ and not to the LSP and thus has lower MET. 
($ii$) The LSP is either Higgsino-like or a Higgsino-wino admixture. From the discussion above, naturalness requires the Higgsinos to be very light and also requires the winos to be fairly light, so this tendency is as expected. Interestingly, winos in these 13 models tend to be even lighter than can be explained by fine-tuning alone, apparently as a result of constraints from the measurement of $b \to s \gamma$, which plays an important role due to the presence of light stops and charginos. The absence of light binos in these models is unsurprising because $M_1$ is not strongly biased towards light values by the fine-tuning requirement, so that randomly chosen bino masses are unlikely to be below the Higgsino mass in such a small sample. Additionally, a light bino LSP may be excluded in some cases by overclosure of the universe, although the presence of other light states may allow for mixing or co-annihilation allowing for this constraint to be avoided. Interestingly, in most (12/13) models, five of the six electroweak gauginos lie below the 
$\tilde t_1$ and/or $\tilde b_1$. As we will discuss in detail below, this results in complex decay patterns for the light stop and/or sbottom that are difficult to observe in the standard LHC searches.
($iii$) All models have a chargino with a mass below $\simeq 270$ GeV. Additionally, the mass splittings between the electroweak gauginos are typically small, frequently leading to soft decay products.
($iv$) The lightest stop (sbottom) has a mass in the range $0.32-1.10~(0.40-1.70)$ TeV. In some models, the lightest sbottom is lighter than the lightest stop. ($v$) The slepton masses are essentially randomly distributed throughout the spectrum.
($vi$) Not shown in these Figures is the result that all these models have $M_A> 460$ GeV along with $\tan \beta > 13.5$.  

Since the presence of light stops and sbottoms is a generic feature of SUSY models with low fine-tuning, the sensitivity of collider searches to these sparticles is a topic of current investigation~\cite{LHCGEN3}. We therefore turn our attention to the phenomenology of 3$^{rd}$ generation squarks in our low fine-tuning models. Particularly, we consider the cascade decays resulting from the presence of multiple electroweak gauginos below the stop and sbottom masses. To be specific, we will first consider model 2403883, shown in detail in Fig.~\ref{fig:spec0}, as the `prototypical' example of a low fine-tuning model with light winos and Higgsinos, and then generalize our observations to the rest of the low fine-tuning models with light winos and Higgsinos.

\begin{figure}
\centering
\includegraphics[width=6.5in]{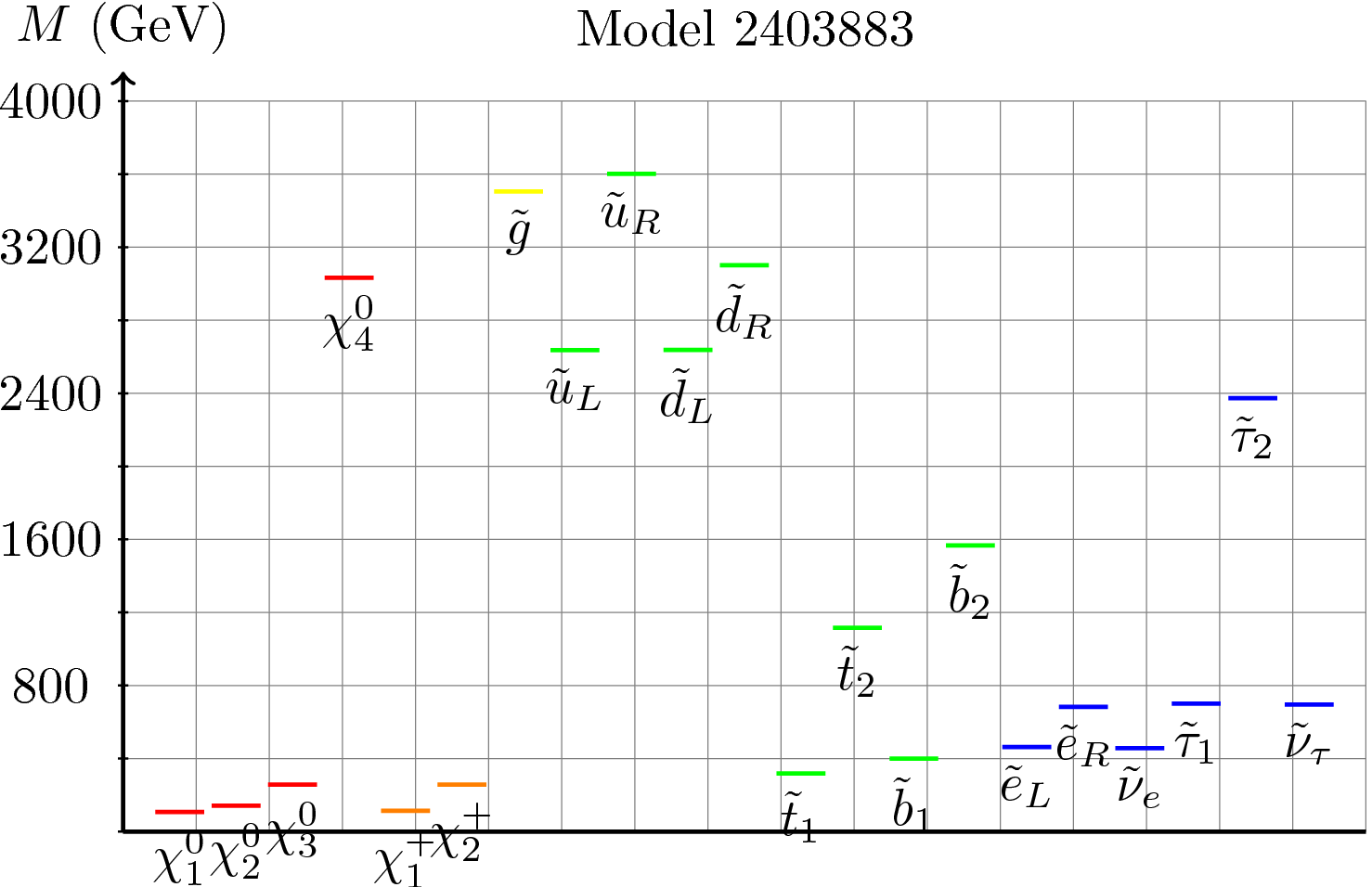}
\vspace*{0.5cm}
\caption{Sparticle mass spectrum of a neutralino LSP pMSSM model, 2403883, which satisfies $m_h=125\pm2$ GeV, $\Delta < 100$, and all current search constraints.}
\label{fig:spec0}
\end{figure}

\begin{figure}
\centering
\subfloat{\includegraphics[width=3.5in]{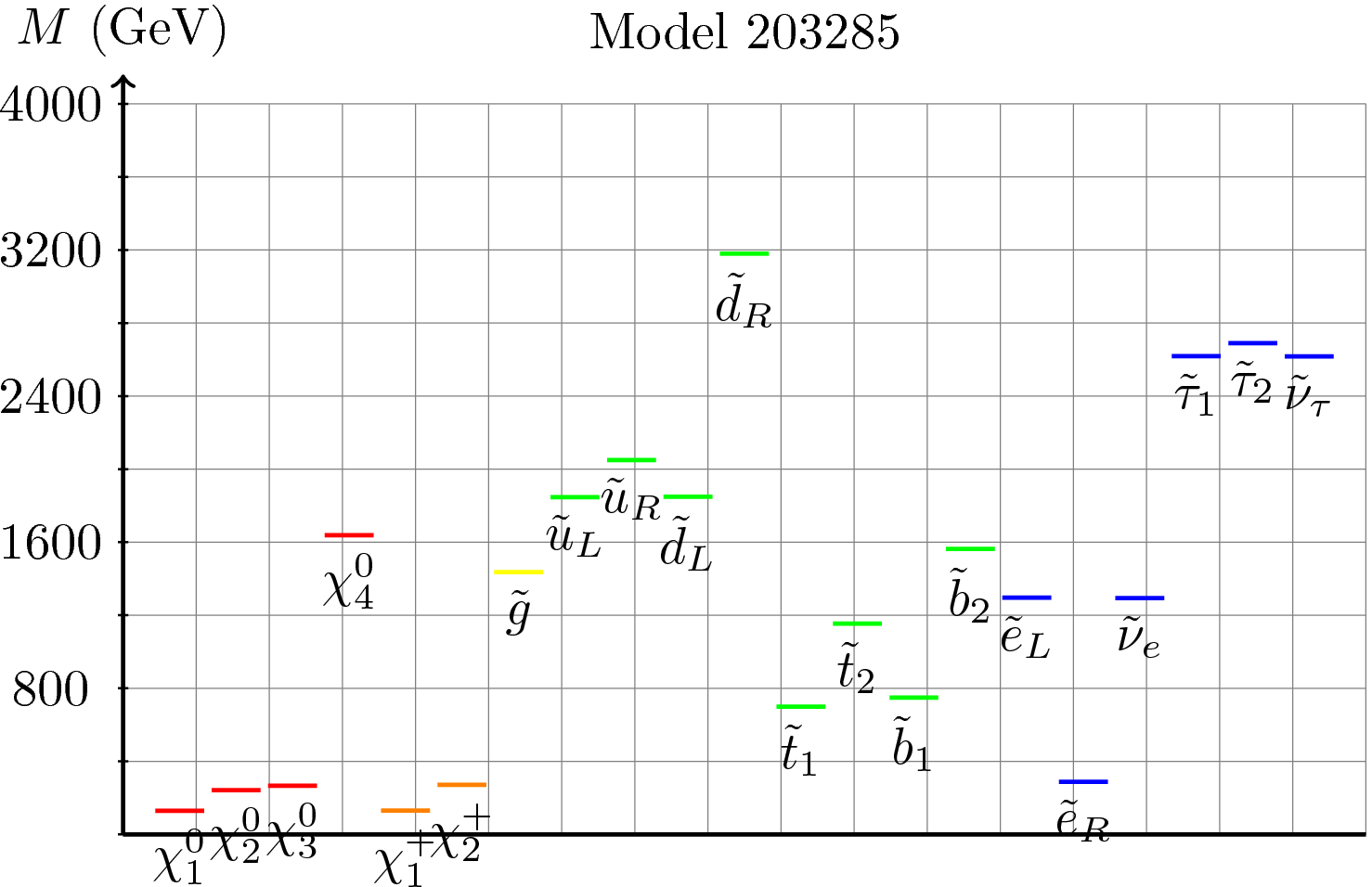}} ~
\subfloat{\includegraphics[width=3.5in]{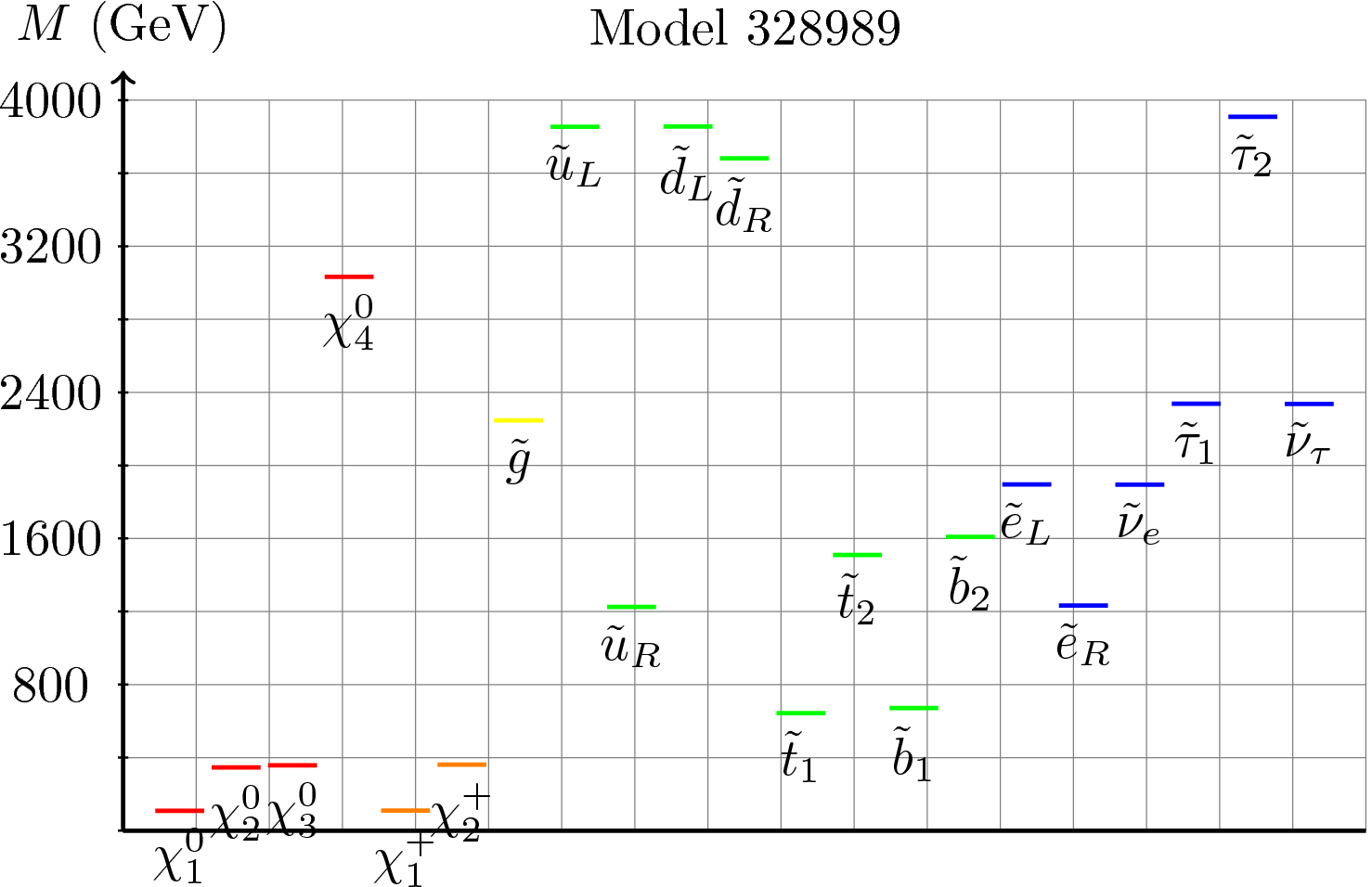}} \\
\subfloat{\includegraphics[width=3.5in]{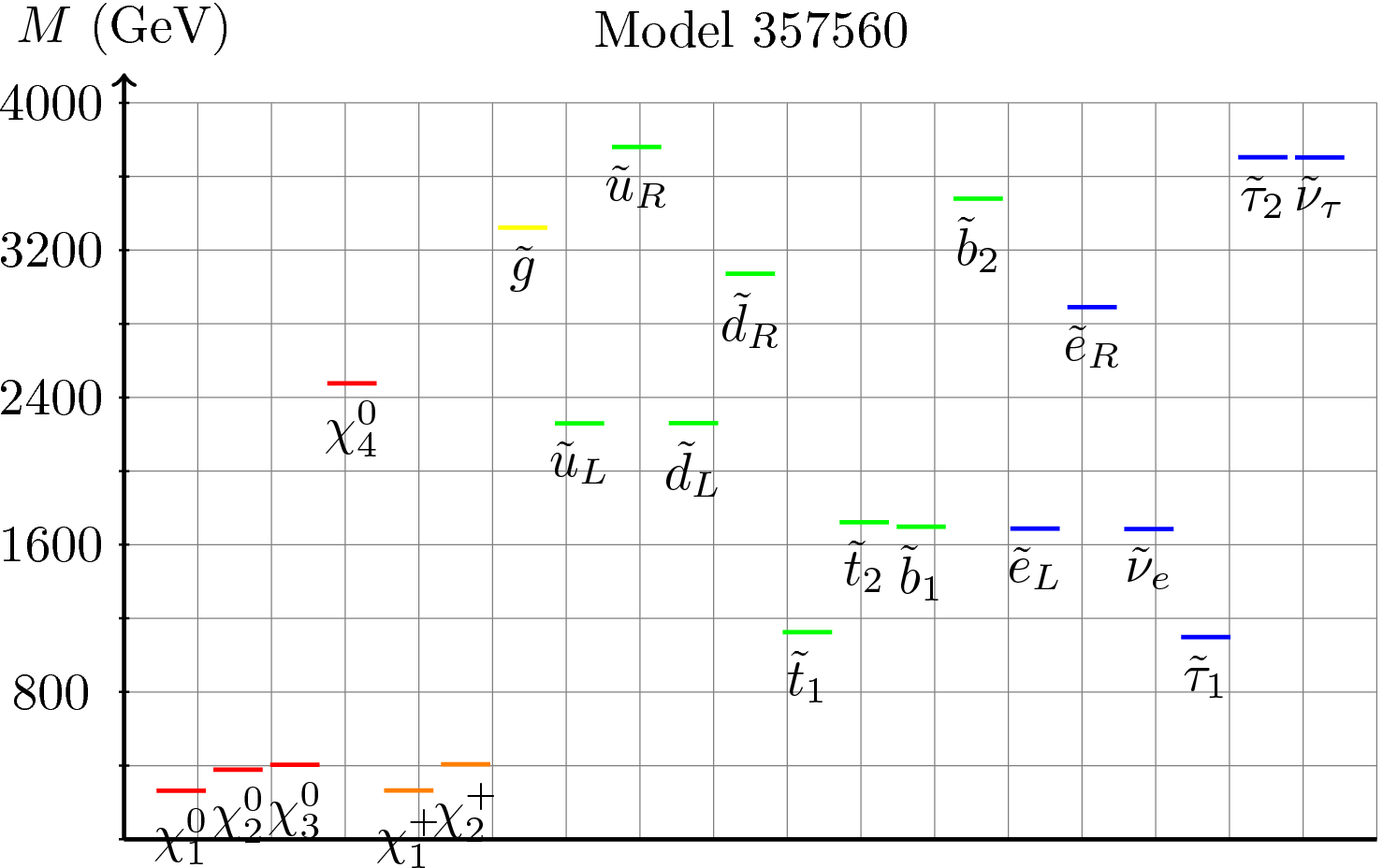}}~
\subfloat{\includegraphics[width=3.5in]{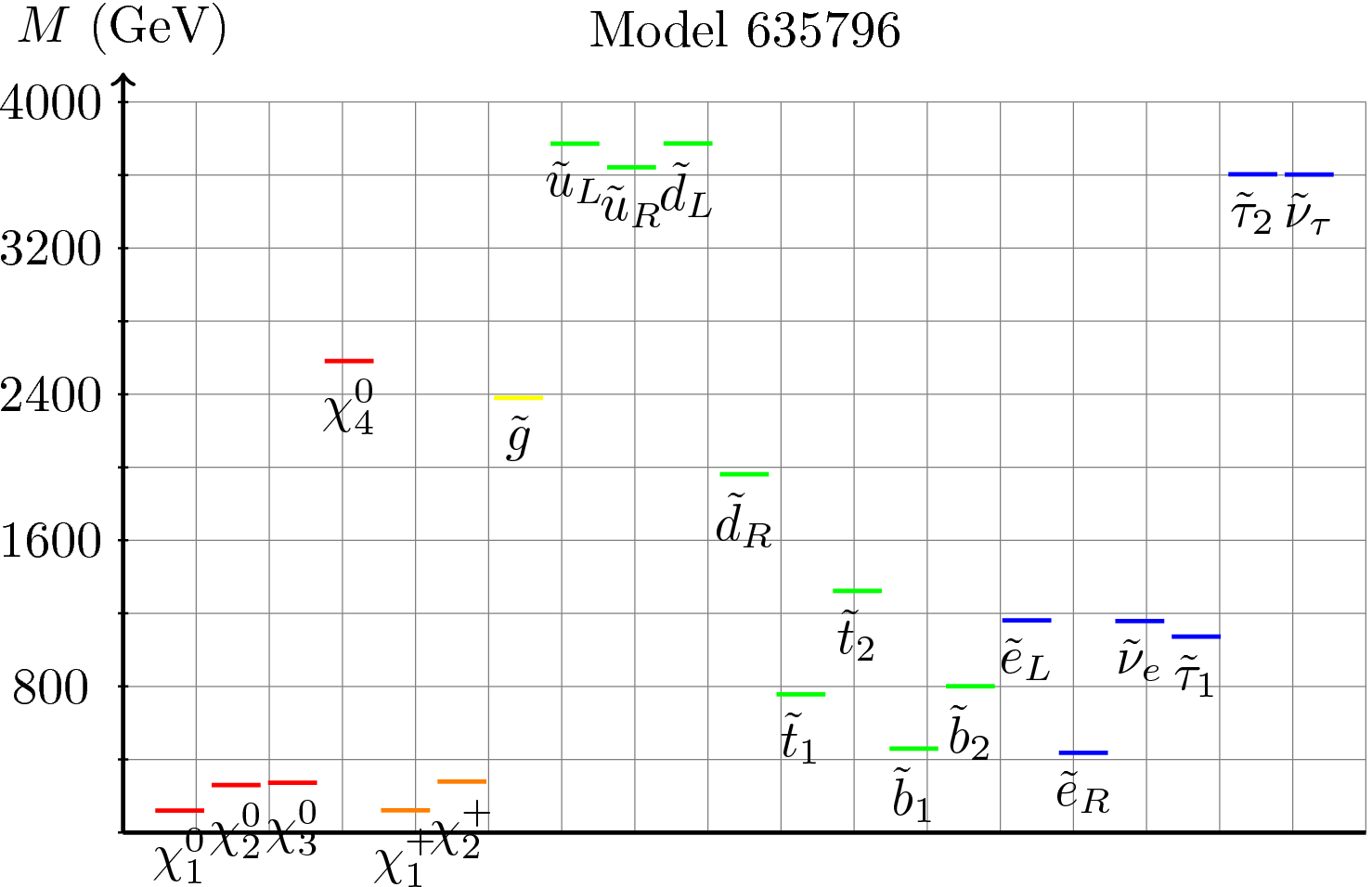}}
\vspace*{0.5cm}
\caption{Sparticle mass spectra of four neutralino LSP pMSSM models which satisfy $m_h=125\pm2$ GeV, $\Delta < 100$, and all current search constraints.}
\label{fig:spec1}
\end{figure}

\begin{figure}
\centering
\subfloat{\includegraphics[width=3.5in]{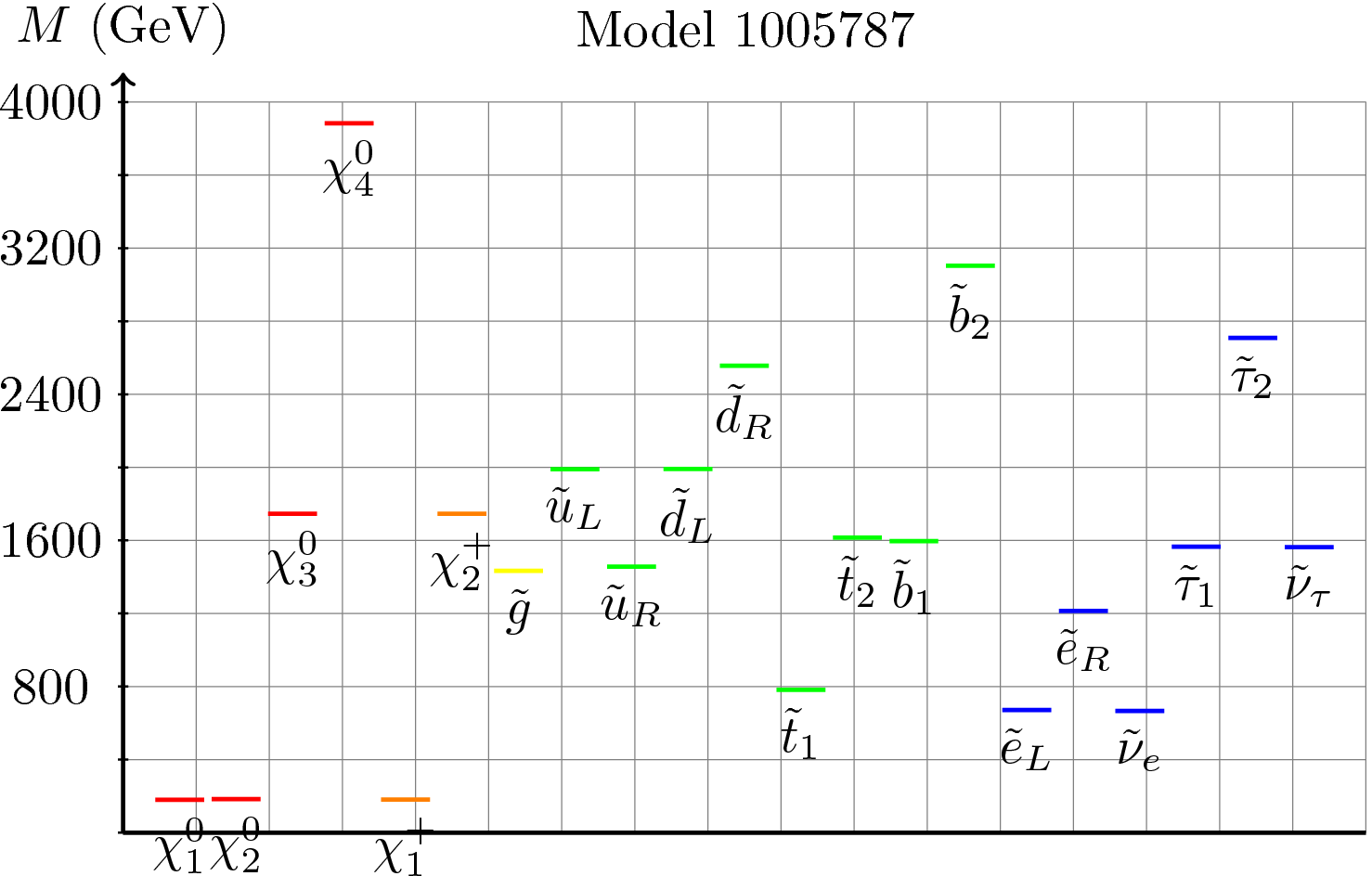}} ~
\subfloat{\includegraphics[width=3.5in]{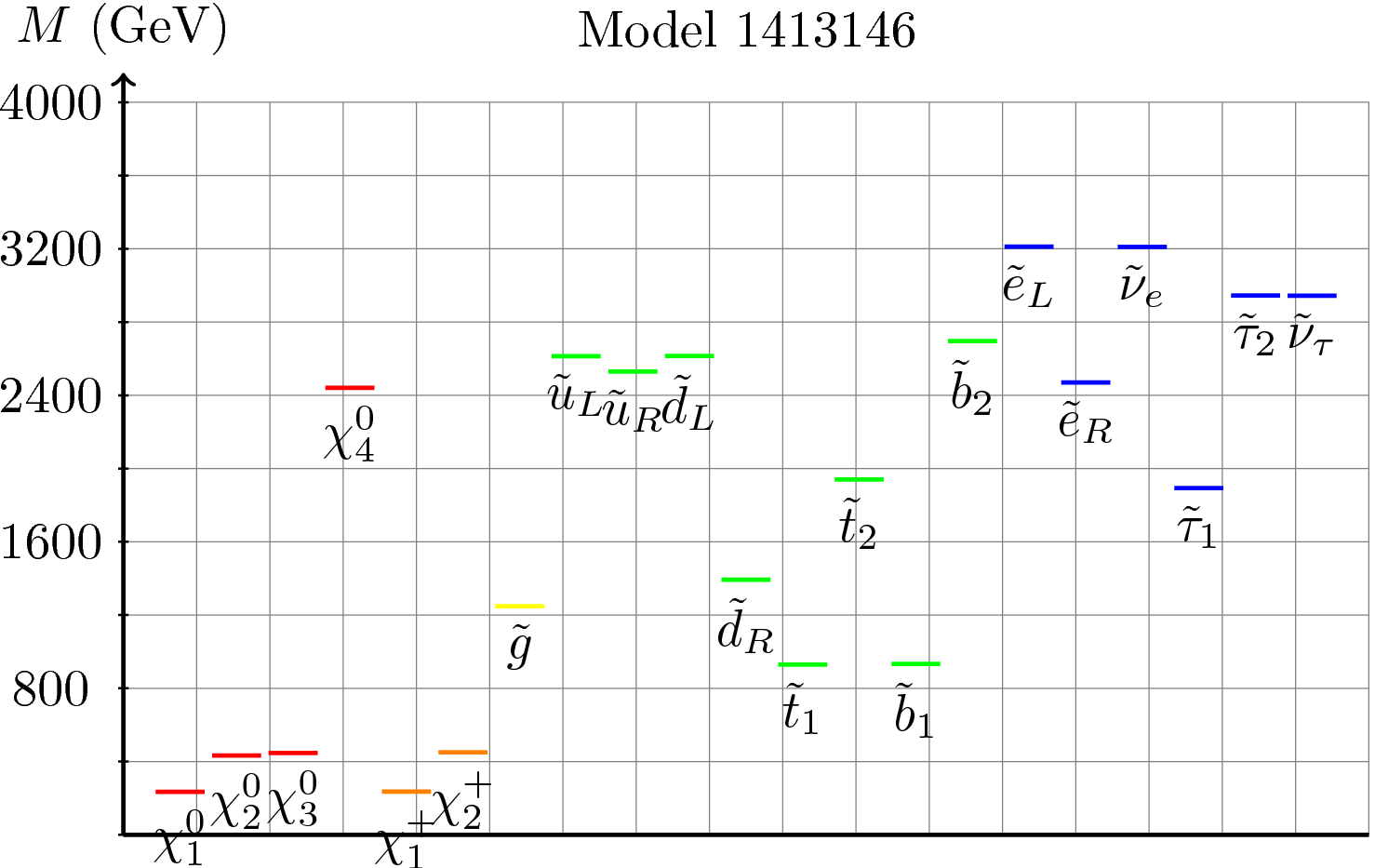}} \\
\subfloat{\includegraphics[width=3.5in]{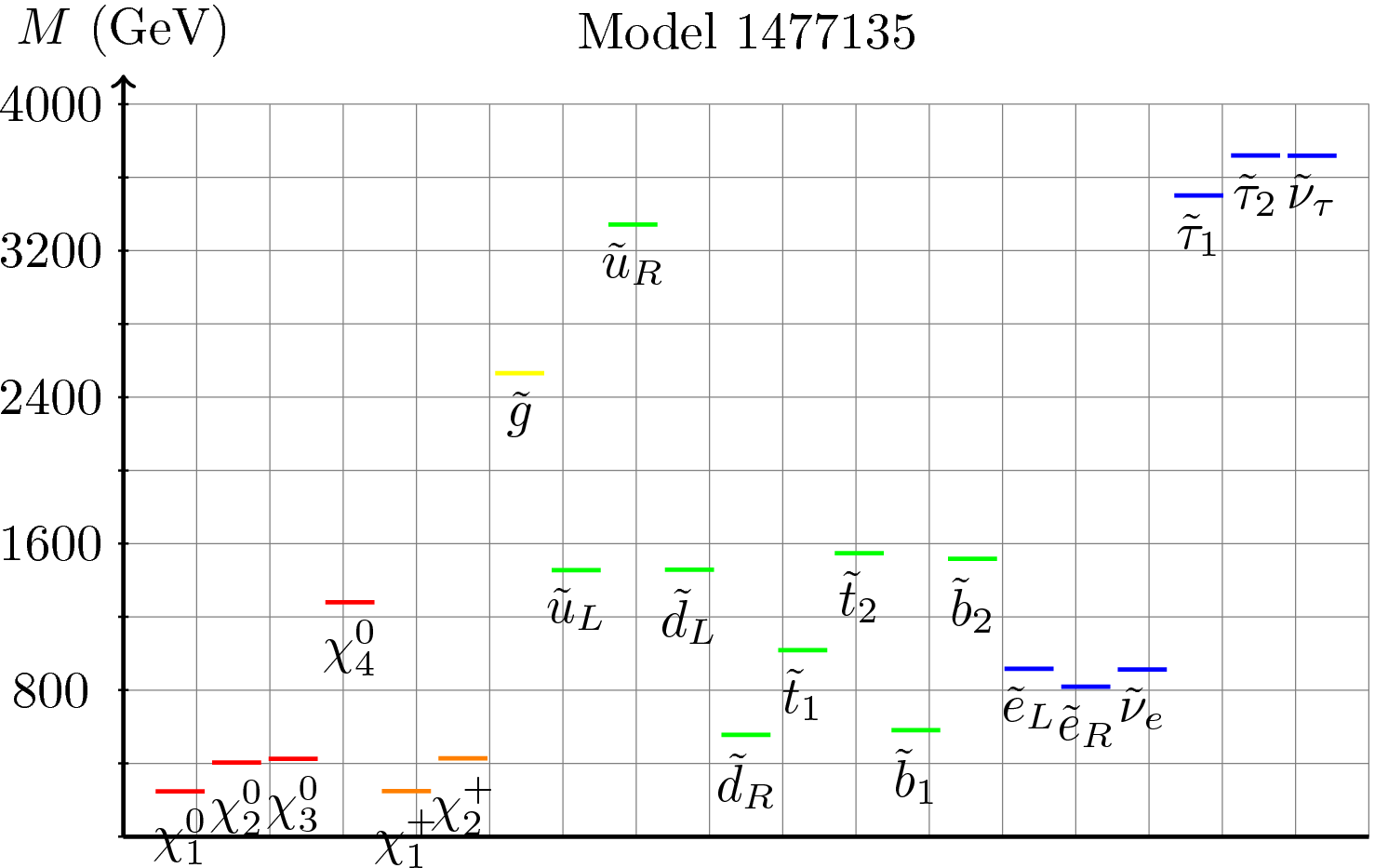}}~
\subfloat{\includegraphics[width=3.5in]{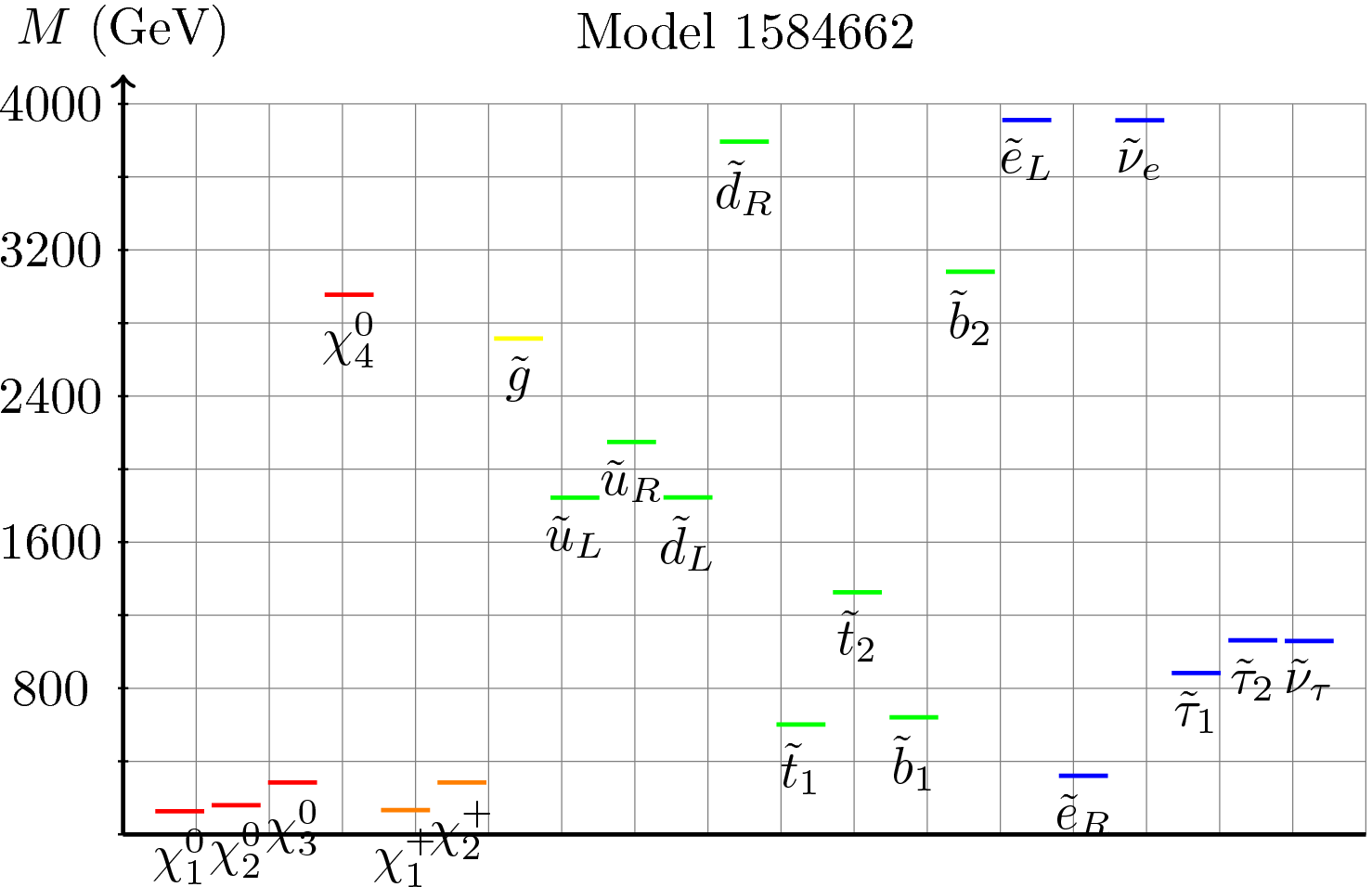}}
\vspace*{0.5cm}
\caption{Same as the previous Figure.}
\label{fig:spec2}
\end{figure}

\begin{figure}
\centering
\subfloat{\includegraphics[width=3.5in]{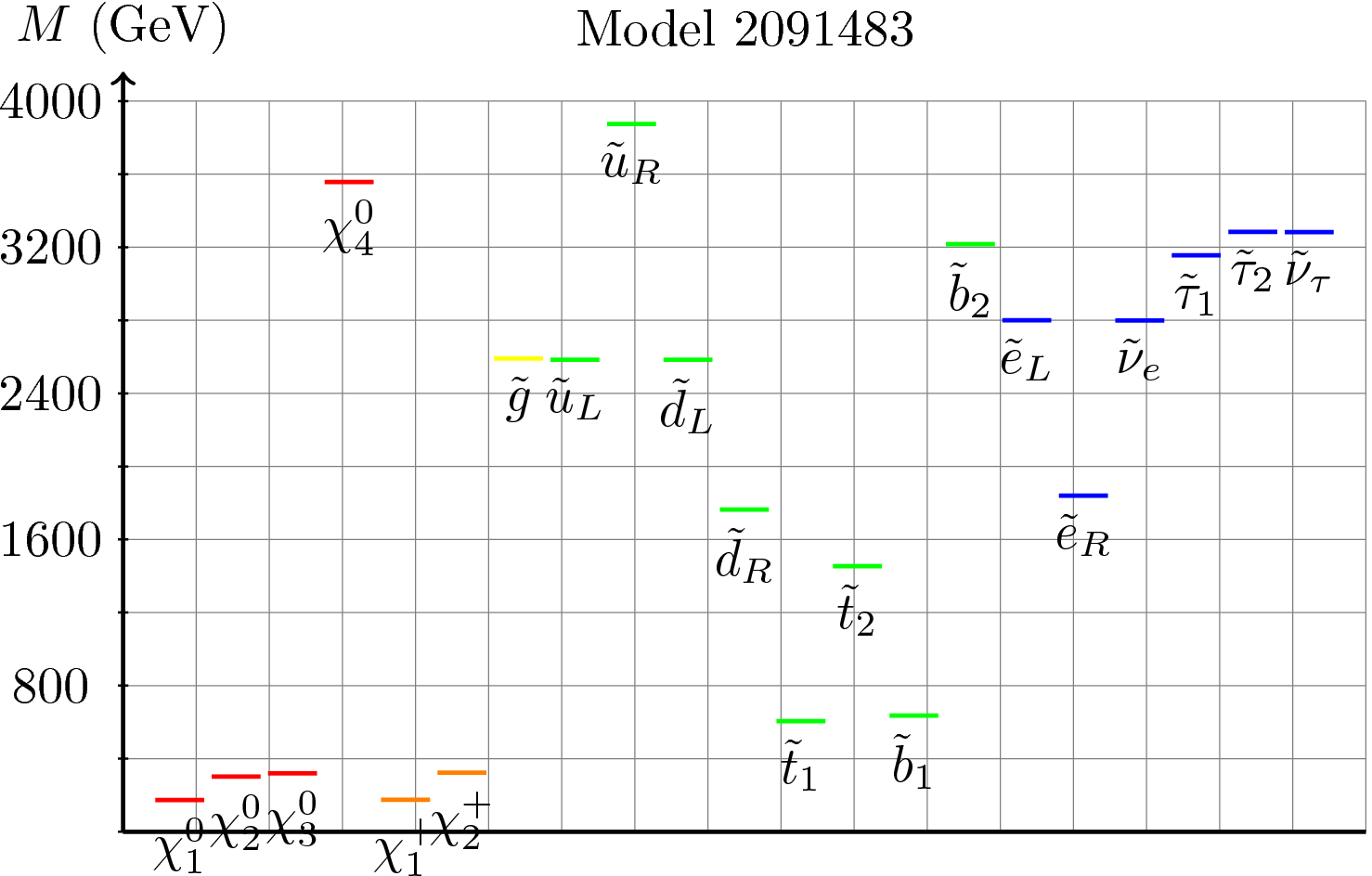}} ~
\subfloat{\includegraphics[width=3.5in]{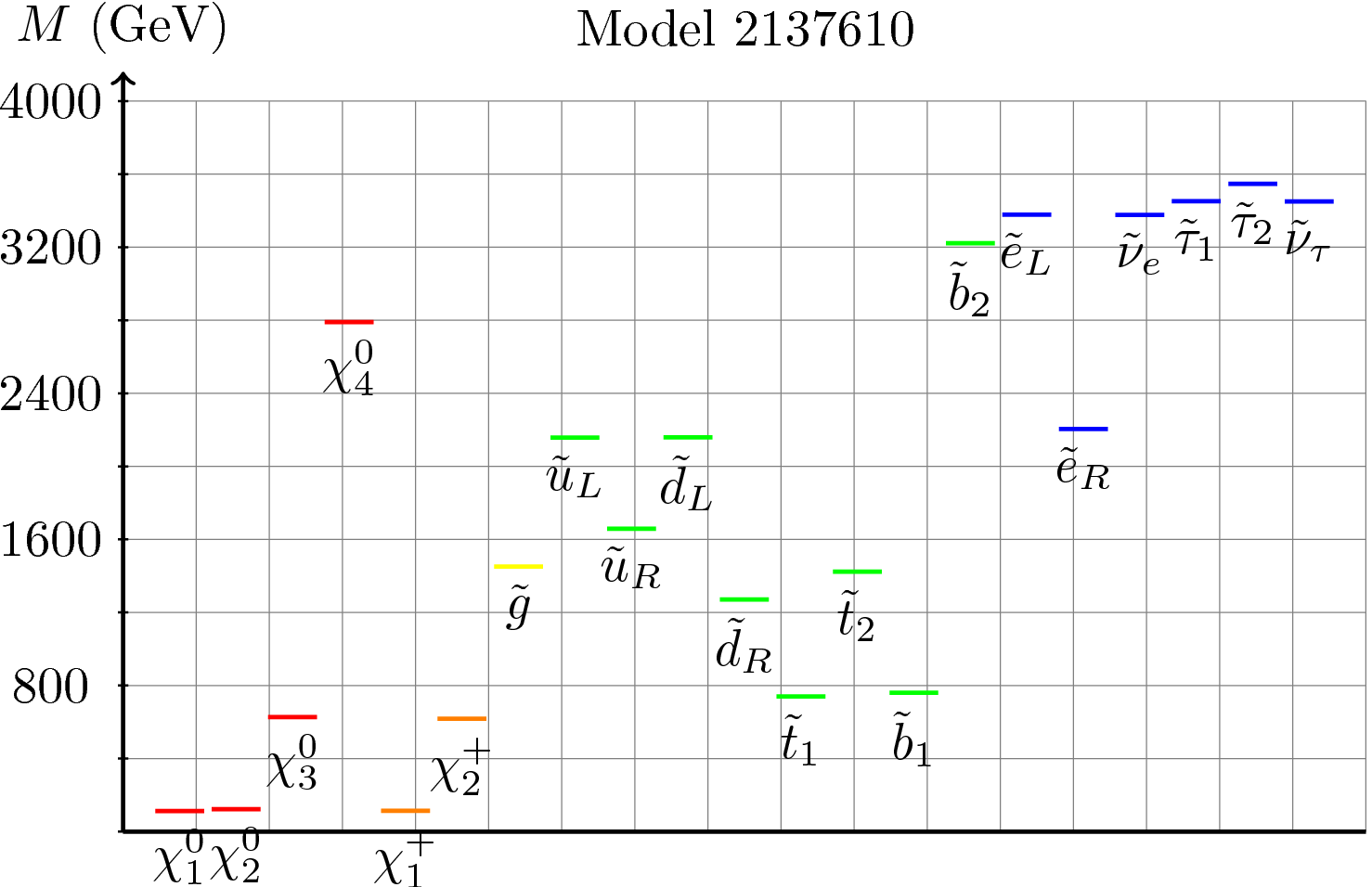}} \\
\subfloat{\includegraphics[width=3.5in]{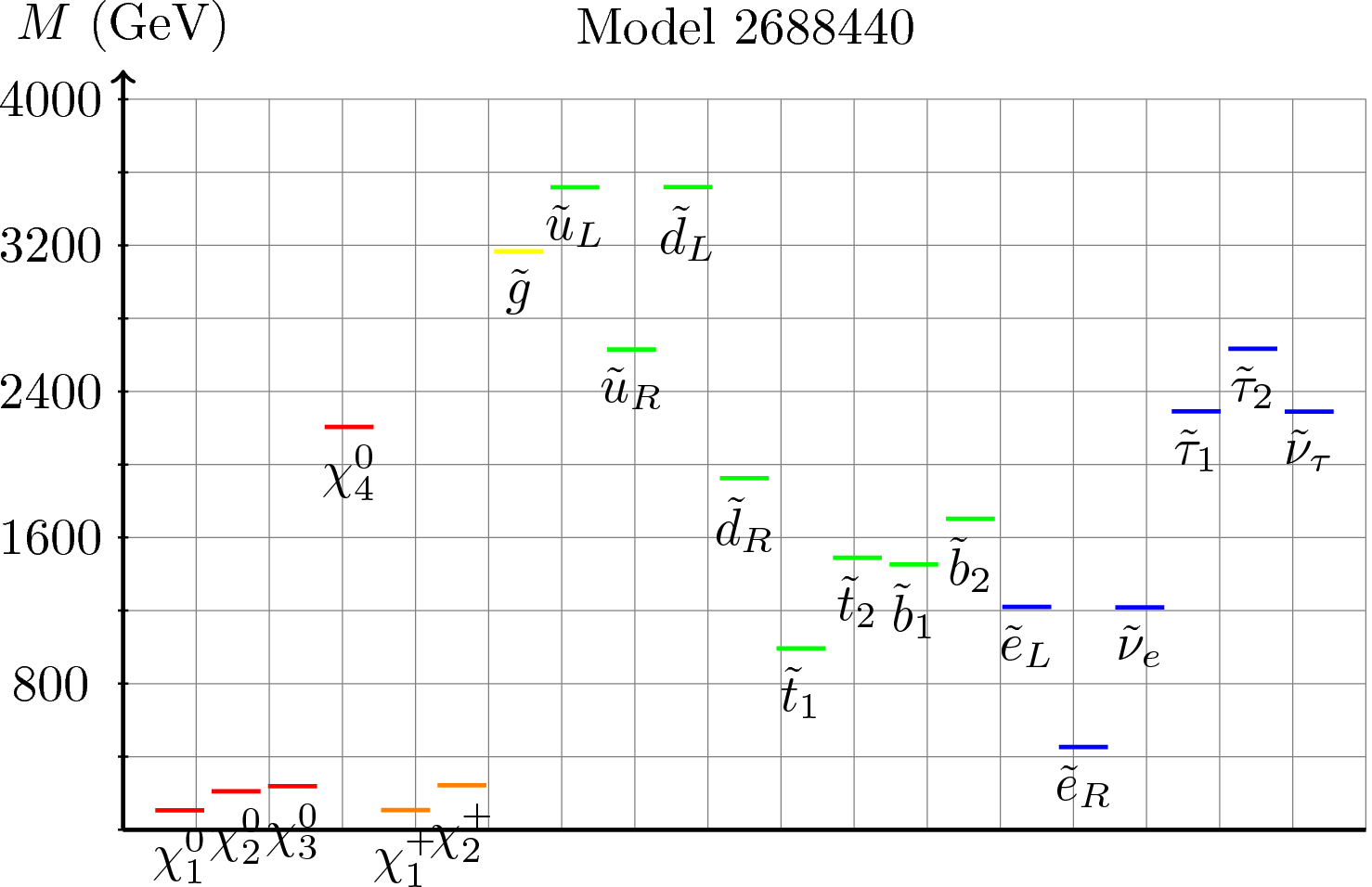}}~
\subfloat{\includegraphics[width=3.5in]{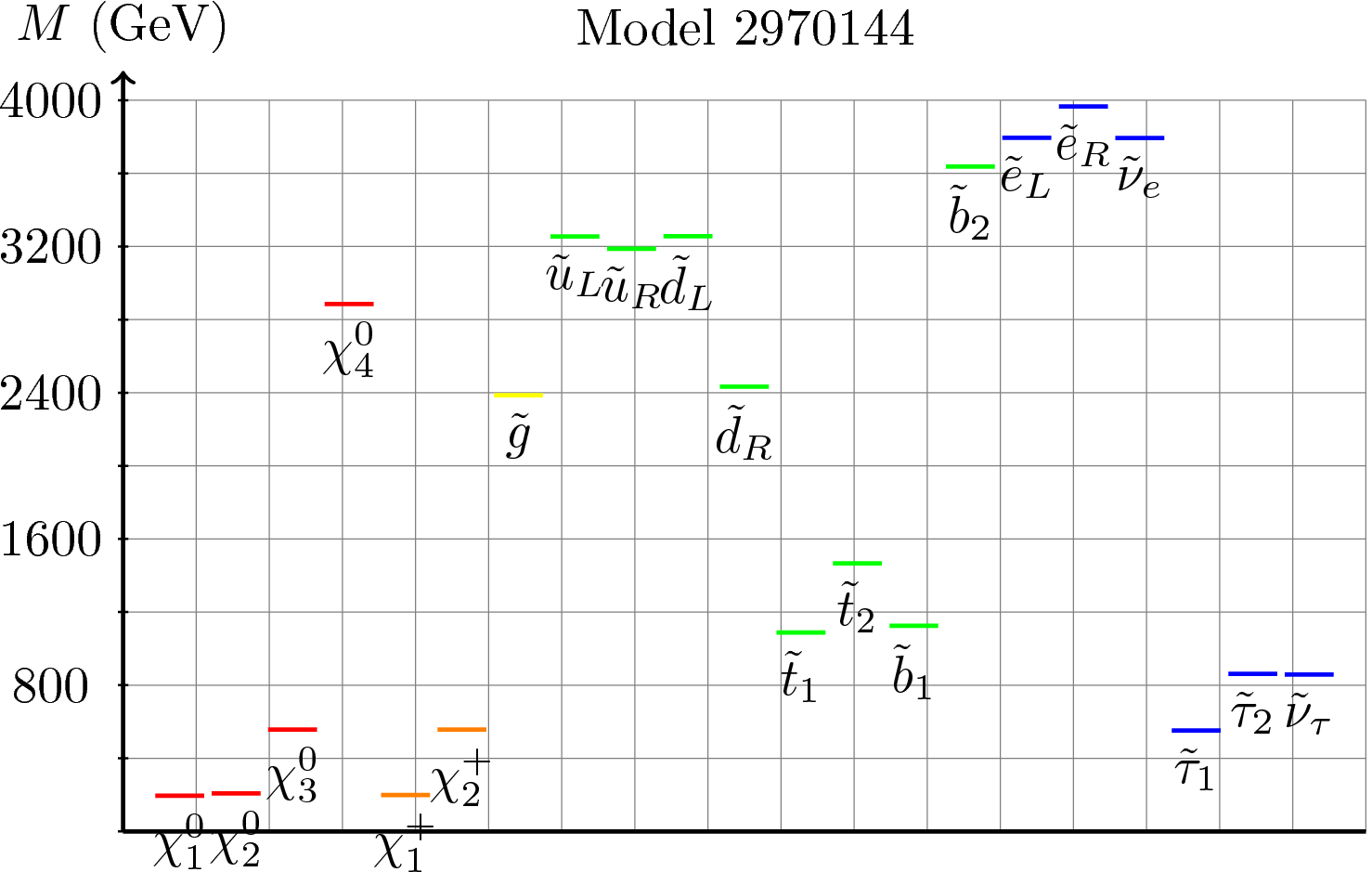}}
\vspace*{0.5cm}
\caption{Same as the previous Figure.}
\label{fig:spec3}
\end{figure}

The decay patterns of a typical light stop and sbottom are displayed in Figs.~\ref{fig:stops1} and ~\ref{fig:sbottoms1}, respectively, for the prototypical model 2403883. In this model, both the Higgsinos and winos (and thus 2 charginos and 3 neutralinos) are lighter than the stop, which is itself split from the heavier sbottom by roughly the $W$ mass. The various gaugino states are mixed to a substantial extent, such that the LSP is a wino-Higgsino admixture. As noted above, the presence of {\it both} light Higgsinos and light winos is unsurprising since $\mu$ must be very small and small values of $M_2$ are favored by fine-tuning and $b \to s \gamma$. The plethora of electroweak gauginos below the stop and sbottom masses leads to a large multiplicity of channels by which both the stop and sbottom can decay, eventually producing the LSP at the end of the decay chain. 

To understand the rates for the various decay paths in these Figures, let us momentarily focus on the light stop decays shown in Fig.~\ref{fig:stops1}.
The uppermost path in this figure shows the decay $\tilde t_1 \to b\tilde \chi_2^+$ with a branching fraction of $\simeq 24\%$. The 258 GeV $\tilde \chi_2^+$ can then decay to $W^+\tilde \chi_2^0(\tilde \chi_1^0)$ with a branching fraction of $\simeq 23~(38)\%$, or to $\tilde \chi_1^+Z(h)$ with a branching fraction of $\simeq 29~(10)\%$. If the $\tilde \chi_2^0$ is produced from the stop decay, it can then further decay to $W^{\pm^*} \tilde \chi_1^{\mp}$ 
with a branching fraction of $59\%$, or to $Z^*(\gamma) \tilde \chi_1^0$ with a branching fraction of $37~(4)\%$. Note that a $W^*(Z^*)$ indicates that the $W(Z)$ in a given decay is virtual since the relevant mass 
splitting is below the required $\simeq 80~(91)$ GeV. As shown in Figs.~\ref{fig:stops1} and ~\ref{fig:sbottoms1}, the various decay channels for the $\tilde t_1$ and $\tilde b_1$ have qualitatively similar branching fractions, meaning that no single mode is strongly dominant.  Therefore substantial branching fraction penalties will apply to any given LHC SUSY search channel. In particular, the typical LHC searches for $\tilde t_1 \tilde t_1^* \to t \bar t + \mathrm{MET}$ would face a branching fraction penalty of $\sim 6.3\%$, while the corresponding $\tilde b_1 \tilde b_1^*\to b \bar b + \mathrm{MET}$ search would face
a similar branching fraction penalty of $3.6\%$. Since the standard simplified treatments of third generation squark searches typically assume branching fractions of $100\%$, these branching fraction penalties will seriously degrade the reach of any one particular channel. Covering all of the various possible decay paths shown in these figures to ensure discovery would therefore require combining a large set of individual search channels at the LHC.

\begin{figure}
\centering
\includegraphics[width=6.5in]{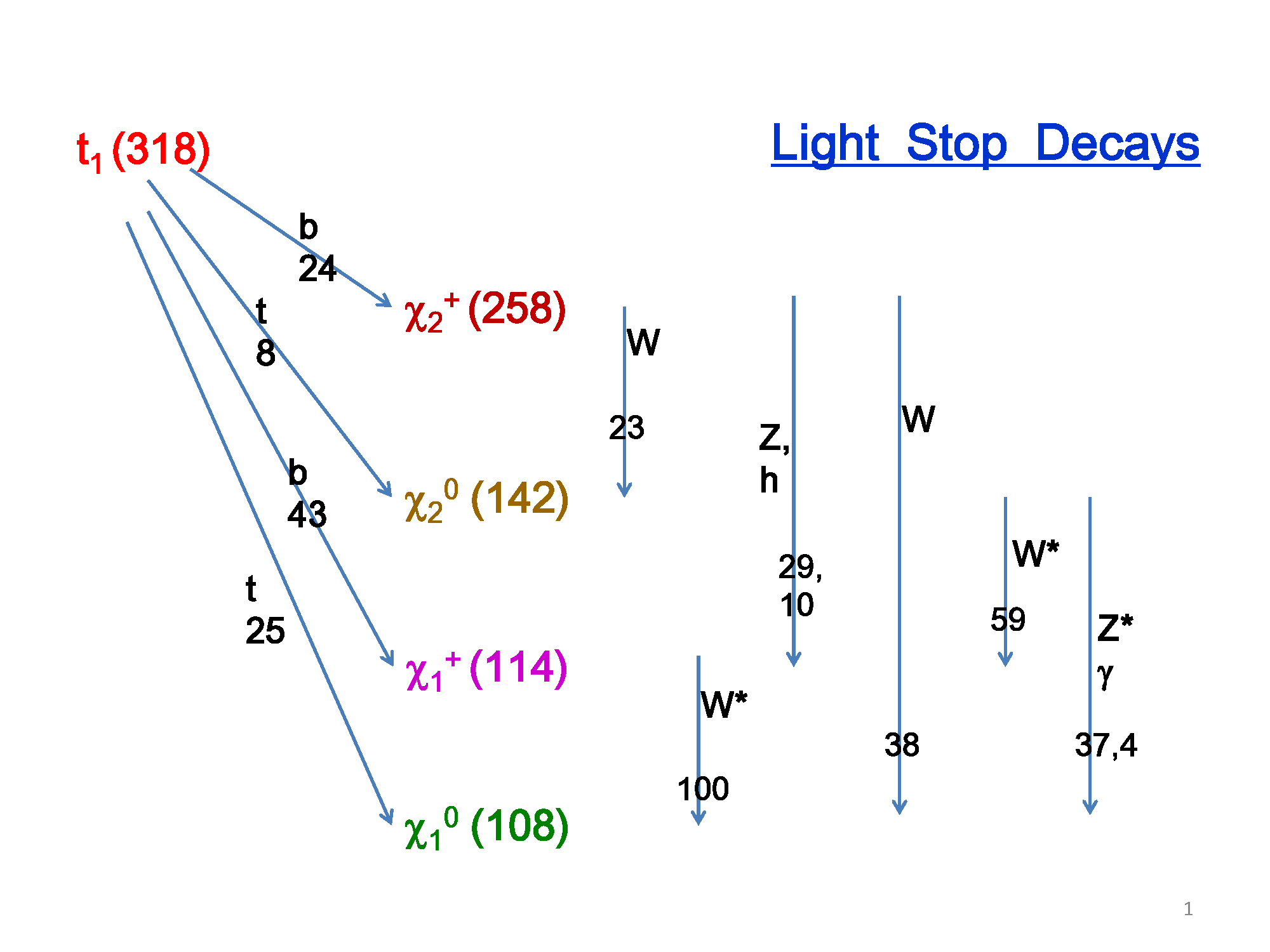}
\vspace*{0.5cm}
\caption{Sample light stop decay pattern for model 2403883. The numbers in parentheses label the sparticle masses and the other numbers indicate the branching fractions in percent 
for the various decay paths as described in the text.}
\label{fig:stops1}
\end{figure}

Naturally, these numerical details are quite specific to this particular model (2403883). If we were to consider the other 11 models with both Higgsinos and winos below the light stop/sbottom, we would find them to have qualitatively similar decay chains but different branching fractions. This is unsurprising since these branching fractions depend upon both the details of the sparticle mass splittings, via the obvious kinematic factors, and also on the gaugino mixing parameters, which strongly affect the third generation sfermion/fermion couplings. However, we generally find that no single channel is strongly dominant and therefore that a variety of decay patterns will be important for LHC searches. The same basic pattern is also observed to apply to the single surviving gravitino LSP model with $\Delta < 100$, as shown in Fig.~\ref{fig:stops1g}.

\begin{figure}
\centering
\includegraphics[width=6.5in]{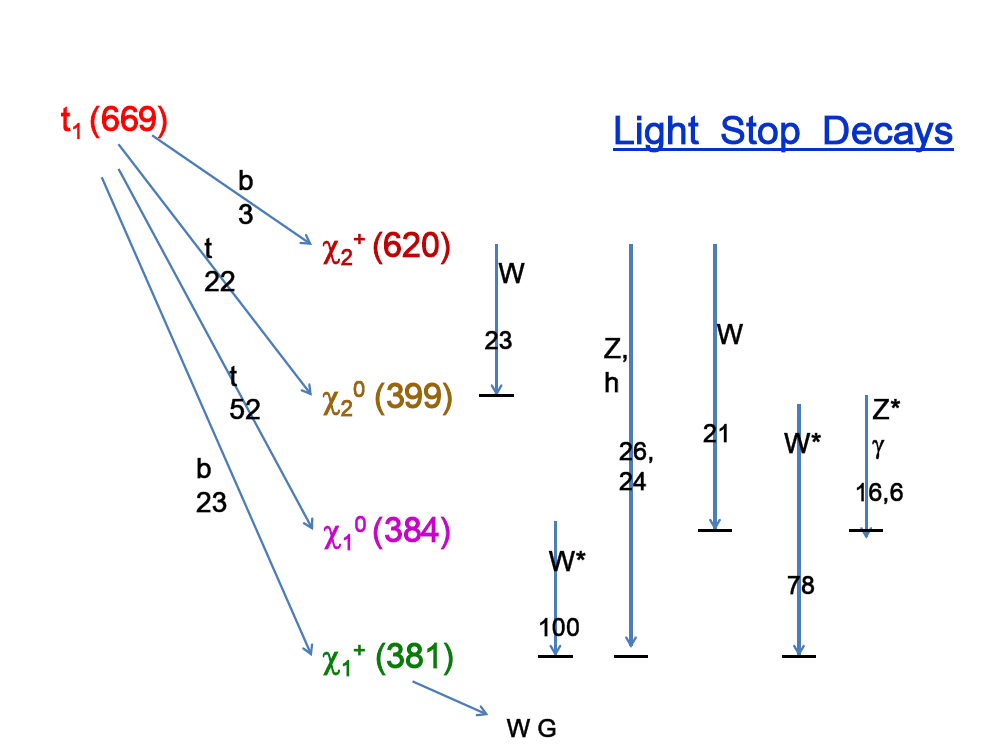}
\vspace*{0.5cm}
\caption{Same as the previous Figure, but now for model 439032 in the gravitino LSP model set.}
\label{fig:stops1g}
\end{figure}

In a single model (1005787, shown in Fig.~\ref{fig:spec2}), the decays of the stops and sbottoms are simplified since only the Higgsinos lie below the lightest stop/sbottom. As a result, only  
the decay modes $\tilde t_1 \to t \tilde \chi_{1,2}^0$ (with a branching fraction of $23~(25)\%$) and $\tilde t_1 \to b\tilde \chi_1^+$ (with a branching fraction of 53\%), and the small mass splittings among the Higgsino states 
(below $\simeq$ a few GeV), are relevant. In this model, $\tilde \chi_1^+$ decays to the LSP plus an off-shell $W^*$, while $\tilde \chi_2^0$ decays to $\tilde \chi_1^0$ via $W^*, Z^*$ and on-shell photons.

Due to the low statistics associated with the $\Delta < 100$ model sample, it is instructive to loosen this restriction to the bound of $\Delta < 120$ to see whether the patterns observed above continue to hold for a larger sample. In this case there are 50 (5) neutralino (gravitino) LSP models. Almost all of the 50 neutralino LSP models have at least 5 electroweak gauginos below the $\tilde{t}_1$/$\tilde{b}_1$; in 4 cases all 6 electroweak gauginos are light with the heaviest one being nearly pure bino. The five gravitino LSP models also have either 5 or 6 electroweak gauginos below the $\tilde{t}_1$/$\tilde{b}_1$, in addition to the gravitino itself. Considering the impact of LHC searches \cite{CahillRowley:2012cb}, 15 of the 50 neutralino LSP models are excluded by the ATLAS MET searches at 7 TeV, and 2 of the remaining models are excluded by the latest limit on $B_s \to \mu^+ \mu^-$~\cite{update}. We additionally estimate that $\sim$ 2-3 of the 5 gravitino LSP models will be excluded by collider search constraints; this will be considered in~\cite{future}.

\begin{figure}
\centering
\includegraphics[width=6.5in]{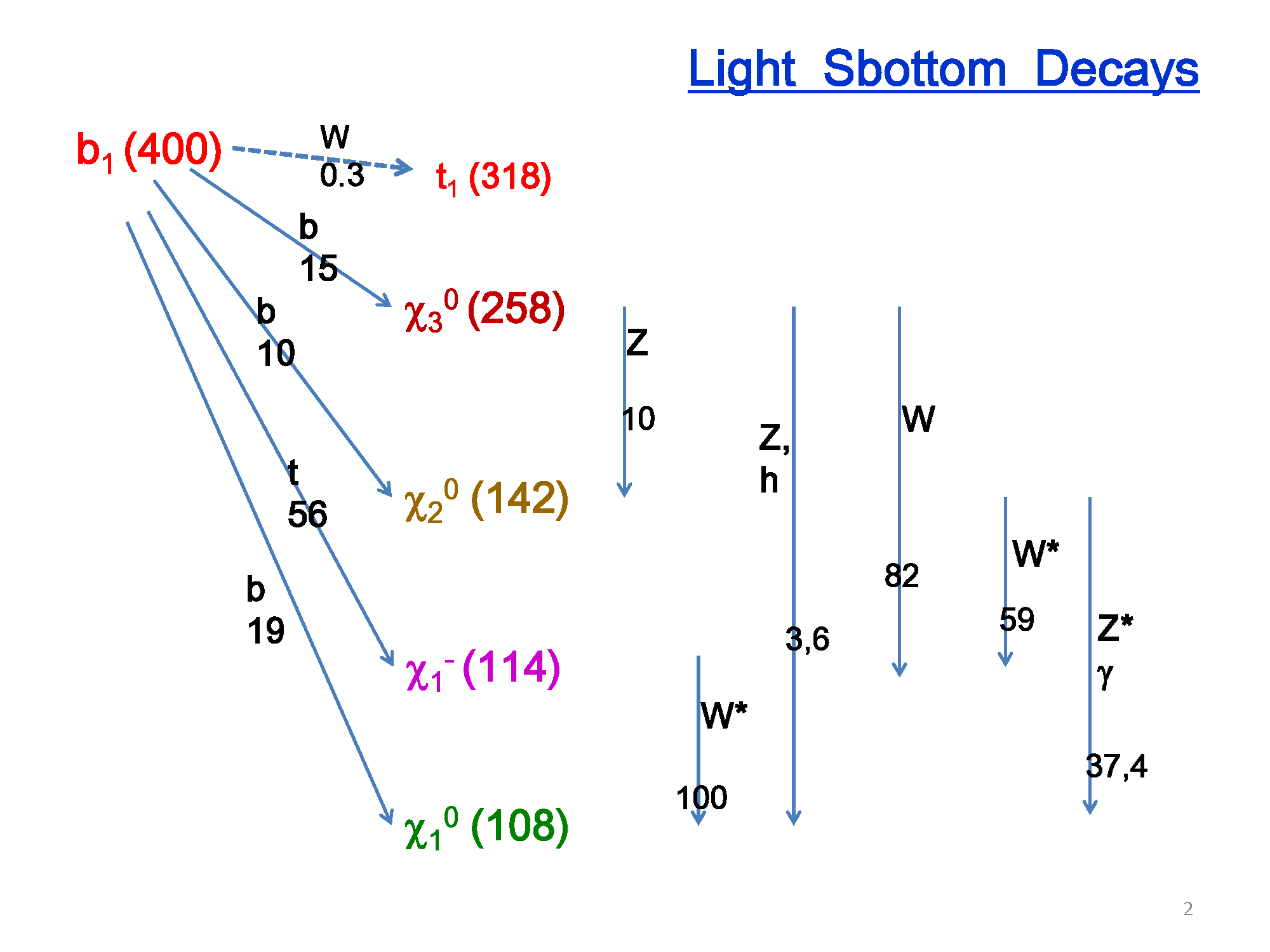}
\vspace*{0.5cm}
\caption{Same as Fig.~\ref{fig:stops1} but now for the light sbottom in the same model (2403883).}
\label{fig:sbottoms1}
\end{figure}

Another window into our low fine-tuning models could come from dark matter direct detection experiments. The LSPs of our 13 low-FT models are all light Higgsinos or Higgsino-wino admixtures, so their dark matter properties are generally similar. Particularly, they tend to have a relatively high spin-independent (SI) DM direct detection cross-section. However, their relic density is predicted to be quite low due to efficient annihilation through a virtual $Z$ boson, making direct detection considerably more difficult and preventing the models from being excluded by current data. Figure~\ref{fig:imps1} displays the LSP relic density vs SI cross-section, and shows that the low-FT models are clustered at high values of the SI cross-section and low values of the relic density. Since constraints from DM direct detection experiments (particular XENON100 and the next generation XENON1T) are expected to improve substantially in the coming years, we can ask whether they will be able to discover or exclude our low-FT models.  Figure~\ref{fig:imps2} shows the SI direct detection cross-section, re-scaled by the ratio of the LSP density to the total DM density, for the low fine-tuning models as a function of the neutralino LSP mass, along with the current~\cite{Aprile:2011hi} and anticipated future~\cite{Xenon1T} limits from XENON. We see that XENON1T is expected to be able to exclude 8 of the 13 low fine-tuning models.
As an aside, we believe it may be possible to create a low-FT model with the correct neutralino LSP relic density by lowering $M_1$ so that the lightest neutralino was mixed or mostly bino. Such a model would have at least 4 and possibly all 6 of the electroweak gauginos below the stop and sbottom masses (with the bino at the bottom of the gaugino spectrum), further increasing the challenges faced by collider searches. 

\begin{figure}
\centering
\includegraphics[width=6.5in]{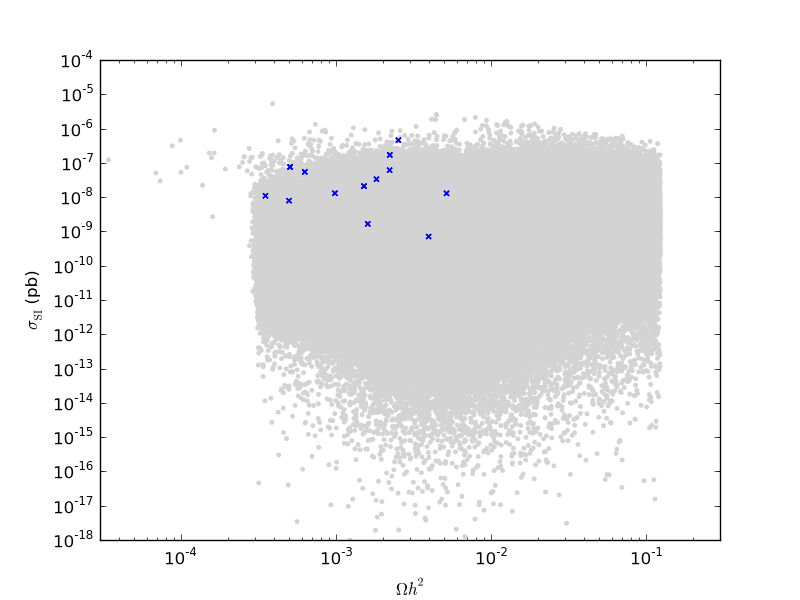}
\vspace*{0.5cm}
\caption{Spin-independent cross section versus LSP relic density for models in the neutralino pMSSM model set (gray dots). The 13 `low-FT' models are indicated by blue crosses.}
\label{fig:imps1}
\end{figure}

\begin{figure}
\centering
\includegraphics[width=6.5in]{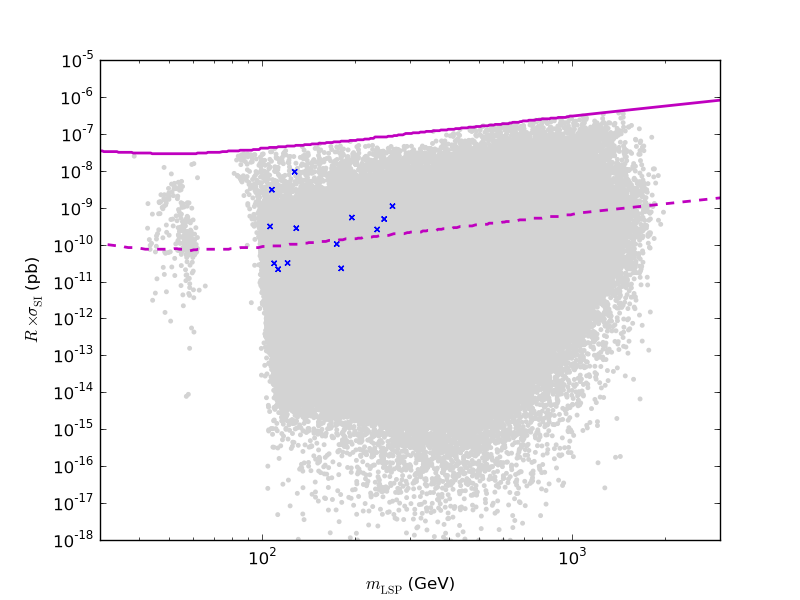}
\vspace*{0.5cm}
\caption{Scaled spin-independent cross section versus LSP mass for models in the neutralino pMSSM model set (gray dots). The 13 `low-FT' models are indicated by blue crosses. The current limit from XENON100 (purple solid line) and the projected limit from XENON1T (purple dashed line) are also shown.}
\label{fig:imps2}
\end{figure}

\section{Discussion and Conclusions}
\label{conc}

Although signals for SUSY have not yet been observed in the LHC data, the discovery of a (broadly speaking) SM-like Higgs boson in the $\sim 125$ GeV mass region has important implications 
for any SUSY scenario. Within the pMSSM, the large amount of parameter freedom allows one to satisfy the LHC SUSY search constraints while also generating the large radiative corrections that are 
necessary to obtain a light $h$ mass in this range. In this paper, we began our analysis by demonstrating that the ease of obtaining such a mass is somewhat sensitive to the choice of the LSP, \ie, whether it is the gravitino or the lightest neutralino. The basic reason for this is that, trivially, the value of the LSP mass provides a lower limit on the masses of all 
the other sparticles which can then enter into the relevant loop corrections. Since gravitino LSPs can be far lighter than neutralino LSPs, gravitino models generally have somewhat lighter sparticle 
spectra the neutralino LSP models. Thus given the scan ranges employed in the generation of our two model sets (which are, in fact, identical except for the gravitino mass itself), we find that 
$\sim 19.4\%$ of the neutralino LSP models lead to a value of $m_h=125\pm 2$ GeV, whereas only $\sim 9.0\%$ of the gravitino LSP models are consistent with this value.

Next we turned to the properties of the light Higgs in both of our pMSSM model sets; though qualitatively similar, the detailed properties of the Higgs were shown to depend on the choice of the LSP. 
Once the relevant Higgs mass range selection of $125 \pm 2$ GeV has been applied to our model set, the most interesting observable is currently $R_{\gamma\gamma}$ (the ratio of rates for the channel $gg\to h\to\gamma\gamma$ in the pMSSM to that of the SM) whose distributions peak near 
unity in both model sets, although the corresponding shape of the distribution of values is found to be quite different. For example, for neutralino (gravitino) LSP models with a Higgs in this mass range, 
23.1\% (5.3\%) predict a value $R_{\gamma\gamma}> 1$. Overall, the various $R_{XX}$ observables, where $XX$ denotes the possible Higgs decay channels, associated with the gluon fusion production of the light Higgs are generally found to be highly correlated in both model sets although, again, the shapes of the distributions of their values differ depending on the nature of the LSP. In almost all 
cases the models in both sets lie within the decoupling regime so that many of the Higgs partial widths are actually within a few percent of their corresponding SM values. The only exception to this is 
the $h\to b\bar b$ partial width, where decoupling can be far slower due to large radiative corrections driven by sbottom mixing. These large non-decoupling effects in the $b\bar b$ final 
state (together with the decoupling elsewhere) are essentially responsible for the nature of our results.  

One important issue associated with a Higgs mass in this range is that of fine-tuning: roughly speaking, stop masses and, more importantly, stop mixing must be large to generate a sufficiently heavy Higgs mass. 
However, if these quantities are too large, they will also generate large values of fine-tuning. Here, we first showed that in both model sets, models with light stops ($\lesssim 300$ GeV) could still achieve $m_h=125\pm 2$ GeV (with or without the additional $R_{\gamma\gamma}> 1$ requirement) as long as the value of $|X_t/M_S|$ was sufficiently large. We then calculated the 
contribution to fine-tuning arising from each of the 19 pMSSM model parameters, in many cases beyond the leading term, via the LL and NLL beta-functions for the general MSSM RGE equations. Some of these 
various contributions are exactly zero in the pMSSM framework, while others are found to be quite small. In general we found that including the NLL fine-tuning contributions (in the cases where they are large) can soften the fine-tuning constraints obtained at LL. Clearly, the value of $|\mu|$ itself is the biggest driver of fine-tuning due to its appearance 
already at tree-level in the usual $Z$-Higgs mass relationship. Following $\mu$, the three weak-scale parameters in the stop mass matrix, $A_t, M_{Q3,u3}$, as well as $M_2$, play the most important roles in 
determining the overall amount of fine-tuning, particularly after the Higgs mass constraint is imposed. Requiring a value of $\Delta < 100~(120)$, together with $m_h=125\pm 2$ GeV, yields only 15 (50) neutralino 
LSP models and only 1 (5) gravitino LSP models. Applying the various SUSY and non-SUSY LHC analyses to the neutralino LSP models reduced the number of surviving models to 13 (33); LHC constraints on the gravitino LSP models are more complicated and will be considered in a future work~\cite{future}. These small numbers are not very surprising since no assumptions about fine-tuning were built into the parameter scan ranges as part of the model generation process.  Nonetheless, it is of interest that there are models contained in our previously generated sets that satisfy all of the data with low fine-tuning.

The 13 low-FT neutralino LSP models are particularly interesting and their characteristics were examined in detail.  They were found to share a number of important features: ($i$) both a light stop and a light sbottom were present in all cases with the 
stop generally being lighter than the sbottom. ($ii$) In 12/13 cases, both the winos as well as the Higgsinos were lighter than the stop. These states were found to be highly mixed. ($iii$) Gluinos in 
these models were found to be moderately heavy while the first and second generation squarks were generally more massive. This same pattern was observed in the larger set of neutralino models with $\Delta < 120$. 
There it was also observed that in $\sim 10\%$ of the cases the bino could also lie below the stop though it was always the heaviest neutralino and not the LSP. ($iv$) A similar mass spectrum was also 
observed for the 5 surviving gravitino models with $\Delta < 120$. These mass patterns have important implications for light stop and sbottom searches at the LHC. Since so many electroweak gauginos 
are lighter than the lightest stop and sbottom, these squarks will have rather complex decay patterns and will likely not be amenable to simplified model treatments. Conventional LHC stop and sbottom searches usually 
assume that one decay mode is dominant, \eg, $\tilde t(\tilde b)_1\to t(b)\tilde \chi_1^0$ and thus could miss models with these features. Although such simple decays do occur in our models, there is usually a substantial branching fraction price to pay for this channel. More than likely, both  
stops and sbottoms will cascade decay down to the LSP via a number of intermediate chargino and neutralino states with a substantial range of possible branching fractions. Clearly all of these channels need 
to be investigated at the LHC in order to fully cover this general scenario. 

In summary, we have identified a corner of parameter space in the pMSSM that is consistent with a $125\pm2$ GeV SM-like Higgs boson, has a low amount of fine-tuning of order 1\%,
and has managed to escape detection (thus far) at the LHC.  These natural models contain light stop and sbottom squarks that have complicated decay chains.  In light of the LHC results, this
is a very attractive scenario that should be pursued further.

With the discovery of a Higgs-like particle at 125 GeV, hopefully the appearance of the 3$^{rd}$ generation superpartners at the 8 TeV run of the LHC is not too far away.

\section*{Acknowledgments}

The authors are grateful for discussions with H. Haber.

\appendix
\section*{Appendix}
\label{sec:appendix}

This Appendix contains the expressions for all of the $LL$ and $NLL$ contributions to the $Z_i$ parameters not provided in the main text.

The LL contribution to FT from the parameter $\mu$ is given by

\begin{equation}
Z_\mu^{LL}=Z_\mu^{TL}\left[1+{X\over 16\pi^2}(3y_t^2+3y_b^2+y_\tau^2-3g^2-g_Y^2)\right] \,.
\end{equation}
The LL contributions arising from $M_{L3}$ and $M_{e3}$ are given by 
\begin{equation}
Z_{M_{L3}}^{LL}={X\over 4\pi^2M_Z^2}{M_{L3}^2\over t_\beta^2-1}[2y_\tau^2+g_Y^2(1+t_\beta^2)]\,,
\end{equation}
and
\begin{equation}
Z_{M_{e3}}^{LL}={X\over 4\pi^2M_Z^2}{M_{e3}^2\over t_\beta^2-1}[2y_\tau^2-g_Y^2(1+t_\beta^2)]\,,
\end{equation}
whereas that for $M_{d3}$ is given by
\begin{equation}
Z_{M_{d3}}^{LL}={X\over 4\pi^2M_Z^2}{M_{d3}^2\over t_\beta^2-1}[6y_b^2-g_Y^2]\,.
\end{equation}
At LL the contribution from $M_{Q3}$ is given by
\begin{equation}
Z_{M_{Q3}}^{LL}={X\over 4\pi^2M_Z^2}{M_{Q3}^2\over t_\beta^2-1}\left[12y_b^2-2g_Y^2-(12y_t^2+2g_Y^2)t_\beta^2\right]\,,
\end{equation}
whereas at NLL we obtain
\begin{equation}
Z_{M_{Q3}}^{NLL}={X^2\over 128\pi^4M_Z^2}{M_{Q3}^2\over t_\beta^2-1}[C_1-C_2t_\beta^2]\,,
\end{equation}
where we have defined
\begin{eqnarray}
C_1 & \equiv & 2y_b^2(32g_s^2-4g_Y^2/3-36y_b^2-12y_t^2)\,,\nonumber\\
C_2 & \equiv & 2y_t^2(32g_s^2+8g_Y^2/3-36y_t^2-12y_b^2)\,,.
\end{eqnarray}
Similarly, for $M_{u3}$ at LL we find
\begin{equation}
Z_{M_{u3}}^{LL}={X\over 8\pi^2M_Z^2}{M_{u3}^2\over t_\beta^2-1}[4g_Y^2-(12y_t^2-4g_Y^2)t_\beta^2]\,,
\end{equation}
while at NLL the corresponding expression is given by
\begin{equation}
Z_{M_{u3}}^{NLL}={X^2\over 128\pi^4M_Z^2}{M_{u3}^2\over t_\beta^2-1}[-12y_b^2y_t^2-2y_t^2(32g_s^2+8g_Y^2/3-36y_t^2-6y_b^2)t_\beta^2]\,.
\end{equation}

\newpage

%
\def\IJMP #1 #2 #3 {Int. J. Mod. Phys. A {\bf#1},\ #2 (#3)}
\def\MPL #1 #2 #3 {Mod. Phys. Lett. A {\bf#1},\ #2 (#3)}
\def\NPB #1 #2 #3 {Nucl. Phys. {\bf#1},\ #2 (#3)}
\def\PLBold #1 #2 #3 {Phys. Lett. {\bf#1},\ #2 (#3)}
\def\PLB #1 #2 #3 {Phys. Lett. B {\bf#1},\ #2 (#3)}
\def\PR #1 #2 #3 {Phys. Rep. {\bf#1},\ #2 (#3)}
\def\PRD #1 #2 #3 {Phys. Rev. D {\bf#1},\ #2 (#3)}
\def\PRL #1 #2 #3 {Phys. Rev. Lett. {\bf#1},\ #2 (#3)}
\def\PTT #1 #2 #3 {Prog. Theor. Phys. {\bf#1},\ #2 (#3)}
\def\RMP #1 #2 #3 {Rev. Mod. Phys. {\bf#1},\ #2 (#3)}
\def\ZPC #1 #2 #3 {Z. Phys. C {\bf#1},\ #2 (#3)}

\end{document}